\documentclass{amsart}

\newif\ifdraft
\draftfalse

\usepackage[margin = 0.85in]{geometry}
\usepackage[titletoc,title]{appendix}
\usepackage{algorithm}
\usepackage{algpseudocode}
\usepackage{booktabs}
\usepackage{multirow}
\usepackage{amsmath,amssymb,amsfonts,amsaddr}
\usepackage{tikz}
\usepackage{pgfplots}
\usepackage{pdfpages}
\usepackage{ifthen}
\usepackage{etoolbox}
\usepackage{url}

\usepackage{color}
\definecolor{myred}{rgb}{0.85, 0.0, 0.09}
\definecolor{mygreen}{rgb}{0.17, 0.57, 0.13}
\definecolor{myblue}{rgb}{0.11, 0.38, 0.65}
\newcommand{\cred}[1]{\ifdraft {\bf \color{myred}{#1}} \else #1 \fi}
\newcommand{\cgreen}[1]{\ifdraft {\bf \color{mygreen}{#1}} \else #1 \fi}
\newcommand{\cblue}[1]{\ifdraft {\bf \color{myblue}{#1}} \else #1 \fi}

\usepackage{listings}
\usepackage[pdftex]{hyperref}

\definecolor{mGreen}{rgb}{0,0.6,0}
\definecolor{mGray}{rgb}{0.5,0.5,0.5}
\definecolor{mPurple}{rgb}{0.58,0,0.82}
\definecolor{backgroundColour}{rgb}{0.95,0.95,0.92}
\definecolor{code}{rgb}{0.7, 0, 0.4}
\definecolor{todo}{rgb}{0.83,0.27,0.25}
\definecolor{info}{rgb}{0.25,0.74,0.86}
\definecolor{c1}{rgb}{0.28,0.52,0.82}
\definecolor{c2}{rgb}{0.64,0.80,0.31}
\definecolor{c3}{rgb}{0.25,0.74,0.86}
\definecolor{c4}{rgb}{0.52,0.38,0.69}
\definecolor{c5}{rgb}{0.52,0.38,0.69}

\newcommand{\tbl}[2]{\centering{#2}\caption{#1}}
\newcommand{\fig}[2]{\centering{#2}\caption{#1}}
\newcommand{\code}[1]{\texttt{\small\color{code} #1}}

\newcommand{\Hcurl}{\ensuremath{H(\mathrm{curl})}}
\newcommand{\Hdiv}{\ensuremath{H(\mathrm{div})}}
\newcommand{\elt}{K}

\begin{document}

  \title{MFEM: A Modular Finite Element Methods Library}

  \author{Robert Anderson, Julian Andrej, Andrew Barker, Jamie Bramwell, Jean-    Sylvain Camier, Jakub Cerveny, Veselin Dobrev, Yohann Dudouit, Aaron Fisher,    Tzanio Kolev, Will Pazner, Mark Stowell, Vladimir Tomov}
  \address{
    Lawrence Livermore National Laboratory, Livermore, USA
  }

  \author{Ido Akkerman}
  \address{Delft University of Technology, Netherlands}

  \author{Johann Dahm}
  \address{IBM Research -- Almaden, Almaden, USA}
  \author{David Medina}
  \address{Occalytics, LLC, Houston, USA}
  \author{Stefano Zampini}
  \address{King Abdullah University of Science and Technology, Thuwal, Saudi Arabia}

\title{MFEM: A Modular Finite Element Methods Library}

\begin{abstract}
MFEM is an open-source, lightweight, flexible and scalable C++ library for modular
finite element methods that features arbitrary high-order finite element meshes
and spaces, support for a wide variety of discretization approaches and emphasis
on usability, portability, and high-performance computing efficiency.  MFEM's
goal is to provide application scientists with access to cutting-edge
algorithms for high-order finite element meshing, discretizations and linear
solvers, while enabling researchers to quickly and easily develop and test
new algorithms in very general, fully unstructured, high-order, parallel and
GPU-accelerated settings.  In this paper we describe the underlying algorithms
and finite element abstractions provided by MFEM, discuss the software
implementation, and illustrate various applications of the library.
\end{abstract}

\maketitle

\section{Introduction} \label{sec:intro}
The Finite Element Method (FEM) is a powerful discretization technique that uses
general unstructured grids to approximate the solutions of many partial
differential equations (PDEs). It has been exhaustively studied, both
theoretically and in practice, in the past several decades \cite{ciarlet78,
MR911477, hpbook-dem, Brenner2008, hpbook-solin, fischerbook, DG, nurbsbook}.

\textit{MFEM} is an open-source, lightweight, modular and scalable software
library for finite elements, featuring arbitrary high-order finite element
meshes and spaces, support for a wide variety of discretization approaches and
emphasis on usability, portability, and high-performance computing (HPC)
efficiency \cite{mfem}. The MFEM project performs mathematical research and
software development that aims to enable application scientists to take
advantage of cutting-edge algorithms for high-order finite element meshing,
discretizations, and linear solvers. MFEM also enables researchers and
computational mathematicians to quickly and easily develop and test new research
algorithms in very general, fully unstructured, high-order, parallel settings.
The MFEM source code is freely available via Spack, OpenHPC, and GitHub,
\url{https://github.com/mfem}, under the open source BSD license.

In this paper we provide an overview of some of the key mathematical algorithms
and software design choices that have enabled MFEM to be widely applicable and
highly performant, while retaining relatively small and lightweight code base
(see \autoref{sec:hpc} and \autoref{sec:apps}). MFEM's main capabilities and their
corresponding sections in the paper are outlined in the following text.

MFEM is distinguished from other finite element packages, such as deal.II
\cite{dealII85}, FEniCS \cite{fenics2015}, DUNE \cite{dune2016}, FreeFem++
\cite{freefem2012}, Hermes \cite{hermes2008}, libMesh \cite{libMesh2006}, FETK
\cite{fetk2001}, NGSolve \cite{ngsolve}, etc., by a unique combination of features, including its
massively parallel scalability, HPC efficiency, support for arbitrary high-order
finite elements, generality in mesh type and discretization methods, support for
GPU acceleration and the focus on maintaining a clean, lightweight code base.
The continued development of MFEM is motivated by close work with a variety of
researchers and application scientists. The wide applicability of the library is
illustrated by the fact that in recent years it has
been cited in journal articles, conference papers, and preprints covering
topology optimization for additive manufacturing, compressible shock
hydrodynamics, reservoir modeling, fusion-relevant electromagnetic simulations,
space propulsion thrusters, radiation transport, space-time discretizations,
PDEs on surfaces, parallelization in time, and algebraic multigrid methods. A
comprehensive list of publications making use of MFEM can be found at
\url{https://mfem.org/publications}.

Conceptually, MFEM can be viewed as a finite element toolbox that provides the
building blocks for developing finite element algorithms in a manner similar to
that of MATLAB for linear algebra methods (\autoref{sec:overview}). MFEM
includes support for the full high-order de Rham complex \cite{Arnold2006}:
$H^1$-conforming, discontinuous ($L^2$), \Hdiv-conforming, \Hcurl-conforming and
NURBS finite element spaces in 2D and 3D (\autoref{ssec:derham}), as well as
many bilinear, linear, and nonlinear forms defined on them, including linear
operators such as gradient, curl, and embeddings between these spaces. It
enables the quick prototyping of various finite element discretizations
including: Galerkin methods, mixed finite elements, discontinuous Galerkin (DG),
isogeometric analysis, hybridization, and discontinuous Petrov-Galerkin
approaches (\autoref{ssec:discr}).

MFEM contains classes for dealing with a wide range of mesh types: triangular,
quadrilateral, tetrahedral, hexahedral, prismatic as well as mixed meshes,
surface meshes and topologically periodic meshes (\autoref{sec:meshes}). It has
general support for mesh refinement and optimization including local conforming
and non-conforming adaptive mesh refinement (AMR) with arbitrary-order hanging
nodes, powerful node-movement mesh optimization, anisotropic refinement,
derefinement, and parallel load balancing (\autoref{sec:adaptivity}). Arbitrary
element transformations allowing for high-order mesh elements with curved
boundaries are also supported. Some commonly used linear solvers, nonlinear
methods, eigensolvers, and a variety of explicit and implicit Runge-Kutta time
integrators are also available.

MFEM supports Message Passing Interface (MPI)-based parallelism throughout the
library and can readily be used as a scalable unstructured finite element
problem generator (\autoref{ssec:parallel}). \cred{Starting with} version 4.0, MFEM offers
initial support for GPU acceleration, and programming models, such as CUDA,
OCCA, RAJA and OpenMP (\autoref{ssec:gpu}). MFEM-based applications have been
scaled to hundreds of thousands of cores. The library supports efficient
operator partial assembly and evaluation for tensor-product high-order elements
(\autoref{ssec:pa}). A serial MFEM application typically requires minimal
changes to transition to a scalable parallel version of the code where it can
take advantage of the integrated scalable linear solvers from the {\em hypre}
library, including the BoomerAMG, AMS, and ADS solvers (\autoref{ssec:solvers}).
Both the serial and parallel versions can make use of high-performance,
\textit{partial assembly} kernels, described in further detail in
\autoref{ssec:gpu}.

Comprehensive support for a number of external numerical libraries, e.g., PETSc
\cite{petsc-user-ref}, SuperLU \cite{Li2005}, STRUMPACK \cite{Ghysels2016},
SuiteSparse \cite{Davis2006}, SUNDIALS \cite{hindmarsh2005sundials}, and PUMI
\cite{pumi} is also included, which gives access to many additional linear and
nonlinear solvers, preconditioners, and time integrators. MFEM's meshes and
solutions can be visualized with its lightweight native visualization tool GLVis
\cite{glvis}, as well as with \cred{ParaView and} the VisIt \cite{VisIt,VisIt_web} visualization and
analysis tool (\autoref{ssec:vis}).

MFEM is used in a number of applications in the U.S.\ Department of Energy,
academia, and industry (\autoref{sec:apps}). The object-oriented design of the
library separates the mesh, finite element, and linear algebra abstractions,
making it easy to extend and adapt to the needs of different simulations.
The MFEM code base is relatively small and is written in highly
portable C++, using a limited subset of the language. This reduces the entry
barrier for new contributors and makes it easy to build the library on early-access
HPC systems with vendor compilers that may not be mature. The serial version of MFEM has no required
external dependencies and is straightforward to build on Linux, Mac, and
Windows. The MPI-parallel version requires only two third-party libraries ({\em
hypre} \cite{hypre} and METIS \cite{metis,metis_web}) and is easy to build with
an MPI compiler.

\cgreen{MFEM's development grew out of a need for robust, flexible, and efficient
simulation algorithms for physics and engineering applications at Lawrence
Livermore National Laboratory (LLNL). The initial open-source release of the
library was in 2010, followed by version 1.2 in 2011 that added MPI
parallelism. Versions 2.0, 3.0 and 3.4 released in 2011, 2015 and 2018 added
new features such as arbitrary high-order spaces, non-conforming AMR, HPC
miniapps and mesh optimization. An important milestone was the initial GPU
support added in MFEM-4.0, which was released in May 2019. The latest version is
4.1, released in March 2020.

MFEM is being actively developed on GitHub with contributions from many
users and developers worldwide. Users can report bugs and connect with the MFEM
developer community via the GitHub issue tracker at \url{https://github.com/mfem/mfem/issues}.
Details on testing, continuous integration, and how to contribute to the project
can be found in the top-level \code{README} and \code{CONTRIBUTING.md} files in
the MFEM repository.}

\section{Finite Element Abstractions} \label{sec:overview}
To illustrate some of the functionality of MFEM, we consider the model
Poisson problem with homogeneous boundary conditions:
\begin{equation}
\label{eq:strongpoisson}
\begin{gathered}[b]
\text{Find $u : \Omega \to \mathbb{R}$ such that} \\
\begin{aligned}[t]
-\Delta u &= f \quad \text{ in } \Omega \\
u &= 0 \quad \text{ on } \Gamma
\end{aligned}
\end{gathered}
\end{equation}
where $\Omega \subset \mathbb{R}^d$ is the domain of interest, $\Gamma$ is its
boundary, and $f : \Omega \to \mathbb{R}$ is the given source. The solution to
this problem lies in the infinite dimensional space of admissible solutions (cf.\ e.g.\
\cite{Brenner2008})
\begin{equation} \label{eq:inf-dim-space}
   V = \{ v \in H^1(\Omega), v = 0 \text{ on } \Gamma \}.
\end{equation}
To discretize \eqref{eq:strongpoisson}, we begin by defining a mesh of the
physical domain $\Omega$. The mesh is represented in MFEM using a \code{Mesh}
object. Once the mesh is given, we may define a finite dimensional subspace $V_h
\subset V$, represented in MFEM by \code{FiniteElementSpace}. The approximate
solution $u_h \in V_h$ is found by solving the corresponding finite element
problem:
\begin{equation}
\label{eq:weakpoisson}
\begin{gathered}[b]
\mbox{Find $u_h \in V_h$ such that} \\
\int_{\Omega} \nabla u_h \cdot \nabla v_h = \int_{\Omega} f\, v_h
   \qquad \forall v_h \in V_h.
\end{gathered}
\end{equation}
This can be written equivalently as
\begin{equation}
\label{eq:abstractbvp}
\begin{gathered}[b]
\mbox{Find $u_h \in V_h$ such that}\\
a(u_h,v_h) = l(v_h) \qquad \forall v_h \in V_h,
\end{gathered}
\end{equation}
where the bilinear form $a(\cdot, \cdot)$ and linear form $l(\cdot)$ are defined
by
\begin{align}
a(u,v) &= \int_{\Omega} \nabla u \cdot \nabla v,
   \label{eq:abstractforms:bilin} \\
l(v) &= \int_{\Omega} f\, v. \label{eq:abstractforms:lin}
\end{align}
These types of forms are represented in MFEM by the classes \code{BilinearForm}
and \code{LinearForm}, respectively. These forms are expressed as sums of terms
defined by classes derived from \code{BilinearFormIntegrator} and
\code{LinearFormIntegrator}, respectively (see \autoref{ssec:discr}). In the
example considered here, the bilinear form \eqref{eq:abstractforms:bilin} has
one term of type \code{DiffusionIntegrator} and the linear form
\eqref{eq:abstractforms:lin} has one term of type \code{DomainLFIntegrator}.
\cblue{Functions such as $f$, and any material coefficients, are represented as
classes derived from \code{Coefficient}, \code{VectorCoefficient}, or
\code{MatrixCoefficient}. Note that due to performance considerations, linear
and bilinear forms in MFEM are described using sub-classes of the above classes
and not with a domain-specific language.}

After defining basis functions $\varphi_j$ for the space $V_h$, the finite
element problem \eqref{eq:weakpoisson} may be rewritten as
\begin{equation}
\begin{gathered}[b]
\mbox{Find coefficients $c_j$ such that} \\
\sum_j c_j \int_{\Omega} \nabla \varphi_j \cdot \nabla \varphi_i
   = \int_{\Omega} f\, \varphi_i.
\end{gathered}
\end{equation}
Defining the linear algebra objects
\begin{align}
   A_{ij} &= \int_\Omega \nabla \varphi_j \cdot \nabla \varphi_i, \\
   b_i    &= \int_\Omega f \varphi_i, \\
   x_i    &= c_i,
\end{align}
we arrive at the discrete system of linear equations
\begin{equation} \label{eq:lin-system}
   Ax = b.
\end{equation}
By calling the \code{FormLinearSystem} method, the \code{BilinearForm} object
representing $a(\cdot, \cdot)$ is transformed into an \code{Operator} object
representing the linear operator $A$, and the \code{LinearForm} object
representing $l(\cdot)$ is transformed into a \code{Vector} object representing
$b$. After the linear system \eqref{eq:lin-system} has been solved, the
resulting \code{Vector} object may be used to define a \code{GridFunction}
representing the discrete solution $u_h \in V_h$ by means of the method
\code{RecoverFEMSolution} (see \autoref{ssec:fels}).

This simple example illustrates some of the core concepts and classes in the
MFEM example. A more comprehensive description of MFEM's capabilities, including
extensions to other discretization techniques, parallelization, mesh adaptivity
and GPU acceleration are described in the following sections.

\section{Meshes} \label{sec:meshes}
The main mesh classes in MFEM are: \code{Mesh} for a serial mesh and
\code{ParMesh} for an MPI-distributed parallel mesh. The class \code{ParMesh} is
derived from \code{Mesh} and extends the local mesh representation
(corresponding to the inherited \code{Mesh} data and interface) with data and
functionality representing the mesh connections across MPI ranks (see
\autoref{ssec:parallel}).

In this section we describe the internal representation aspects of these two
classes. Mesh input and output functionality is described in
\autoref{ssec:vis}, and mesh manipulation capabilities (refinement,
derefinement, etc.) will be described later in \autoref{sec:adaptivity}.
\subsection{Conforming Meshes} \label{ssec:mesh-conforming}
The definition of a serial (or a local component in parallel), unstructured,
conforming mesh in MFEM consists of two parts: {\em topological} (connectivity)
data and {\em geometric} (coordinates) data.

The \emph{primary} topology data are: a list of vertices, list of elements, and
list of boundary elements. Each element has a type (triangle, quad,
tetrahedron, etc.), an attribute (an integer used to identify subdomains and
physical boundaries), and a tuple of vertex indices. Boundary elements are
described in the same way, with the assumption that they define elements with
dimension one less than the dimension of the regular elements. Any additional
topological data --- such as edges, faces, and their connections to the
elements, boundary elements and vertices --- is derived internally from the
primary data.

\begin{figure}[!hbt]
  \fig{
    Left: The mapping $\Phi$ from the reference element $\hat{\elt}$ to a
    bi-cubic element $\elt$ in physical space with high-order nodes shown as
    black dots.  Right: Example of a highly deformed high-order mesh from a
    Lagrangian hydrodynamics simulation (see \autoref{ssec:hydro}).
    \label{figure:ho_elements}
  }{
    \raisebox{4mm}{\includegraphics[height=.2\textwidth]{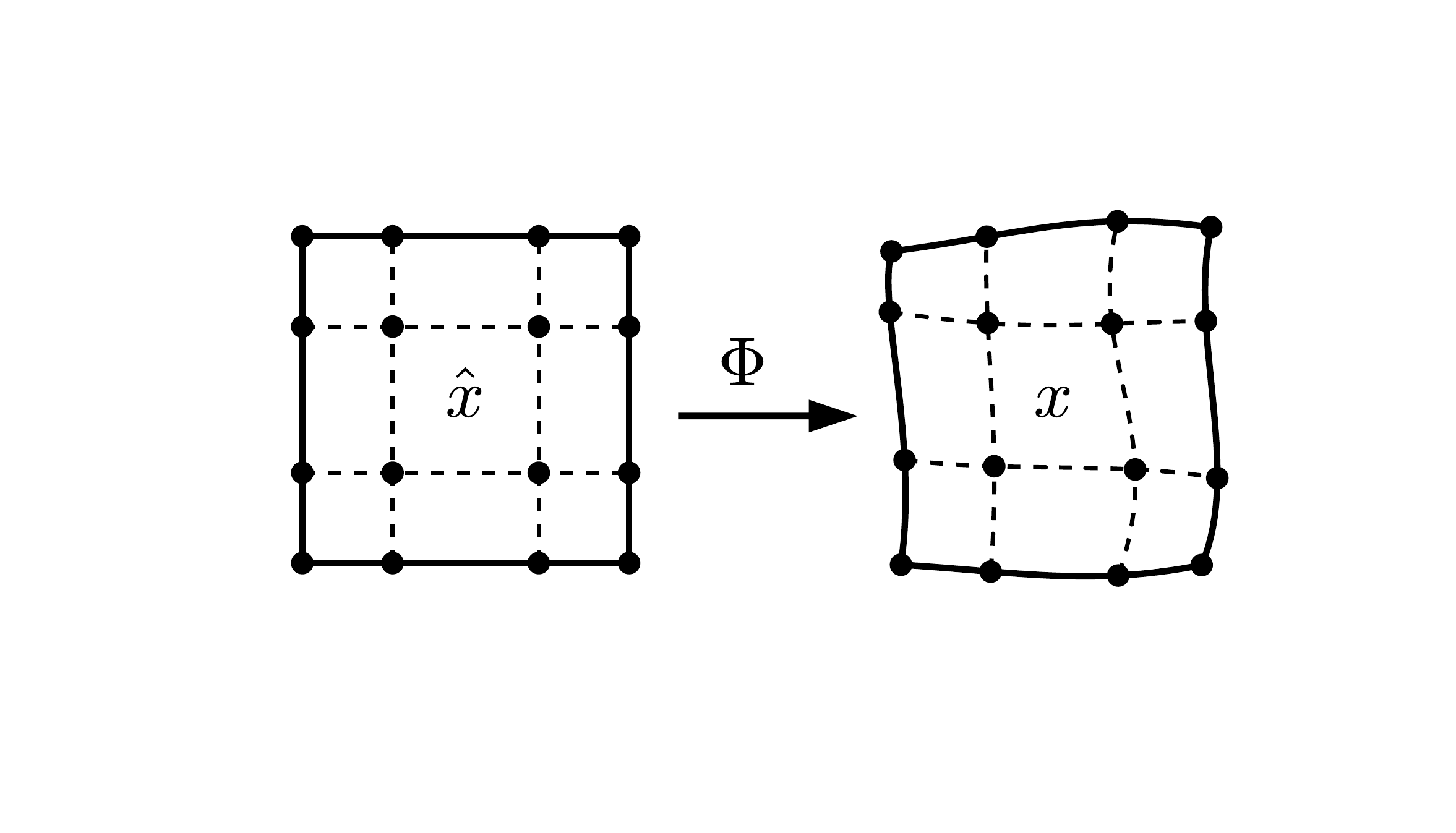}}
    \hspace{-59mm}{\Large $\hat{\elt}$}\hspace{36mm}{\Large $\elt$}\hspace{15mm}
    \hspace{6mm}
    \includegraphics[height=1.55in]{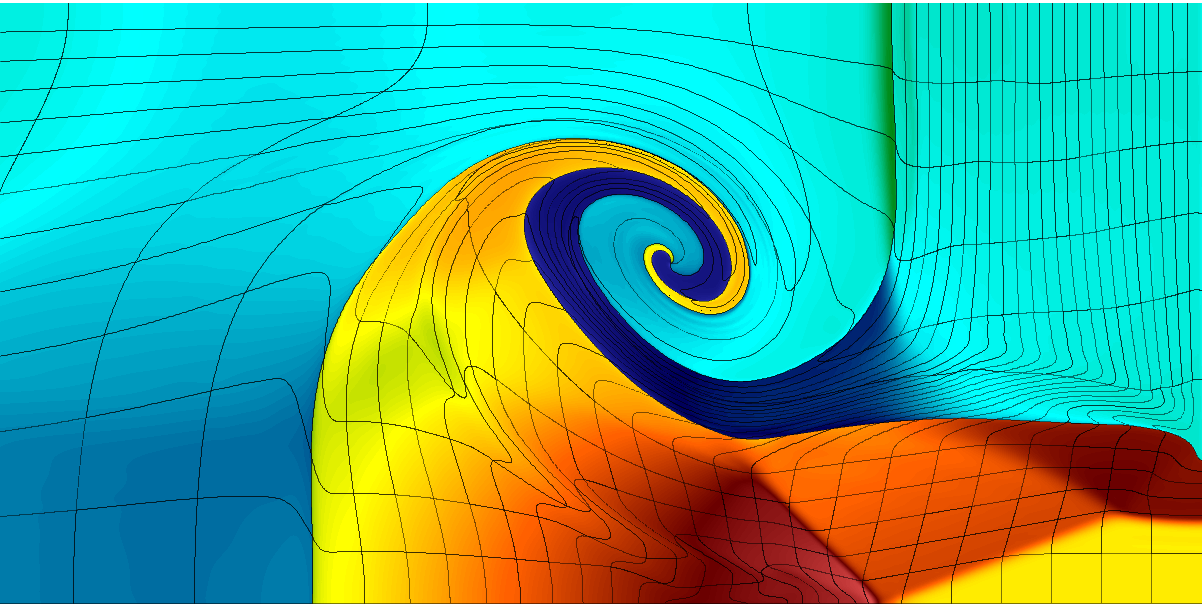}
  }
\end{figure}

The geometric locations of the mesh entities can be described in one of two
ways: (1) by the coordinates of all vertices, and (2) by a \code{GridFunction}
called nodal grid function, or simply \emph{nodes}. Clearly, the first approach
can only be used when describing a \emph{linear} mesh. In the second case, the
\code{GridFunction} class is the same class that MFEM uses to describe any
finite element function/solution. In particular, it defines (a) the basis
functions mapping each reference element to physical space, and (b) the
coefficients multiplying the basis functions in the finite element expansion\cred{---}we
refer to these as nodal coordinates, control points, or nodal degrees of
freedom (DOFs) of the mesh. The nodal geometric description is much richer than
the one based only on the vertex coordinates: it allows nodes to be associated
not only with the mesh vertices but with the edges, faces, and the interiors of
the elements (see \autoref{figure:ho_elements}).

The exact shape of an element is defined through a mapping $\Phi \equiv
\Phi_{\elt}: \hat{\elt} \to \elt$ from the reference element $\hat{\elt}$, as
shown in \autoref{figure:ho_elements}. The mapping $\Phi$ itself is defined in
terms of the basis functions $\{w_i(\hat{x})\}_{i=1}^N$, typically polynomials,
and the local nodal coordinates $\mathbf{x}_{\elt}$ which are extracted/derived from the
global nodal vector $\mathbf{x}$,
\begin{equation}
\label{mesh_eq_x}
  x(\hat{x}) = \Phi(\hat{x}) =
    \sum_{i=1}^N \mathbf{x}_{\elt,i} w_i(\hat{x}) \,.
\end{equation}
Both $\{w_i\}$ and $\{\mathbf{x}_{\elt}\}$ are defined from the geometric mesh
description --- either the vertex coordinates with linear (bilinear for
quadrilaterals, or trilinear for hexahedra) polynomials, or the nodal
\code{GridFunction} with its respective definition of basis functions and node
coordinates.  Typically, the basis functions $\{w_i\}$ are scalar functions and
the coefficients $\{\mathbf{x}_{\elt,i}\}$ are small vectors of the same
dimension as $x \in \elt \equiv \Phi(\hat{\elt})$.  In MFEM, the mapping $\Phi$,
for a particular element $\elt$, is represented by the class
\code{ElementTransformation}. The element transformation for an element $\elt$
can be obtained directly from its \code{Mesh} object using the method
\code{GetElementTransformation(k)}, where \code{k} is the index of the element
$\elt$ in the mesh. Once constructed, the \code{ElementTransformation} object
can be used for computing the physical coordinates of any reference point, the
Jacobian matrix of the mapping, the integration weight associated with the
change of the variables from $\elt$ to $\hat{\elt}$, etc. All of these
operations generally depend on a reference point of interest which is typically
a quadrature point in a quadrature rule. This motivates the use of the class
\code{IntegrationPoint} to represent reference points.

Note that MFEM meshes distinguish between the dimension of the reference space
of all regular elements (reference dimension) and the dimension of the space
into which they are mapped (spatial dimension). This way, surface meshes are
naturally supported with reference dimension of 2 and space dimension of 3,
see e.g. \autoref{fig_examples}.

\subsection{Non-Conforming Meshes} \label{ssec:mesh-nc}
Non-conforming meshes, also referred to as meshes with hanging nodes, can be
viewed as conforming meshes (as described above) with a set of constraints
imposed on some of their vertices. Assuming a linear mesh, the requirement is that each
constrained vertex has to be the convex combination of a set of parent vertices.
Note that, in general, the parent vertices of a constrained vertex can be
constrained themselves. However, it is usually required that all the
dependencies can be uniquely resolved and all constrained vertices can be
expressed as linear combinations of non-constrained ones,
see \autoref{ssec:nc-amr} for more details.

The need for such non-conforming meshes arises most commonly in the local
refinement of quadrilateral and hexahedral meshes. In such scenarios, an element
that is refined shares a common entity (edge or face that the first element
needs to refine) with another element that does not need to refine the shared
entity. To restrict the propagation of the refinement, the first element
introduces one or more constrained vertices on the shared entity and constrains
them in terms of the vertices of the shared entity. The goal of the constraint
is to ensure that the refined sub-entities introduced by the refinement of the
first element are completely contained inside the original shared entity. In
simpler terms, the goal is to make sure that the mesh remains ``watertight'',
i.e.\ there are no gaps or overlaps in the refined mesh.

When working with high-order curved meshes, or high-order finite element spaces
on linear non-conforming meshes, one has to replace the notion of constrained
vertices with constrained degrees of freedom. The goal of the constraints is
still the same: ensure there are no gaps or overlaps in the refined mesh. In the
case of high-order spaces, the goal is to ensure that the constrained
non-conforming finite element space is still a subspace of the discretized
continuous space, $H^1$, $\Hdiv$, etc.  High-order finite elements are further
discussed in \autoref{ssec:ho}.

The observation that a non-conforming mesh can be represented as a conforming
mesh plus a set of linear constraints on some of its nodes, is the basis for the
handling of non-conforming meshes in MFEM. Specifically, the \code{Mesh} class
represents the topology of the conforming mesh (which we refer to as the ``cut''
mesh) while the constraints on the mesh nodes are explicitly imposed on the
nodal \code{GridFunction} which contains both the unconstrained and the
constrained degrees of freedom. In order to store the additional information
about the fact that the mesh is non-conforming, the \code{Mesh} class stores a
pointer to an object of class \code{NCMesh}. For example, \code{NCMesh} stores
the full refinement hierarchy along with all parent-child relations for
non-conforming edges and faces, while \code{Mesh} simply represents the current
mesh consisting of the leaves of the full hierarchy, see \cite{2018-AMR}.

Notable features of the \code{NCMesh} class include its ability to perform both
isotropic and anisotropic refinement of quadrilateral and hexahedral meshes
while supporting an arbitrary number of refinements across a single edge or face
(i.e.\ arbitrary level of hanging nodes).

\subsection{NURBS Meshes} \label{ssec:mesh-nurbs}
Non-Uniform Rational B-Splines (NURBS) are often used in geometric modeling. In
part, this is due to their capability to represent conic sections exactly. In
the last decade, the use of NURBS discrete functions for PDE discretization has
also become popular and is often referred to as IsoGeometric Analysis (IGA), see
\cite{nurbsbook}.

In principle, the construction of NURBS meshes and discrete spaces is very
similar to the case of high-order polynomials. For example, a NURBS mesh can be
viewed as a quadrilateral (in 2D) or hexahedral (in 3D) mesh where the basis
functions are tensor products of 1D NURBS basis functions. However, an important
distinction is that the nodal degrees of freedom are no longer associated with
edges, faces, or vertices. Instead, the nodal degrees of freedom (usually called
control points in this context) can participate in the description of multiple
layers of elements --- a fact that follows from the observation that NURBS basis
functions have support (i.e.\ are non-zero) inside of blocks of
$(k+2)\times(k+2)$ (2D) and $(k+2)\times(k+2)\times(k+2)$ (3D) elements, with
$k$ the continuity of the NURBS space.

In MFEM, NURBS meshes are represented internally through the class
\code{NURBSExtension} which handles all NURBS-specific implementation details
such as constructing the relation between elements and their degrees of freedom.
However, from the user perspective, a NURBS mesh is still represented by the
class \code{Mesh} (with quadrilateral or hexahedral elements) which, in this
case, has a pointer to an object of type \code{NURBSExtension} and a nodal
\code{GridFunction} that defines the appropriate NURBS basis functions and
control points. Most MFEM examples can directly run on NURBS meshes, and some of
them also support IGA discretizations. As of version 3.4, MFEM can also handle
variable-order NURBS, see the examples in the \code{miniapps/nurbs} directory.

\subsection{Parallel Meshes} \label{ssec:mesh-parallel}
As mentioned in the beginning of this section, an MPI-distributed parallel mesh
is represented in MFEM by the class \code{ParMesh} which is derived from class
\code{Mesh}. The data structures and functionality inherited from class
\code{Mesh} represent the local (to the MPI task) portion of the mesh. Note that
each element in the global mesh is assigned to exactly one MPI rank, so
different processors cannot own/share the same element; however they can share
mesh entities of lower dimensions: faces (in 3D), edges (in 2D and 3D), and
vertices (in 3D, 2D, and 1D).

The standard way to construct a \code{ParMesh} in MFEM is to start with a serial
\code{Mesh} object and a partitioning array that assigns an MPI rank to each
element in the mesh. By default, the partitioning array is constructed using the
METIS graph partitioner \cite{metis,metis_web} where mesh elements are the
vertices of the partitioned graph, and the graph edges correspond to the
internal faces (3D), edges (2D) and vertices (1D) connecting two adjacent mesh
elements.

Given the partitioning array, each shared entity can be associated with a unique
set of processors, namely, the set of processors that share that entity. Such
sets of processors are called \emph{processor groups} or simply \emph{groups}.
Each MPI rank constructs its own set of groups and represents it with an object
of class \code{GroupTopology} which represents the communication connections of
each rank with its (mesh) neighbors. Inside each group one of the processors is
selected as the \emph{master} for the group. This choice must be made
consistently by all processors in the group. For example, MFEM assigns the
processor with the lowest rank in the group to be the master.

In order to maintain a consistent mesh description across processors, it is
important to ensure that shared entities are described uniformly across all MPI
tasks in the shared entity group. For example, since \code{ParMesh} does not
define a global numbering of all vertices, a shared triangle with local vertex
indices $(a,b,c)$ on processor $A$ must be described on processor $B$ as
$(x,y,z)$ such that the shared vertex with index $x$ on processor $B$ is the
same as the shared vertex with index $a$ on processor $A$, and similarly for the
indices $y$ and $z$. This uniformity must be ensured during the construction of
the \code{ParMesh} object and maintained later, e.g.\ during mesh refinement.

For this reason, shared entities are stored explicitly (as tuples of local
vertex indices) on each processor. In addition, the shared entities are ordered
by their dimension (vertices, edges, faces) and by their group, making it easier
to maintain consistency across processors.

The case of parallel non-conforming meshes is treated similarly to the serial
case: the \code{ParMesh} object is augmented by an object of class
\code{ParNCMesh} which inherits from \code{NCMesh} and provides all required
parallel functionality. In this case, the parallel partitioning is performed
using a space-filling curve instead of using METIS. This is discussed in more
detail later in \autoref{ssec:nc-amr}.

The case of parallel NURBS meshes is also treated similarly to the serial case:
the \code{ParMesh} object is augmented with an object of class
\code{ParNURBSExtension} which inherits from \code{NURBSExtension}. Note that,
currently, MFEM does not support parallel refinement of NURBS meshes.

\section{Finite Element Discretization} \label{sec:fespace}
In this section, we introduce and describe the main classes (in addition to the
mesh classes described in \autoref{sec:meshes}) required for the full definition
of any finite element discretization space: the class \code{FiniteElement} with
its derived classes, the class \code{FiniteElementCollection} with its derived
classes, and finally the class \code{FiniteElementSpace}. In addition, we
describe the class \code{GridFunction} which represents a particular discrete
function in a finite element space.
\subsection{Finite Elements} \label{ssec:finite-elements}
The concept of a \emph{finite element} is represented in MFEM by the abstract
base class \code{FiniteElement}. The main characteristics of the class are the
following.

\textbf{Reference element.} This is the precise definition of the reference
geometric domain along with descriptions of its vertices, edges, faces, and how
they are ordered. As previously mentioned, this information is included in
class \code{Geometry}. In the \code{FiniteElement} class, this information is
represented by a specifier of type \code{Geometry::Type}. This data member can
be accessed via the method \code{GetGeomType()}. The respective dimension of
the reference element can be accessed via the method \code{GetDim()}.

\textbf{Map type.} This is an integer given by one of the constants:
\code{VALUE}, \code{INTEGRAL}, \code{H\_DIV}, and \code{H\_CURL} defined in the
\code{FiniteElement} class. These constants represent one of the four ways a
function on the reference element $\hat{\elt}$ can be transformed into a
function on any physical element $\elt$ through a transformation
$\Phi:\hat{\elt}
\to \elt$. The four choices are:
\begin{itemize}
\item[\code{VALUE}]
This map-type can be used with both scalar- and vector-valued functions on the
reference element: assume that $\hat{u}(\hat{x})$, $\hat{x}\in\hat{\elt}$ is a
given function, then the transformed function $u(x)$, $x\in\elt$ is defined by
\[
u(x) = \hat{u}(\hat{x}) \,, \qquad\text{where}\qquad x = \Phi(\hat{x})\,.
\]
\vspace{1mm}
\item[\code{INTEGRAL}]
This map-type can be used with both scalar- and vector-valued functions on the
reference element: assume that $\hat{u}(\hat{x})$, $\hat{x}\in\hat{\elt}$ is a
given function, then the transformed function $u(x)$, $x\in\elt$ is defined by
\[
u(x) = \frac{1}{w(\hat{x})} \hat{u}(\hat{x}) \,, \qquad\text{where}\qquad
  x = \Phi(\hat{x})\,,
\]
and $w(\hat{x})$ is the transformation weight factor derived from the Jacobian
$J(\hat{x})$ of the transformation $\Phi(\hat{x})$, which is a matrix of
dimensions $d \times \hat{d}$ (where $\hat{d} \le d$ are the dimensions of the
reference and physical spaces, respectively):
\[
w(\hat{x}) =
\begin{cases}
\det(J(\hat{x})) & \text{ when $\hat{d}=d$, i.e.\ $J$ is square} \\
\det(J(\hat{x})^t J(\hat{x}))^{\frac{1}{2}} & \text{ otherwise.}
\end{cases}
\]
This mapping preserves integrals over mapped subsets of $\hat{\elt}$ and $\elt$.
\vspace{1mm}
\item[\code{H\_DIV}]
This map-type can be used only with vector-valued functions on the reference
element where the number of the vector components is $\hat{d}$, i.e.\ the
reference element dimension: assume that $\hat{u}(\hat{x})$,
$\hat{x}\in\hat{\elt}$ is such a function, then the transformed function $u(x)$,
$x\in\elt$ is defined by
\[
u(x) = \frac{1}{w(\hat{x})} J(\hat{x}) \hat{u}(\hat{x}) \,,
  \qquad\text{where}\qquad x = \Phi(\hat{x})\,,
\]
and $w(\hat{x})$ and $J(\hat{x})$ are as defined above. This is the Piola
transformation used for mapping $\Hdiv$-conforming basis functions
\cite{Boffi2013}. This mapping preserves the integrals of the normal component
over mapped $(\hat{d}-1)$-dimensional submanifolds of $\hat{\elt}$ and $\elt$.
\vspace{1mm}
\item[\code{H\_CURL}]
This map-type can be used only with vector-valued functions on the reference
element where the number of the vector components is $\hat{d}$, i.e.\ the
reference element dimension: assume that $\hat{u}(\hat{x})$,
$\hat{x}\in\hat{\elt}$ is such a function, then the transformed function
$u(x)$, $x\in\elt$ is defined by
\[
u(x) =
\begin{cases}
J(\hat{x})^{-T} \hat{u}(\hat{x}) &
  \text{ when $\hat{d}=d$, i.e.\ $J$ is square} \\
J(\hat{x}) [J(\hat{x})^t J(\hat{x})]^{-1} \hat{u}(\hat{x}) &
  \text{ otherwise,}
\end{cases}
  \qquad\text{where}\qquad x = \Phi(\hat{x})\,,
\]
and $w(\hat{x})$ and $J(\hat{x})$ are as defined above. This is the Piola
transformation used for mapping $\Hcurl$-conforming basis functions
\cite{Boffi2013}. This mapping preserves the integrals of the tangential
component over mapped 1D paths.
\end{itemize}
There is a connection between the way a function is mapped and how its gradient,
curl or divergence is mapped: if a function is mapped with the \code{VALUE} map
type, then its gradient is mapped with \code{H\_CURL}; if a vector function is
mapped with \code{H\_CURL}, then its curl is mapped with \code{H\_DIV}; and
finally, if a vector function is mapped with \code{H\_DIV}, then its divergence
is mapped with \code{INTEGRAL}.

The map type can be accessed with the method \code{GetMapType()}. In MFEM, the
map type also determines the type of basis functions used by the
\code{FiniteElement}: scalar (for \code{VALUE} or \code{INTEGRAL} map types) or
vector (for \code{H\_CURL} or \code{H\_DIV} map types).

\textbf{Degrees of freedom.} The number of the degrees of freedom in a
\code{FiniteElement} can be obtained using the method \code{GetDof()} which is
also the number of basis functions defined by the finite element. Each degree of
freedom $i$ has an associated point in reference space, called its node ($i$-th
node). For many scalar finite elements (referred to as \textit{nodal} finite
elements in MFEM), evaluating the $j$-th basis function at the $i$-th node gives
$\delta_{ij}$ (the Kronecker delta). However, this is not true in general for all
finite element types.

The basis functions can all be evaluated simultaneously at a single reference
point, given as an \code{IntegrationPoint}, using the virtual method
\code{CalcShape()} for scalar finite elements or \code{CalcVShape()} for vector finite elements.
Similarly, based on the specific finite element type, the gradient, curl, or
divergence of the basis functions can be evaluated with the
method \code{CalcDShape()},
\code{CalcCurlShape()}, or \code{CalcDivShape()}, respectively.

In order to simplify the construction of a global enumeration for the DOFs, each
local DOF in a \code{FiniteElement} is associated with one of its vertices,
edges, faces, or the element interior. Then the local DOFs are ordered in the
following way: first all DOFs associated with the vertices (in the order defined
by the reference element), then all edge DOFs following the order and
orientation of the edges in the reference element, and then similarly the face
DOFs, and finally, the interior DOFs. This local ordering is then easier to
translate to the global mesh level where global DOFs are numbered in a similar
manner but now traversing all mesh vertices first, then all mesh edges, then all
mesh faces, and finally all element interiors.

For vector finite elements, in addition to the node, each DOF $i$ has an associated
$\hat{d}$-dimensional vector, $\vec{r}_i$. For DOF $i$, associated with a
non-interior entity (usually edge or face) the vector $\vec{r}_i$ is chosen to
be either normal or tangential to the face/edge based on its map type:
\code{H\_DIV} or \code{H\_CURL}, respectively. The role of these associated
vectors is to define the basis functions on the reference element, so that
evaluating the $j$-th vector basis function at the $i$-th node and then
computing the dot product with the vector $\vec{r}_i$ gives $\delta_{ij}$.
Note that the vectors $\vec{r}_i$ have to be scaled appropriately in order to
preserve the rotational symmetries of the basis functions.

\begin{figure}
  \fig{
    \cgreen{Linear, quadratic and cubic $H^1$ finite elements and their respective
    $\Hcurl$, $\Hdiv$ and $L^2$ counterparts in 2D. Note that the MFEM degrees
    of freedom for the Nedelec ($ND$) and Raviart-Thomas ($RT$) spaces are not
    integral moments, but dot products with specific vectors in specific points
    as shown above.}
  }{
    \includegraphics{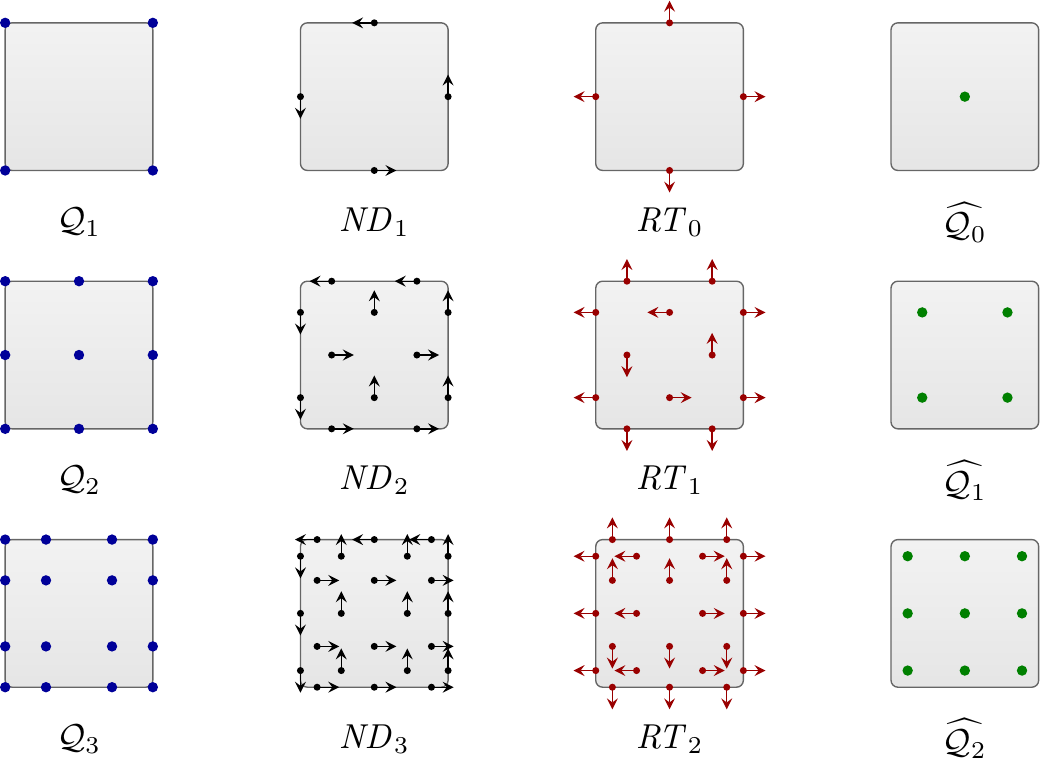}
  }
\label{fig:ho}
\end{figure}

The main classes derived from the base \code{FiniteElement} class are the
arbitrary order $H^1$-conforming (with class names beginning with \code{H1}),
the $L^2$-conforming (i.e.\ discontinuous, with class names beginning with
\code{L2}), the $\Hcurl$-conforming (with class names beginning with \code{ND},
short for \emph{Nedelec}), and the $\Hdiv$-conforming (with class names
beginning with \code{RT}, short for \emph{Raviart-Thomas}) finite elements. All
of these elements are defined for all reference element types where they make
sense. \cblue{These elements can be used with several types of bases, including
the nodal Lagrange basis at Gauss-Lobatto or uniform points (or Gauss-Legendre
points for $L^2$ finite elements) and the Bernstein basis.} For an illustration\cred{,}
see \autoref{fig:ho}.

In addition to the methods for evaluating the basis functions and their
derivatives, class \code{FiniteElement} introduces a number of other useful
methods. Among these are: methods to support mesh refinement:
\code{GetLocalInterpolation()} and \code{GetTransferMatrix()}; methods to
support finite element interpolation/projection: \code{Project()} (scalar and
vector version), \code{ProjectMatrixCoefficient()}; and methods to support the
evaluation of discrete operators such as embedding, gradient, curl, and
divergence: \code{ProjectGrad()}, \code{ProjectCurl()}, etc.

In order to facilitate programming independent of the mesh type, while
simultaneously defining any required permutations of DOFs shared by neighbor
elements in the process of mapping global DOFs to local DOFs, MFEM introduces
the abstract base class \code{FiniteElementCollection}. Its main functionality
is to (1) define a specific finite element for every mesh entity type, and (2)
define a permutation for the DOFs on any mesh entity type (face, edge), based on the
\emph{orientation} of that entity relative to any other possible orientations;
these orientations correspond to the different permutations of the
vertices of the entity, as seen from the points of view of adjacent elements.

The main classes derived from \code{FiniteElementCollection} are the arbitrary
order \code{*\_FECollection} classes where the \code{*}-prefix is one of
\code{H1}, \code{L2}, \code{ND}, or \code{RT} which combine the appropriate
finite element classes with the respective prefix for all different types of
reference elements. Note that in the case of \code{RT\_FECollection}, the
regular (non-boundary) elements use \code{RT\_*} finite elements, however, the
edges (2D) or faces (3D) use \code{L2\_*} elements with \code{INTEGRAL}
map-type. In addition to these ``standard'' \code{FiniteElementCollection}s,
MFEM defines also \emph{interfacial} collections used for defining spaces on the
mesh skeleton/interface which consists of all lower-dimensional mesh entities,
excluding the regular full-dimension mesh elements. These collections can be
used to define discrete spaces for the traces (on the mesh skeleton) of the
regular $H^1$, $\Hcurl$, and $\Hdiv$ spaces.

\subsection{Finite Element Spaces} \label{ssec:fe-spaces}
In MFEM, the mathematical concept (or definition) of a discrete finite element
function space is encapsulated in the class \code{FiniteElementSpace}. The two
main components for constructing this class are a \code{Mesh} and a
\code{FiniteElementCollection} which provides sufficient information in order to
determine global characteristics such as the total number of DOFs and the
enumeration of all the global DOFs. In the \code{FiniteElementSpace}
constructor, this enumeration is generated and stored as an object of class
\code{Table} which represents the mapping: for any given element index $i$,
return the ordered list of global DOF indices associated with element $i$. The
order of these global DOFs in the list corresponds exactly to the local ordering
of the local DOFs as described by the \code{FiniteElement}. The specific
\code{FiniteElement} object associated with an element $i$ can be obtained by
first looking up the reference element type in the \code{Mesh} and then querying
the \code{FiniteElementCollection} for the respective \code{FiniteElement}
object. Thus, the \code{FiniteElementSpace} can produce the basis functions for
any mesh element and the global indices of the respective local DOFs.

The global DOF numbering is created by first enumerating all DOFs associated
with all vertices in the mesh; then enumerating all DOFs associated with all
edges in the mesh --- this is done, edge by edge, choosing a fixed direction on
each edge and listing the DOFs on the edge following the chosen direction; next,
the DOFs associated with faces are enumerated --- this is done face by face,
choosing a fixed orientation for each face and following it when listing the
DOFs on the face; finally, all DOFs associated with the interiors of all mesh
elements are enumerated, element by element. Various renumbering schemes, such
as \cite{gecko}, are also supported to improve the cache locality.

An additional parameter in the construction of a \code{FiniteElementSpace} is
its \emph{vector dimension} which represents, mathematically, a Cartesian power
(i.e.\ number of components) applied to the space defined by the
\code{FiniteElement} basis functions. The additional optional parameter,
\code{ordering}, of the \code{FiniteElementSpace} constructor, determines how
the components are ordered globally: either \code{Ordering::byNODES} (default)
or \code{Ordering::byVDIM}; the \emph{vector} DOF (\code{vdof}) index $k$
corresponding to the (scalar) DOF $i$ in component $j$ is given by $k = i + j
N_d$ in the first case ($N_d$ is the number of DOFs in one component), and $k =
j + i N_c$, in the second ($N_c$ is the number of components).

\subsection{Discrete de Rham Complex} \label{ssec:derham}

\begin{figure}
  \fig{
    Continuous de Rham complex in 3D and example physical fields that can be
    represented in the respective spaces.
  }{
    \includegraphics{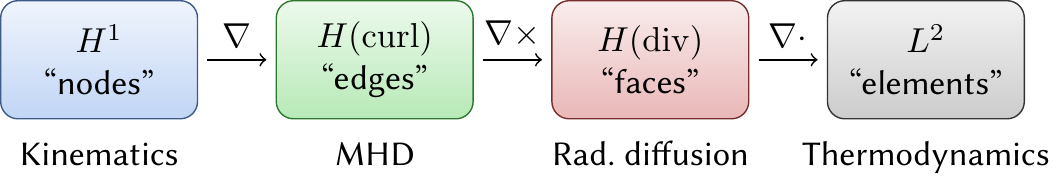}
  }
  \label{fig:deRham}
\end{figure}

The {\em de Rham complex} \cite{arnold:acta,nedelec:1980} is a compatible
multi-physics discretization framework that naturally connects the solution
spaces for many common PDEs. It is illustrated in \autoref{fig:deRham}.  The
finite element method provides a compatible approach to preserve the de Rham
complex properties on a fully discrete level.
In MFEM, constructing a \code{FiniteElementSpace} using the
\code{*\_FECollection} with \code{*} replaced by \code{H1}, \code{ND},
\code{RT}, or \code{L2}, creates the compatible discrete finite element space
for the continuous $H^1$, $\Hcurl$, $\Hdiv$, or $L^2$ space, respectively.  Note
that the order of the space is simply a parameter in the constructor of the
respective \code{*\_FECollection}, see \autoref{fig:ho}.

The finite element spaces in the de Rham sequence are the natural discretization
choices respectively for: kinematic variables (e.g., position, velocity),
electromagnetic fields (e.g., electric field in magnetohydrodynamics (MHD)),
diffusion fluxes (e.g., in flux-based radiation-diffusion) and thermodynamic
quantities (e.g., internal energy, density, pressure). MFEM includes full
support for the de Rham complex at arbitrary high order, on arbitrary order
meshes, as illustrated for example in the first four example codes that come
with the MFEM distribution, see \autoref{ssec:examples}.

Finite element functions are represented by the class \code{GridFunction}.
A \code{GridFunction} is the list of DOFs for a discrete function in a
particular \code{FiniteElementSpace}, so it could be used both on a linear
algebra level (as a \code{Vector} object), or on the finite element level (as a
piecewise-smooth function on the computational mesh). Grid functions are primal
vectors, see \autoref{ssec:fels}, that are used to represent the finite element
approximate solution. They contain methods for interpolation of continuous data
(\code{ProjectCoefficient}), evaluation of integrals and errors
(\code{ComputeL2Error}), as well as many linear algebra operations that are
inherited from the \code{Vector} class.

\subsection{High-Order Spaces} \label{ssec:ho}
High-order methods are playing an increasingly important role in computational
science due to their potential for better simulation accuracy and favorable
scaling on modern architectures
\cite{hpbook-solin,fischerbook,hpbook-dem,brown2010ens,ceed}. MFEM supports
arbitrary-order elements, and provides efficient implementations of specialized
algorithms designed to control the algorithmic complexity with respect to the
polynomial order, see \autoref{ssec:pa}.

\subsection{Input/Output and Visualization} \label{ssec:vis}
MFEM provides integrations with several external tools for easy and accurate
visualization of finite element meshes and grid functions, including arbitrary
high-order meshes and fields. These integrations are based on sampling of the
geometry and grid function data on a reference space lattice via the
\code{GeometryRefiner}. (One example of its use is the {\it Shaper} miniapp in
\code{miniapps/meshing}.) MFEM can also provide accurate gradients enabling
better surface normal vector computations.

Two of the visualization tools with which MFEM has been integrated are GLVis
\cite{glvis} and VisIt \cite{VisIt,VisIt_web}. GLVis is MFEM's lightweight {\em
in-situ} visualization tool that directly uses MFEM classes for OpenGL
visualization supporting interactive refinement of the reference-space sampling
and uses accurate gradients for surface normals. VisIt is a comprehensive data
analysis framework developed at LLNL, which includes native MFEM support via an
embedded copy of the library. The sampled data in this case is controlled by a
{\em multi-resolution} slider and is treated as low-order refined information
so all VisIt functionality can be used directly. Various file formats are
supported, including in-memory remote visualization via socket connection in
the case of GLVis.

For mesh I/O, there are two MFEM native ASCII formats: one for generic
(non-NURBS) meshes, and one that is specific for NURBS meshes. These are the
default formats used when writing a mesh to a C++ output stream
(\code{std::ostream}) or when calling the \code{Print()} method of class
\code{Mesh} or \code{ParMesh}. Note that the cross-processor connectivity in a
parallel mesh is lost when using the \code{Print()} method which, however, is
not required for visualization purposes. To save a parallel mesh with all
cross-processor connections, one can use the method \code{ParMesh::ParPrint()}.

Other \emph{input} formats supported by class \code{Mesh} are: Netgen
\cite{netgen_ngsolve_web,ngsolve}, TrueGrid \cite{TrueGrid}, unstructured VTK
\cite{VTK4}, Gmsh (linear elements only) \cite{Geuzaine2009}, and Exodus format
(produced by the Cubit mesh generator, among others) \cite{Shemon2014}. Class
\code{Mesh} also provides output support for the unstructured VTK format through
the method \code{PrintVTK()}.

For more comprehensive input/output, where a mesh is stored with any number of
finite element solution fields, MFEM defines the base class
\code{DataCollection} along with several derived classes:
\code{VisItDataCollection}: writes an additional \code{.mfem\_root} file that
can be opened by the MFEM plugin in VisIt \cite{VisIt,VisIt_web};
\code{SidreDataCollection}: a set of data formats based on the Sidre component
of LLNL's Axom library \cite{Axom} which, in particular, supports binary I/O
and can also be opened by VisIt; and \code{ConduitDataCollection}: a set of
data formats based on LLNL's Conduit library \cite{conduit_web} which also
supports binary I/O and can be opened by VisIt. Note that the class
\code{VisItDataCollection} uses the default ASCII format to save the mesh and
finite element solution fields. The class \code{ParaViewDataCollection} can be
used to output XML data in ParaView's ``VTU'' format, using either ASCII or
compressed binary format. In addition to standard low-order output,
\code{ParaViewDataCollection} also supports ParaView's high-order Lagrangian
elements.

\section{Finite Element Operators} \label{sec:feoperators}
\subsection{Discretization Methods} \label{ssec:discr}

MFEM includes the abstractions and building blocks to \emph{discretize}
equations; that is, the process by which the linear system is formed from a PDE,
choice of basis functions, and mesh.
As discussed in \autoref{sec:overview}, before discretizing a linear PDE using
the finite element method, it is converted into a variational form like
\eqref{eq:abstractbvp} consisting of a bilinear and a linear form. In MFEM,
they are represented by the classes \code{BilinearForm} and \code{LinearForm},
respectively. Depending on the PDE, each of these forms consists of one or more
terms, called {\em integrators} in MFEM. The process of describing the PDE in
MFEM consists of defining a \code{BilinearForm} and a \code{LinearForm} and then
adding integrators to them by calling their \code{Add*Integrator} methods,
e.g.\ \code{AddDomainIntegrator} or \code{AddBoundaryIntegrator}. The main
parameter for these methods is an instance of an integrator: a subclass of the
abstract base classes \code{BilinearFormIntegrator} and
\code{LinearFormIntegrator}. An extensive list of the integrators defined in
MFEM can be found at \url{https://mfem.org/fem}. Note that this design is
extensible since it allows users to implement and use their own integrators.

There are many different approaches for expressing a given PDE in variational
form which, in turn, give rise to different finite element methods for the same
given PDE.
MFEM's included examples illustrate some of these different methods. For
example, a very common approach for discretizing Poisson's equation is to use
$H^1$ elements of any order and spatial dimension, where the basis functions are
continuous across element interfaces. This is illustrated in Example 1. This is
the most straightforward discretization of the equation, but there are many
other approaches possible. For instance, Example 8 and Example 14 solve the same
PDE with discontinuous Petrov-Galerkin (DPG) \cite{2018-SISC-DPG} and
discontinuous Galerkin (DG) discretizations, respectively, see
\autoref{ssec:examples}. The examples include interactive documentation
(in \code{examples/README.html} or online at \url{https://mfem.org/examples})
organized by the different discretization methods available in the library and
are the fastest route to learn about MFEM's capabilities.

Examples 3-5 show a wide range of the discretization capability of MFEM and many
of the possible finite elements. Example 3 solves the second-order definite
Maxwell equation using the $\Hcurl$ Nedelec finite elements with the curl-curl
and mass bilinear form integrators. Example 4 progresses down the de Rham
sequence, and solves a \cred{second-order} definite equation with a Neumann boundary
condition using $\Hdiv$ Raviart-Thomas finite elements and div-div and mass
bilinear form integrators. Example 5 uses a mixed $\Hdiv$ and $L^2$ (DG)
discretization of a Darcy problem, solving these together in a $2\times 2$ block
bilinear form. These three examples are just a few of the many examples included
with the library, but they show a wide range of the finite elements and
discretization approaches possible along the de Rham sequence.

On the meshing side, there are also many different approaches. As described
above in \autoref{sec:meshes}, MFEM supports arbitrary-order meshes, which can
be topologically periodic or assigned boundary tags. However, MFEM also
includes an extension to its \code{Mesh} class to generate basis functions from
non-uniform Rational B-splines (NURBS), see \autoref{ssec:mesh-nurbs}. This
allows for isogeometric analysis, where the basis is refined without changing
the geometry or its parametrization \cite{nurbsbook}.

MFEM includes various ordinary differential equation (ODE) solvers that can be
used in conjunction with the finite elements and bilinear forms to discretize
the time derivative terms. Many ODE solvers are distributed with the library:
various implicit and explicit Runge-Kutta (RK) methods including
singly-diagonal implicit versions (SDIRK), and symplectic methods.
Additionally, MFEM supports time integration with the SUNDIALS (SUite of
Nonlinear and DIfferential/ALgebraic equation Solver) library, which provides
many additional ODE solvers; explicit and fully-implicit time stepper solvers
(TS) \cite{abhyankar2018petscts} from PETSc are also supported. Finally, MFEM's
ODE solvers can be extended by inheriting from the abstract base class
\code{ODESolver}.

\subsection{Finite Element Linear Systems} \label{ssec:fels}
One of the main operations that MFEM performs is the construction of a linear
system of the form \eqref{eq:lin-system} given a finite element description of
problem such as in \eqref{eq:abstractbvp}. Performing this task while supporting
distributed memory architectures, high-order basis functions, non-conforming
meshes, or more general basis function types introduces complications that
require careful treatment. To manage these complexities MFEM makes use of
abstractions which clearly separate finite element concepts from linear algebra
concepts.

MFEM's linear algebra objects include \code{Vector} and \code{SparseMatrix} in
serial and \code{HypreParVector} and \code{HypreParMatrix} in parallel. Parallel
linear algebra via PETSc is supported via the classes \code{PetscParVector}
and \code{PetscParMatrix}; the latter also provides on-the-fly conversion routines
between {\em hypre} and PETSc parallel data formats. The finite element objects include
\code{(Par)GridFunction}, \code{(Par)LinearForm}, and \code{(Par)BilinearForm}.
For convenience \code{(Par)GridFunction} and \code{(Par)LinearForm} inherit from
the \code{Vector} class and can therefore be used as vectors, and similarly
\code{(Par)BilinearForm} can be used as a matrix.

\begin{figure}
  \fig{Graphical depiction of the relationship between the finite element
    bilinear and linear form objects, and the linear algebra matrices
    and primal/dual vectors in MFEM.}
     {\includegraphics{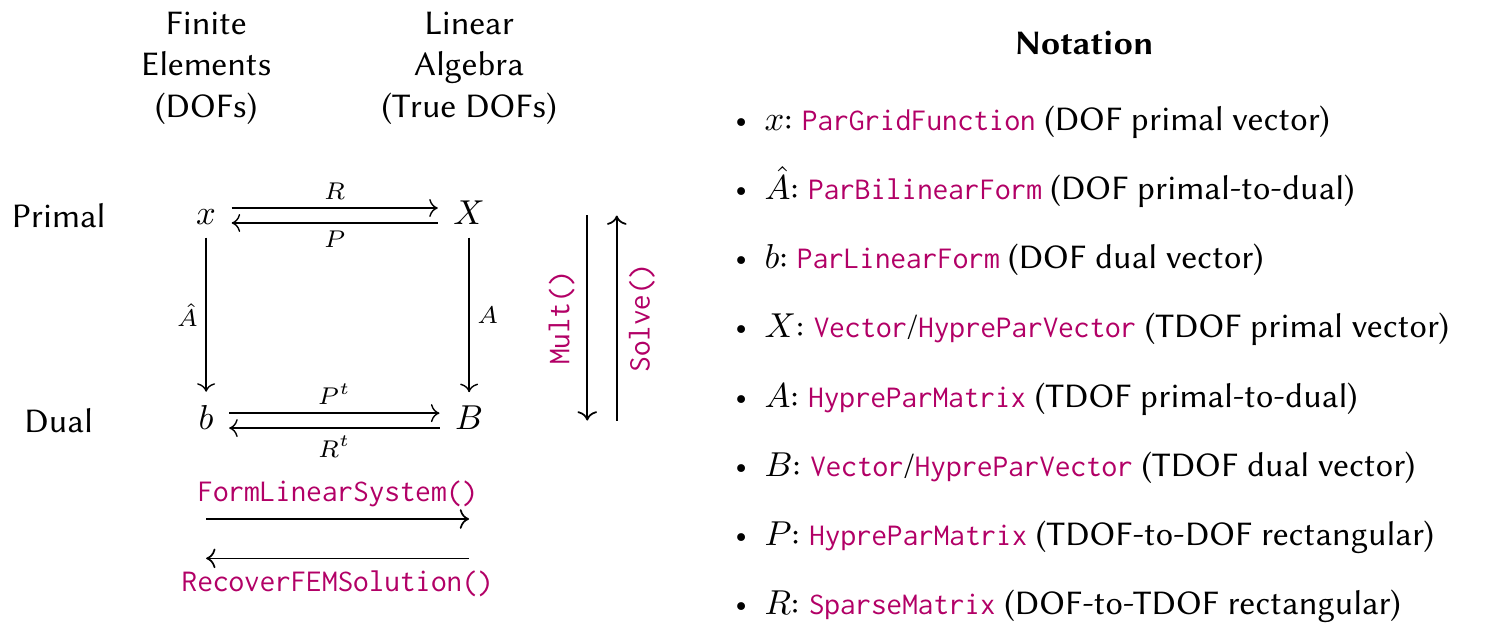}}
  \label{fig:fem-la-1}
\end{figure}

\autoref{fig:fem-la-1} illustrates the relationship between the finite element
and linear algebra objects in MFEM. The \code{ParGridFunction} object contains,
among other things, all of the degrees of freedom needed to interpolate field
values within every element contained in the local portion of the computational
mesh, denoted by $x$ in \autoref{fig:fem-la-1}.  $X$ in \autoref{fig:fem-la-1}
is a linear algebra \code{Vector} (or \code{HypreParVector}) related to this
\code{ParGridFunction} but potentially quite different. $X$ represents a
non-overlapping, parallel decomposition of the \textit{true} degrees of freedom
of the \code{ParGridFunction} $x$.  For example, some of the degrees of freedom
in the \code{ParGridFunction} may be subject to constraints if they happen to be
shared with neighboring elements in a non-conforming portion of the mesh, or
they may be constrained to match degrees of freedom owned by elements found on
another processor.  Some of the degrees of freedom in the \code{ParGridFunction}
may not even directly contribute to the linear system if static condensation or
hybridization is being used. Thus, the linear algebra \code{Vector} represented
by $X$ may be much smaller than the \code{ParGridFunction} $x$. The $P$ and $R$
operators shown in \autoref{fig:fem-la-1}, called the {\em prolongation} and
{\em restriction} operators, respectively, are created and managed by the
\code{ParFiniteElementSpace} and can be used to map data between the finite
element representation of a field and its linear algebra representation.

The \code{ParGridFunction}, labeled $x$ in \autoref{fig:fem-la-1}, and its
linear algebra counterpart $X$, are called {\em primal} vectors because of
their direct relationship with the finite element expansion of a field.  Indeed
the values stored in $x$ are the expansion coefficients $f_i$ in
\begin{equation}
f(\vec{x})=\sum_i f_i \varphi_i(\vec{x}). \label{eq:fem-exp}
\end{equation}
Conversely, a \code{ParLinearForm}, labeled $b$ in \autoref{fig:fem-la-1}, or
the vector labeled $B$ are {\em dual} vectors.  In this context duality refers
to the fact that {\em dual} vectors map {\em primal} vectors to the set of real
numbers \cite{ciarlet78}.  More importantly, they can be used to map a
\code{ParGridFunction} to a physical quantity of interest.  For example, if we
have a \code{ParGridFunction} $\rho$ representing the mass density of a fluid,
and a \code{ParLinearForm} $v$ such that $v_i=\int_{\Omega}\varphi_i$, i.e. a
\code{ParLinearForm} representing the constant function $1$, then $v\cdot\rho$
would approximate the integral of the density over the computational domain
which would equal the total mass of the fluid in this illustration.  Dual
vectors will be of the same length as their primal counterparts but their
entries have very different meanings.  The relationship between $b$ and $B$ is
complementary to that between $x$ and $X$. Whereas the restriction operator
removes dependent entries from $x$ to produce the shorter vector $X$, the
transpose of the prolongation operator is used to coalesce entries from $b$ to
form those of $B$.  For example $P^t$ will add together entries from $b$ to
sum the contributions from different elements to the basis function integral
over its entire support which will be stored in $B$.

Dual vectors can be created directly by integrating a function times the
appropriate basis functions as occurs inside a \code{(Par)LinearForm} or
indirectly by applying a \code{(Par)BilinearForm} or a system matrix to a primal
vector.  The resulting dual vector should be identical in either case.  Which
scheme is used to create a particular dual vector is usually determined by how
the source terms in the PDE arise.  If the sources are determined by known
functions it is generally most efficient to provide these functions to
a \code{(Par)LinearForm} object and compute the dual vector directly.  If, on
the other hand, the source term is the result of a field represented by
a \code{(Par)GridFunction} it could be more efficient to simply apply
a \code{(Par)BilinearForm} to the appropriate primal vector.

As implied in \autoref{fig:fem-la-1}, the linear algebra operator $A$ can be
computed from the \code{(Par)BilinearForm} $\hat{A}$ as $A=P^t \hat{A} P$; however, many
finite element linear systems require boundary conditions to ensure that they
are non-singular.  To facilitate the application of boundary conditions
the \code{(Par)BilinearForm} class has a \code{FormLinearSystem} method which
prepares the three linear algebra objects, as well as applying boundary
conditions.  In the simplest case this method performs the following operations:
\begin{eqnarray*}
  A &=& P^t \hat{A} P, \\
  X &=& R x, \\
  B &=& P^t b.
\end{eqnarray*}
Further modifications are also performed in order to impose essential boundary
conditions.

The \code{FormLinearSystem} method also supports two more advanced and closely
related techniques for reducing the size of a finite element linear system: {\it
hybridization} and {\it static condensation}, see e.g.
\cite{hpbook-solin,Boffi2013}. Note that hybridization in MFEM is applied to a
single bilinear form, see \cite{2018-Hybridization}, instead of the more
classical hybridization approach applied to mixed finite element
discretizations. These more advanced techniques, which compute only portions of
the solution vector, necessitate a further step of reconstructing the entire
solution vector. The \code{(Par)BilinearForm} class provides a
\code{RecoverFEMSolution} method for exactly this purpose.  Given the partial
solution vector $X$ and the \code{(Par)LinearForm} $b$ this method computes the
full degree of freedom vector $x$ needed to properly represent the solution
field throughout the mesh. Additional details can be found in
\cite{2018-Hybridization}.

\subsection{Operator Decomposition} \label{ssec:decomp}
Finite element operators are typically defined through weak formulations of
partial differential equations that involve integration over a computational
mesh. The required integrals are computed by splitting them as a sum over the
mesh elements, mapping each element to a simple reference element (e.g. the unit
square) and applying a quadrature rule in reference space, see
\autoref{sec:overview}.

This sequence of operations highlights an inherent hierarchical structure
present in all finite element operators where the evaluation starts on {\em
  global (trial) degrees of freedom} on the whole mesh, restricts to {\em
  degrees of freedom on subdomains} (groups of elements), then moves to
independent {\em degrees of freedom on each element}, transitions to independent
{\em quadrature points} in reference space, performs the integration, and then
goes back in reverse order to global (test) degrees of freedom on the whole
mesh.

\begin{figure}[h!]
  \centerline
  {
     \includegraphics[width=0.95\linewidth]{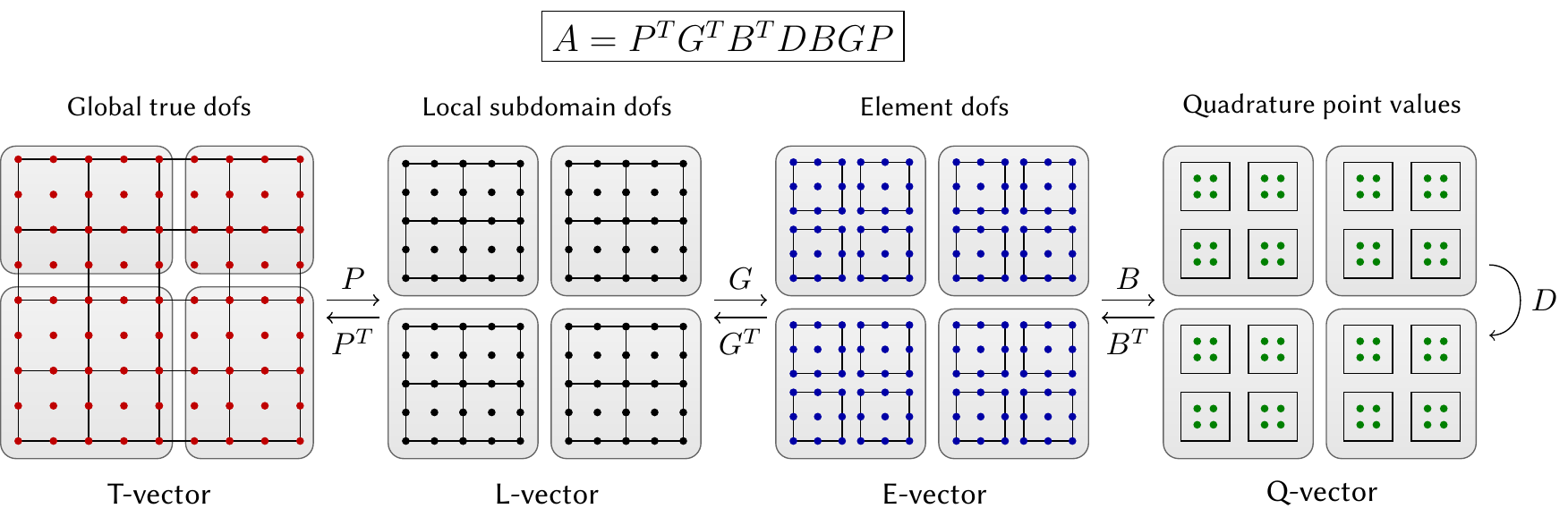}
  }
  \caption
  {
    Fundamental finite element operator decomposition. This algebraically factored
    form is a much better description than a global sparse matrix for high-order
    methods and is easy to incorporate in a wide variety of applications.
    See also the libCEED library in \cite{ceed}.
  }
  \label{fig_op_decom}
\end{figure}

This is illustrated in \autoref{fig_op_decom} for the simple case of a
symmetric linear operator on second order ($Q_2$) scalar continuous ($H^1$)
elements, where we use the notions {\bf T-vector} (true vector), {\bf L-vector}
(local vector), {\bf E-vector} (element vector) and {\bf Q-vector} (quadrature
vector) to represent the sets corresponding to the (true) degrees of freedom on
the global mesh, the split local degrees of freedom on the subdomains, the
split degrees of freedom on the mesh elements, and the values at quadrature
points, respectively. Note that class \code{(Par)GridFunction} represents an
L-vector, and T-vector is typically represented by either \code{HypreParVector}
or \code{Vector}, cf.\ \autoref{fig:fem-la-1}. We remark that although the
decomposition presented in \autoref{fig_op_decom} is appropriate for square,
symmetric linear operators, the generalization of this finite element
decomposition to rectangular and nonlinear operators is straightforward.

One of the challenges with high-order methods is that a global sparse matrix is
no longer a good representation of a high-order linear operator, both with
respect to the FLOPs needed for its evaluation \cite{ORSZAG198070}, as well as
the memory transfer needed for a matrix-vector product (matvec)
\cite{2017-hpc-mat-free-DG, 2019-DealII-mat-free-DG}. Thus, high-order methods
require a new ``format'' that still represents a linear (or more generally,
nonlinear) operator, but not through a sparse matrix.

We refer to the operators that connect the different types of vectors as:
\begin{itemize}
\item Subdomain restriction $P$.
\item Element restriction $G$.
\item Basis (DOFs to quadrature points) evaluator $B$.
\item Operator at quadrature points $D$.
\end{itemize}
More generally, when the test and trial space differ, each space has its own
versions of $P$, $G$ and $B$.

Note that in the case of adaptive mesh refinement (AMR), the restriction $P$
will involve not just extracting sub-vectors, but evaluating values at
constrained degrees of freedom through the AMR interpolation,
see \autoref{ssec:nc-amr}.  There can also be several levels of subdomains
($P_1$, $P_2$, etc.), and it may be convenient to split $D$ as the product of
several operators ($D_1$, $D_2$, etc.).

After the application of each of the first three transition operators, $P$, $G$
and $B$, the operator evaluation is decoupled on their ranges, so $P$, $G$ and
$B$ allow us to ``zoom-in'' to subdomain, element, and quadrature point level,
ignoring the coupling at higher levels.  Thus, a natural mapping of $A$ on a
parallel computer is to split the {\bf T-vector} over MPI ranks in a
non-overlapping decomposition, as is typically used for sparse matrices, and
then split the rest of the vector types over computational devices (CPUs, GPUs,
etc.) as indicated by the shaded regions in \autoref{fig_op_decom}. This is
discussed further in \autoref{ssec:parallel}.

One of the advantages of the decomposition perspective in these settings is that
the operators $P$, $G$, $B$ and $D$ clearly separate the MPI parallelism in the
operator ($P$) from the unstructured mesh topology ($G$), the choice of the
finite element space/basis ($B$) and the geometry and point-wise physics $D$.
These components also naturally fall in different classes of numerical
algorithms -- parallel (multi-device) linear algebra for $P$, sparse (on-device)
linear algebra for $G$, dense/structured linear algebra (tensor contractions)
for $B$ and parallel point-wise evaluations for $D$.

Since the global operator $A$ is just a series of variational restrictions
(i.e.\ transformations $Y \to X^T Y X$) with
$B$, $G$ and $P$, starting from its point-wise kernel $D$, a matrix-vector
product with $A$ can be performed by evaluating and storing some of the
innermost variational restriction matrices, and applying the rest of the
operators ``on-the-fly''. For example, one can compute and store a global matrix
on the {\bf T-vector} level. Alternatively, one can compute and store only the
subdomain ({\bf L-vector}) or element ({\bf E-vector}) matrices and perform the
action of $A$ using matvecs with $P$ or $P$ and $G$. While these options are
natural for low-order discretizations, they are not a good fit for high-order
methods due to the amount of FLOPs needed for their evaluation, as well as the
memory transfer needed for a matvec.

Much higher performance can be achieved by the use of {\em partial assembly}
algorithms, \cgreen{as described in the following section}. In this case, we compute and store only $D$
(or portions of it) and evaluate the actions of $P$, $G$ and $B$ on-the-fly.
Critically for performance, we take advantage of the tensor-product structure of
the degrees of freedom and quadrature points on quadrilateral and hexahedral
elements to perform the action of $B$ without storing it as a matrix.
Implemented properly, the partial assembly algorithm requires the optimal amount
of memory transfers (with respect to the polynomial order) and near-optimal
FLOPs for operator evaluation. It consists of an operator {\em setup} phase,
that evaluates and stores $D$ and an operator {\em apply} (evaluation) phase
that computes the action of $A$ on an input vector. When desired, the setup
phase may be done as a side-effect of evaluating a different operator, such as a
nonlinear residual.  The relative costs of the setup and apply phases are
different depending on the physics being expressed and the representation of
$D$.

\subsection{High-Order Partial Assembly} \label{ssec:pa}
In the traditional finite element setting, the operator is assembled in the form
of a matrix. The action of the operator is computed by multiplying with this
matrix. At high orders this requires both a large amount of memory to store the
matrix, as well as many floating point operations to compute and apply it. By
exploiting the structure shown in \autoref{ssec:decomp} as well as the basis
functions structure, there are options for creating operators that require much
less storage and scale better at high orders. This section introduces partial
assembly and sum factorization \cite{ORSZAG198070,fischerbook}, which reduce
both the assembly storage and number of floating point operations required to
apply the operator, and discusses general algorithm opportunities and challenges
in the MFEM code.

Removing the finite element space restriction operator from the assembly for
domain-based operators\footnote{Domain-based operators correspond to bilinear
forms which use integrals over the problem domain, as opposed to its boundary,
for example.} yields the element-local matrices at the
\textbf{E-vector} level. This storage can lead to faster data access, since the
block is stored contiguously in memory, and applications of the block can be
designed to maximally use the cache.

Partial assembly operates at the \textbf{Q-vector} level, after additionally
removing the basis functions and gradients, $B$, from the assembled operator.
This leaves only the \(D\) operator to store for every element, see
\autoref{ssec:decomp}. This by itself reduces the storage but not the number of
floating point operations required for evaluation. As will be discussed later,
this is key to offloading the operator action to a co-processor that may have
less memory.

As an illustration of partial assembly, consider the decomposition of the mass
matrix evaluated on a single element \(E\)
\begin{equation}
\left(M_E \right)_{ij} = \int_E \rho \, \varphi_j \varphi_i \, dx
\end{equation}
where $\rho$ is a given density coefficient and $\{\varphi_i\}$ are the finite
element basis functions on the element $E$. Changing the variables in the
integral from $E$ to the reference element $\hat{E}$ and applying a quadrature
rule with points \(\{\hat{x}_k\}\) and weights \(\{\alpha_k\}\) yields
\begin{equation}
\left(M_E \right)_{ij} =
   \sum_{k} \alpha_k \left(\rho \circ \Phi\right)(\hat{x}_k) \,
      \hat{\varphi}_j(\hat{x}_k) \hat{\varphi}_i(\hat{x}_k) \,
      \det(J(\hat{x}_k)).
\end{equation}
In the last expression, $\Phi$ is the mapping from the reference element
$\hat{E}$ to the physical element $E$, $J$ is its Jacobian matrix, and
$\{\hat{\varphi}_i\}$ are the finite element basis functions on the reference
element.
Defining the matrix $B$ of basis functions evaluated at quadrature points as
\(B_{ki} = \hat{\varphi}_i(\hat{x}_k)\), the above equation can be rewritten as
\begin{equation}
\left(M_E \right)_{ij} =
   \sum_{k} B_{ki} (D_E)_{kk} B_{kj},\qquad \text{where} \quad
\left(D_E\right)_{kk} =
   \alpha_k \det(J(\hat{x}_k)) \left(\rho \circ \Phi\right)(\hat{x}_k),
   \quad \left(D_E\right)_{kl} = 0, ~ k\ne l.
\end{equation}
Using this definition, the matrix operator can be written simply as
\(M_E = B^t D_E B\). Matrix-vector evaluations are computed as the series of
products by \(B\), \(D_E\), and \(B^t\) without explicitly forming \(M_E\).

For general $B$, its application requires the same order of floating point
operations as applying the fully-assembled \(M_E\) matrix:
\(\mathcal{O}(p^{2d})\) (assuming that the number of quadrature points is
$\mathcal{O}(p^d)$). Taking advantage of the tensor-product structure of the
basis functions and quadrature points on quad and hex elements, \(B_{ki}\) can
be written as
\begin{equation}
B_{ki} = \hat{\varphi}^{1d}_{i_1}\left(\hat{x}^{1d}_{k_1}\right) \ldots
         \hat{\varphi}^{1d}_{i_d}\left(\hat{x}^{1d}_{k_d}\right), \qquad
         k=(k_1,\ldots,k_d), \quad i=(i_1,\ldots,i_d)
\end{equation}
with \(d\) the number of dimensions. In this case the matrix \(B\) itself is
decomposed as a tensor product of smaller one-dimensional matrices
\(B^{1d}_{l j} = \hat{\varphi}^{1d}_{j}\left(\hat{x}^{1d}_{l}\right) \) so that
\begin{equation}
  \label{eqn:paB}
  B_{ki} = B^{1d}_{k_1 i_1} \ldots B^{1d}_{k_d i_d}.
\end{equation}
Applying the series of \(B^{1d}\) matrices reduces the overall number of
floating point operations when applying \(M_E\) to \(\mathcal{O}(p^{d+1})\)
(assuming that the number of 1D quadrature points is $\mathcal{O}(p)$).
This evaluation strategy is often referred to as {\em sum factorization}.

To make this point concrete, consider the application of a quad basis to a
vector \(v\) for interpolation at a tensor product of quadrature points. Without
taking advantage of the structure of the basis, the product takes the form
\begin{equation}
  (B v)_k = \sum_i B_{ki} v_i = \sum_i \hat{\varphi}_{i}(\hat{x}_k) v_i,
\end{equation}
which requires \(\mathcal{O}(p^{2d})\) (\(d=2\)) storage and operations for the
matrix-vector product. When using the alternative form \eqref{eqn:paB} the
operation can be rewritten as
\begin{equation}
  (B v)_k = \sum_i B_{ki} v_i =
     \sum_{i_1,i_2} B^{1d}_{k_1 i_1} B^{1d}_{k_2 i_2} V_{i_1 i_2} =
     \left[ B^{1d} V \left( B^{1d} \right)^t \right]_{k_1 k_2},
\end{equation}
where $V$ is the vector $v$ viewed as a square matrix: $V_{i_1 i_2} = v_i$.
This highlights an interesting aspect of sum factorization: with each smaller
matrix product with \(B^{1d}\), an additional axis is converted from basis
(\(i_j\)) to quadrature (\(k_j\)) indices. The same reasoning can also be
applied to three spatial dimensions. Using the sum factorization approach, the
storage was reduced to \(\mathcal{O}(p^d)\) and the number of operations to
\(\mathcal{O}(p^{d+1})\).

Choosing to store the partially assembled operator instead of the full matrix
affects the solvers that can be used, since the full matrix is not available to
be queried. This means for instance that traditional multigrid solvers are
difficult to apply. These issues are discussed further in
\autoref{ssec:solvers}.

The storage and asymptotic number of floating point operations required for
assembly and evaluation using the different methods are recorded in
\autoref{tab:HO-PA:op-storage-counts}. Sum factorization can be utilized to
reduce the cost of assembling the local element matrices and thus the cost of
full assembly (\textbf{T-vector} level) -- this is shown in the second row of
the table. Furthermore, partial assembly has improved the asymptotic scaling for
high orders in both storage and number of floating point operations for assembly
and evaluation. Therefore, partial assembly is well-suited for high orders.

\begin{table}[b]
  \tbl{Comparison of storage and Assembly/Evaluation FLOPs required for full and partial assembly algorithms on tensor-product element meshes (\cred{quadrilaterals and hexahedra}).
    Here, \(p\) represents the polynomial order of the basis functions and \(d\)
    represents the number of spatial dimensions. The number of DOFs on each element is \(\mathcal{O}(p^{d})\) so the ``sum factorization full assembly'' and ``partial assembly'' algorithms are nearly optimal.}{
  {\def\arraystretch{1.2}
    \begin{tabular}{l@{\hspace{2em}}lll}
      \toprule
      \textbf{Method} & \textbf{Storage} & \textbf{Assembly} & \textbf{Evaluation} \\
      \midrule
      Traditional full assembly + matvec & \(\mathcal{O}(p^{2d})\) & \(\mathcal{O}(p^{3d})\) & \(\mathcal{O}(p^{2d})\) \\
      Sum factorized full assembly + matvec & \(\mathcal{O}(p^{2d})\) & \(\mathcal{O}(p^{2d+1})\) & \(\mathcal{O}(p^{2d})\) \\
      Partial assembly + matrix-free action  & \(\mathcal{O}(p^{d})\) & \(\mathcal{O}(p^d)\) & \(\mathcal{O}(p^{d+1})\)\\
      \bottomrule
    \end{tabular}
    \vspace{\floatsep}
  }}
  \label{tab:HO-PA:op-storage-counts}
\end{table}

There are many opportunities and challenges for parallelization with partial
assembly using sum factorization. At the \textbf{E-vector} level the products
can be applied independently for every element in parallel, which makes partial
assembly with sum factorization a promising portion of the finite element
algorithm to offload to co-processors, such as GPUs. In MFEM, partial assembly
and sum factorization are implemented in the bilinear and nonlinear form
integrators themselves. Specifically, in the base class
\code{BilinearFormIntegrator}, the assembly and evaluation are performed by the
virtual methods \code{AssemblePA} and \code{AddMultPA}, respectively. MFEM
supports partially assembly for the entire de Rham complex, including $H^1,
\Hcurl, \Hdiv,$ and $L^2$ spaces. MFEM currently supports partial assembly for
tensor-product elements (quadrilaterals and hexahedra), for which sum
factorization is most efficient. Partial assembly on simplices and mixed meshes
are partially supported through MFEM's integration with the libCEED library. In
the case of simplices, sum factorization cannot be used for the evaluation of
the action of the $B$ operator, however, other efficient algorithms exist, for
example using the Bernstein basis \cite{Ainsworth2011}.

\section{High-Performance Computing} \label{sec:hpc}
\subsection{Parallel Meshes, Spaces, and Operators} \label{ssec:parallel}
\begin{figure}[t]
\begin{center}
  \includegraphics{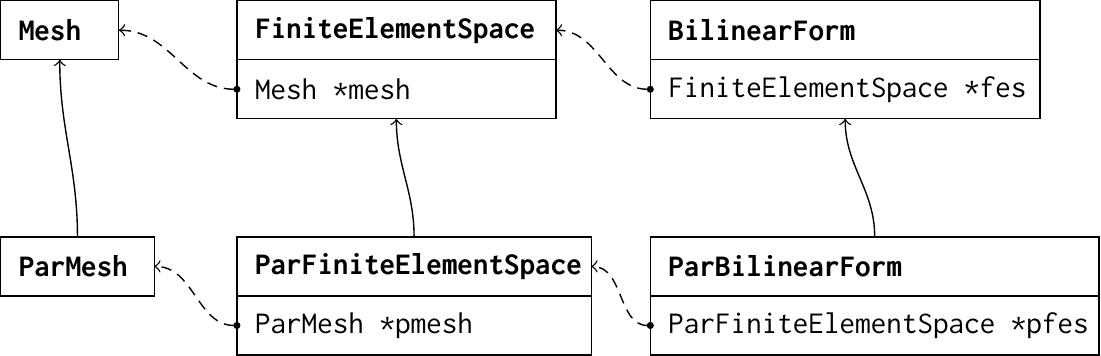}
\end{center}
\caption{Parallel classes inherit from, and partially override, serial classes.}
\label{fig:classes}
\end{figure}

The MFEM design handles large scale parallelism by utilizing the Message Passing
Interface (MPI) library in an additional layer\cred{,} that reuses as
much of the serial code as possible. In terms of object-oriented design, this is
done by sub-classing the serial classes to augment them with parallel logic, see
\autoref{fig:classes}, occasionally overriding small parts of the code using
virtual functions.

\begin{figure}[!hbt]
\begin{center}
  \includegraphics[width=0.3\textwidth]{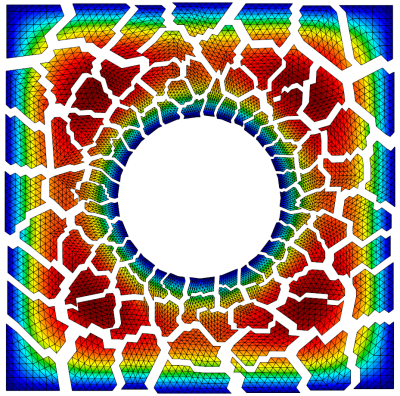}
  \hspace{1.2cm}
  \includegraphics[width=0.37\textwidth]{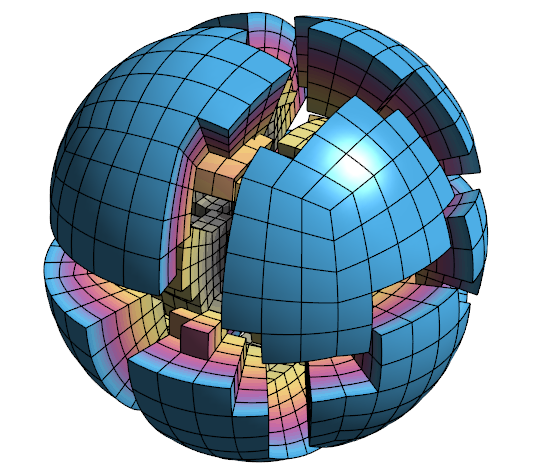}
\end{center}
\caption{
Left: Solving a Poisson problem (parallel example 1, \code{examples/ex1p.cpp}) in parallel on 100 processors with a relatively coarse version of \code{data/square-disc.mesh}.
Right: Unstructured parallel decomposition of a fourth order NURBS mesh of the unit ball on 16 processors.}
\label{fig:ex1p}
\end{figure}

If $K$ is the number of MPI tasks, MFEM decomposes the problem domain
(i.e.\ the mesh) into $K$
parts, with the goal of processing the parts as locally as possible, see
\autoref{fig:ex1p}. The parallel mesh object, \code{ParMesh}, is just a regular
serial \code{Mesh} on each MPI task plus additional information that describes
the geometric entities (faces, edges, vertices) that are shared with other
processors. See \autoref{ssec:mesh-parallel} for more details. The parallel
finite element space, \code{ParFiniteElementSpace} is just a regular serial
\code{FiniteElementSpace} on each task plus a description of the shared degrees
of freedom, grouped in {\em communication groups}. As in the serial case, one of
the main responsibilities of the parallel finite element space is to provide,
via \code{GetProlongationMatrix()}, the prolongation matrix $P$, see
\autoref{ssec:decomp}, which is used for parallel assembly (see below) or
adaptive mesh refinement, see \autoref{sec:adaptivity}. Parallel grid functions,
\code{ParGridFunction}, are just regular \code{GridFunction} objects on {\bf
L-vector} level which can be mapped back and forth to {\bf T-vectors}, e.g. with
the \code{ParallelAverage} and \code{Distribute} methods.

The finite element stiffness matrix at the {\bf L-vector} level,
$A_L \equiv G^T B^T D B G$, has $K$ diagonal blocks and can be assembled without
any parallel communication. The prolongation matrix $P$ is parallel and its
construction requires communication, however part of that communication can be
overlapped with the computation of $A_L$. As a general rule, we try to keep MPI
messages to a minimum and only communicate with immediate neighbors in the
parallel mesh, ideally overlapping communication with computation using
asynchronous MPI calls.

Based on the variational restriction perspective presented
in \autoref{ssec:decomp}, the final parallel assembly is computed with a
parallel $P^t\!A_L P$ triple matrix product, which is performed either with the
{\em hypre} library \cite{hypre} (making use of the \code{RAP}
triple-product kernel which {\em hypre} provides internally for the coarse-grid
operator construction in its algebraic multigrid solvers), or via PETSc routines,
depending on the underlying operator type set via the \code{SetOperatorType}
method of the \code{ParBilinearForm} class.

One of the advantages of handling parallelism by sub-classing the serial finite
element classes is that serial MFEM-based application codes are easily converted
to highly-scalable parallel versions by simply adding the \code{Par} prefix to
the types of finite element variables. To emphasize this point, the MFEM
distribution includes serial and parallel versions of most of its example codes,
so the changes needed to transition between the two are easy to compare.

\subsection{Scalable Linear Solvers} \label{ssec:solvers}
Parallel matrices in MFEM are computed and stored directly in the ParCSR format
of the {\em hypre} library, which gives the user direct access to
high-performance parallel linear algebra algorithms. For example, MFEM uses {\em
hypre}'s matvec routines, as well as the \code{RAP} function, see
\autoref{ssec:parallel}, which has been optimized in {\em hypre} for the
construction of coarse-grid operators in a multigrid hierarchy.

This tight integration with {\em hypre} enables MFEM applications to easily
access the powerful algebraic multigrid (AMG) preconditioner in the library,
which has demonstrated scalability to millions of parallel tasks. All parallel
MFEM examples are using these scalable preconditioners, which only take a line
of code in MFEM. For example the parallel linear system in
\code{examples/ex1p.cpp} is defined by
\begin{lstlisting}[firstnumber=205]
   OperatorPtr A;
   Vector B, X;
   a->FormLinearSystem(ess_tdof_list, x, *b, A, X, B);
\end{lstlisting}
and then {\em hypre}'s BoomerAMG preconditioner can be used with the
preconditioned conjugate gradient (CG) method to solve it simply with

\begin{lstlisting}[firstnumber=212]
   Solver *prec = NULL;
   if (!pa) { prec = new HypreBoomerAMG; }
   CGSolver cg(MPI_COMM_WORLD);
   cg.SetRelTol(1e-12);
   cg.SetMaxIter(2000);
   cg.SetPrintLevel(1);
   if (prec) { cg.SetPreconditioner(*prec); }
   cg.SetOperator(*A);
   cg.Mult(B, X);
\end{lstlisting}

In addition to general black-box solvers, such as BoomerAMG, the MFEM interface
enables access to {\em discretization-enhanced} AMG methods such as the
auxiliary-space Maxwell solver (AMS) \cite{kolev_vassilevski_ams} which is
specifically designed for second-order definite Maxwell problems discretized
with Nedelec \Hcurl-conforming elements, see \autoref{ssec:derham}. The AMS
algorithm needs the discrete gradient operator between the nodal $H^1$ and
the Nedelec spaces, which in MFEM is represented as a
\code{DiscreteLinearOperator} corresponding to an embedding between spaces. This
operator is constructed in general parallel settings (including on surfaces and
mesh skeletons) with the following code from \code{linalg/hypre.cpp}:

\begin{lstlisting}[firstnumber=2856]
   ParDiscreteLinearOperator *grad;
   grad = new ParDiscreteLinearOperator(vert_fespace, edge_fespace);
   if (trace_space)
   {
      grad->AddTraceFaceInterpolator(new GradientInterpolator);
   }
   else
   {
      grad->AddDomainInterpolator(new GradientInterpolator);
   }
   grad->Assemble();
   grad->Finalize();
   G = grad->ParallelAssemble();
\end{lstlisting}
From the user perspective, this is handled automatically given a
\code{FiniteElementSpace} object, and the use of AMS is also a one-liner in
MFEM. This is illustrated in the following excerpt from
\code{examples/ex3p.cpp}, which also shows how static condensation is seamlessly
handled by the preconditioner:
\begin{lstlisting}[firstnumber=196]
   ParFiniteElementSpace *prec_fespace =
      (a->StaticCondensationIsEnabled() ? a->SCParFESpace() : fespace);
   HypreSolver *ams = new HypreAMS(A, prec_fespace);
   HyprePCG *pcg = new HyprePCG(A);
   pcg->SetTol(1e-12);
   pcg->SetMaxIter(500);
   pcg->SetPrintLevel(2);
   pcg->SetPreconditioner(*ams);
   pcg->Mult(B, X);
\end{lstlisting}
Different preconditioning options are also easy to combine as
illustrated in Example 4p which solves an \Hdiv\ problem discretized with
Raviart-Thomas finite elements. Depending on the dimension, and the use of
hybridization or static condensation, see \autoref{ssec:fels}, several different
preconditioning options could be appropriate. All of them can be handled with
the following simple code segment:

\begin{lstlisting}[firstnumber=221]
   if (hybridization) { prec = new HypreBoomerAMG(A); }
   else
   {
      ParFiniteElementSpace *prec_fespace =
         (a->StaticCondensationIsEnabled() ? a->SCParFESpace() : fespace);
      if (dim == 2)   { prec = new HypreAMS(A, prec_fespace); }
      else            { prec = new HypreADS(A, prec_fespace); }
   }
   pcg->SetPreconditioner(*prec);
\end{lstlisting}

MFEM provides easy access to a variety of other iterative and direct solvers.
For example, discretization-enhanced Balancing Domain Decomposition by
\cred{Constraints} (BDDC) solvers from PETSc \cite{zampini2016pcbddc} are
exposed via the \code{PetscBDDCSolver} class. These methods provide
customizable, multilevel preconditioning for various finite element
discretizations, as well as for isogeometric analysis, see
\cite{da2014isogeometric} and \cite{zampini2016robustness} for a recent review.
Examples \code{examples/petsc/ex3p.cpp} and \code{examples/petsc/ex4p.cpp}
construct the BDDC solver for the second-order definite Maxwell equations
\cite{zampini2017adaptive,zampini2017balancing} as well as for the \Hdiv{}
\cite{oh2018bddc} problem, as shown in the below code snippet:

\begin{lstlisting}[numbers=none]
   PetscParMatrix A;
   a->SetOperatorType(Operator::PETSC_MATIS);
   a->FormLinearSystem(ess_tdof_list, x, *b, A, X, B);

   ParFiniteElementSpace *prec_fespace =
      (a->StaticCondensationIsEnabled() ? a->SCParFESpace() : fespace);
   PetscPCGSolver *pcg = new PetscPCGSolver(A);
   PetscPreconditioner *prec = NULL;

   PetscBDDCSolverParams opts;
   opts.SetSpace(prec_fespace);
   prec = new PetscBDDCSolver(A,opts);
   pcg->SetPreconditioner(*prec);
\end{lstlisting}

In addition, the \code{PetscBDDCSolver} class provides support for
preconditioning symmetric indefinite linear systems
\cite{zampini2017multilevel}, as shown in \code{examples/petsc/ex5p.cpp} for the
mixed \Hdiv\ - $L^2$ formulation of the Poisson equation. The same example
showcases MFEM's interface to the generic field-split solver
\code{PetscFieldSplitSolver} in PETSc, which can be used to quickly and easily
prototype block-preconditioning techniques for complicated multi-physics
problems.

With high-order methods, the explicit assembly of finite element matrices
becomes a bottleneck, as discussed in \autoref{ssec:pa}. While matrix-free
(partially assembled) high-order operators offer many benefits, one of their
drawbacks is that the entries of the matrix are not readily available, and thus
purely algebraic preconditioners cannot be used. An ongoing area of research
pursued by the MFEM team is the development of matrix-free preconditioners for
high-order operators. These include matrix-free $h$- and $p$-multigrid methods,
as well as low-order refined preconditioning, which is based on the idea of
preconditioning a spectrally equivalent low-order refined operator obtained by
meshing the nodes of each of the high-order elements
\cite{ORSZAG198070,DevilleMund_1985,Canuto2007,2019LOR,2019CF}.

\subsection{GPU Acceleration} \label{ssec:gpu}
\newcommand{\ins}[2][0.24]{\includegraphics[scale=#1,keepaspectratio]{./figs/#2.pdf}}

\label{sec:org9507bd3}

Version 4.0 of MFEM \cred{introduced} initial support for hardware accelerators, such
as GPUs, as well as programming models and libraries, such as CUDA, OCCA
\cite{occa}, libCEED \cite{libceed}, RAJA \cite{RAJA} and OpenMP in the library.
This support is based on new backends and kernels working seamlessly with a new
lightweight memory spaces manager.  Several of the MFEM example codes and the
Laghos miniapp \cite{laghos_web} (see \autoref{ssec:hydro}) can now take advantage of this GPU
acceleration.

\cgreen{
Given the rapidly changing computing landscape, the MFEM performance portability
approach has been to not commit to a single framework, but instead to support a
variety of different backends, which may differ in the set of features they
actually implement, the technology they use (OCCA, external library such as
libCEED, OpenMP, CUDA, RAJA, HIP), the targeted architecture (Intel, IBM, AMD,
Nvidia), the algorithms to achieve performance, and the implementations of these
algorithms. This flexibility allows generic backends like the core backend of
MFEM, using the macro \code{MFEM\_FORALL} described below, to target all architectures
with a performance emphasis on GPU architectures. The OCCA backend serves a
similar purpose of generic backend using the OCCA just-in-time compilation
technology, but varies in the algorithms used, and more significantly in the
implementation ideas. The libCEED backend uses the libCEED library that itself
contains numerous backends. This modularity over backends increases both the
portability and performance of MFEM algorithms, as different backends provide
the best performance in different scenarios, see \cite{CEED-MS6, bps20}.
}

One main feature of the MFEM performance portability approach is the ability to
select the backends at runtime: e.g. different MPI ranks can choose different
backends (like CPU or GPU) allowing applications to take full advantage of
heterogeneous architectures. Another important aspect of MFEM's approach is the
ability to easily mix CPU-only code with code that utilizes the new backends,
thus allowing for selective gradual transition of existing capabilities.

Most of the kernels are based on a single source, while still offering good
efficiency. For performance critical kernels, where single source does not
provide the best performance, the implementation introduces dispatch points
based on the selected backend and, in some cases, on kernel parameters such as
the finite element order. Many of the linear algebra and finite element
operations can now benefit fully from the new GPU acceleration.

\begin{figure}[t]
\centering
\includegraphics{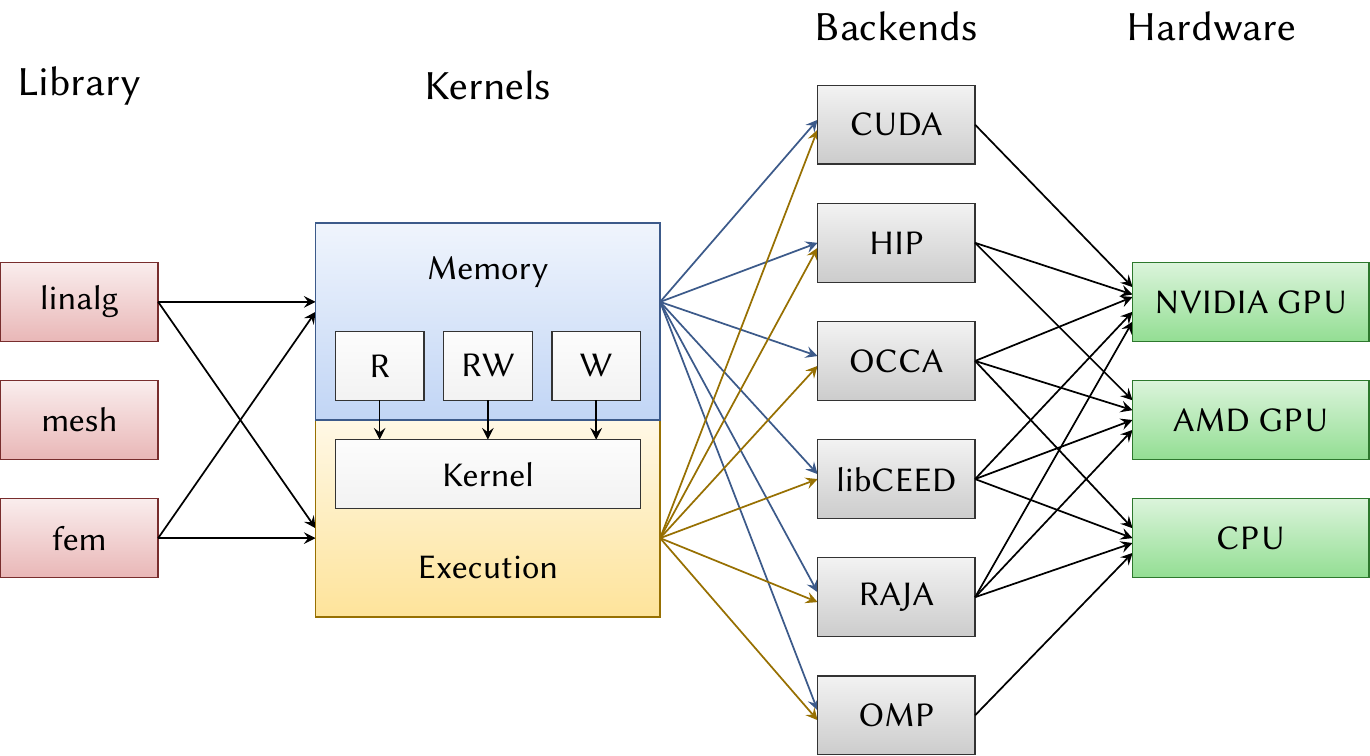}
\caption{\label{fig:org8bb88a8} Diagram of MFEM's modular design for accelerator
  support, combining flexible memory management with runtime-selectable backends
  for executing key finite element and linear algebra kernels.}
\end{figure}

Figure \ref{fig:org8bb88a8} illustrates the main components of MFEM's modular
design for accelerator support. The \textit{Library} side of MFEM (on the left)
represents the software components where new kernels have been added. The
following components have been extended with new accelerated \textit{kernels}:
\begin{itemize}
\item The \code{linalg} directory: most operations in class \code{Vector} and
some operations (e.g.\ matvec) in class \code{SparseMatrix}. Other classes, such
as the Krylov solvers and time-stepping methods, are automatically executed on
the device because they are written in terms of \code{Vector} operations.
\item The \code{mesh} directory: the computation of the so-called
\textit{geometric factors}.
\item \cgreen{The \code{fem} directory: the \textit{mass}, \textit{diffusion},
\textit{convection} (including DG), gradient, divergence, and some $\Hcurl$
\code{BilinearFormIntegrator}s; the \textit{element restriction} and
\textit{quadrature interpolator} operators ($G$ and $B$ on
\autoref{fig_op_decom}) associated with class \code{FiniteElementSpace}; the
matrix-free action of the \code{BilinearForm}, \code{MixedBilinearForm} and
\code{NonlinearForm} classes.}
\end{itemize}

\cgreen{
Note, however, that many of the capabilities in the library are still not
ported to GPU including the mesh refinement/derefinement, a number of the
\code{BilinearFormIntegrator} classes, sparse matrix assembly, error estimation,
integration with external libraries, etc. Some of these missing parts are
currently under development and will become available in the near future.
}

The integration of the kernels has been made at the \emph{for-loop} level.
Existing code has been transformed to use a new \emph{for-loop} abstraction
defined as a set of new \code{MFEM\_FORALL} macros, in order to take
advantage of various \textit{backends} supported via the new macros.  This
approach allows for gradual code transformations that are not too disruptive for
both, MFEM developers and users. Existing applications based on MFEM should
be able to continue to work as before and have an easy way to transition to
accelerated kernels. Another requirement is to allow interoperability with
other software components or external libraries that MFEM could be used in
conjunction with, for instance {\em hypre}, PETSc, and SUNDIALS, among others.

The main challenge in this transition to {\em kernel-centric} implementation is
the need to transform existing algorithms to take full advantage of the
increased levels of parallelism in the accelerators while maintaining good
performances on standard CPU architectures. Another important aspect is the need
to manage memory allocation and transfers between the CPU (host) and the
accelerator (device). In MFEM, this is achieved using a new \code{Memory} class
that manages a pair of host and device pointers and provides a simple interface
for copying or moving the data when needed. An important feature of this class
is the ability to work with externally allocated host and/or device pointers
which is essential for interoperability with other libraries.

\subsubsection*{Lambda-capturing \emph{for-loop} bodies}
\label{sec:orga77ea16}
There are multiple ways to write kernels, but one of the easiest ways,
from the developer's point of view, is to turn \emph{for-loop} bodies
into \emph{kernels} by keeping the bodies unchanged and having a way
to \emph{wrap} and \emph{dispatch} them toward native backends. This
can be easily done for the first outer for-loop using standard
\texttt{C++11} features. However, additional care is required when one
wants to address deeper levels of parallelism. The following listing
illustrates a possible implementation in MFEM of the diffusion setup
(partial assembly) kernel in 2D.

\begin{lstlisting}
void PADiffusionSetup2D(const int Q, const int N, const Array<double> &w,
                        const Vector &j, const double alpha, Vector &y) {
   auto W = w.Read();
   auto J = Reshape(j.Read(), Q*Q, 2, 2, N);
   auto Y = Reshape(y.Write(), Q*Q, 3, N);
   MFEM_FORALL_2D(e, N, Q, Q,
      MFEM_FOREACH_THREAD(qx, x, Q)
         MFEM_FOREACH_THREAD(qy, y, Q) {
            const int q = qx + qy * Q;
            const double J11 = J(q,0,0,e), J21 = J(q,1,0,e);
            const double J12 = J(q,0,1,e), J22 = J(q,1,1,e);
            const double c_detJ = alpha * W[q] / ((J11*J22)-(J21*J12));
            Y(q,0,e) =  c_detJ * (J12*J12 + J22*J22);
            Y(q,1,e) = -c_detJ * (J12*J11 + J22*J21);
            Y(q,2,e) =  c_detJ * (J11*J11 + J21*J21);
         });
}
\end{lstlisting}

The kernel is structured as follows:
\begin{itemize}
\item Lines \texttt{3} to \texttt{5} are the portion of the kernel where the pointers
are requested from the memory manager (presented in the next paragraph) and
turned into tensors with given shapes.
\item Line \texttt{6} holds the \code{MFEM\_FORALL\_2D} \textit{wrapper} of the first
  outer \emph{for-loop}, with the iterator, the range, and the \emph{for-loop} body.
\item Lines \texttt{7} and \texttt{8} allow inner for-loops to be
  mapped to blocks of threads with arbitrary sizes (from 1 to
  thousands): it uses another level of parallelism \cred{within} the lambda
  body for each mesh element.
\item Lines \texttt{9} to \texttt{15} are the core of the computation
  and show how to use the tensors declared before entering the kernel.
  This portion may use \textit{shared} memory as a fast scratch memory
  shared \cred{within} the thread block when supported by the respective
  backend. This kernel is the one used both for the OpenMP and the
  CUDA backends.
\end{itemize}

\subsubsection*{Memory management}
\label{sec:orgf9ab449}
Before entering each kernel, the pointers that will be used in it have to be
requested from the new \code{Memory} class which acts as the \textit{frontend}
of the internal lightweight MFEM memory manager. Access to the pointers stored
by the \code{Memory} class is requested using three modes: \code{Read}-only,
\code{Write}-only, and \code{ReadWrite}. These access types allow the memory
manager to seamlessly copy or move data to the device when needed. Portions of
the code that do not use acceleration (i.e.\ run on CPU) need to request access
to the \code{Memory} using the \textit{host} versions of the three access
methods: \code{HostRead}, \code{HostWrite}, and \code{HostReadWrite}. The use of
these access types allows the memory manager to minimize memory transfers
between the host and the device. The pointers returned by the three access
methods can be reshaped as \textit{tensors} with given dimensions using the
function \code{Reshape} which then allows for easy multi-dimensional indexing
inside the computational kernels.

In addition to holding the host and device pointers, the
\code{Memory} class keeps extra metadata in order to keep track of the
usage of the different memory spaces. For example, if a vector currently
residing in device memory is temporarily needed on the host where it will not be
modified (e.g.\ to save the data to a file), the host code can use
\code{HostRead} to tell the memory manager to copy the data to the host while
also telling it that the copied data will not be modified; using this
information, the memory manager knows that a subsequent call to, say,
\code{Read} will not require a memory copy from host to device.

\subsubsection*{Transitioning applications to GPUs}
Porting existing codes to GPUs can be relatively simple in some cases. The first
step is to configure an \code{mfem::Device} object, e.g. using a string from a
command-line option. The next step is typically to enable the partial assembly
mode in the \code{(Par)BilinearForm} object(s). Since in this mode the fully
assembled sparse matrix is not available, one has to switch to suitable
matrix-free solvers. In cases when an application uses MFEM at a lower level,
e.g. to implement some algorithm on an element level, porting to GPU will be
more involved. For such cases, the user will typically need to learn more about
the \code{MFEM\_FORALL} macros and the memory management.

Some current limitations in the GPU support are: not all pre-defined integrators
in MFEM have been ported to GPU; full assembly on GPU (which may be of interest
for low-orders) is also not available. \cred{As pointed out earlier,} these and many other missing components
are being actively developed and will become immediately available in the MFEM
source repository when completed.

\subsubsection*{Results}
\label{sec:org50f3c54}
\begin{figure}[!h]
\centering
\includegraphics{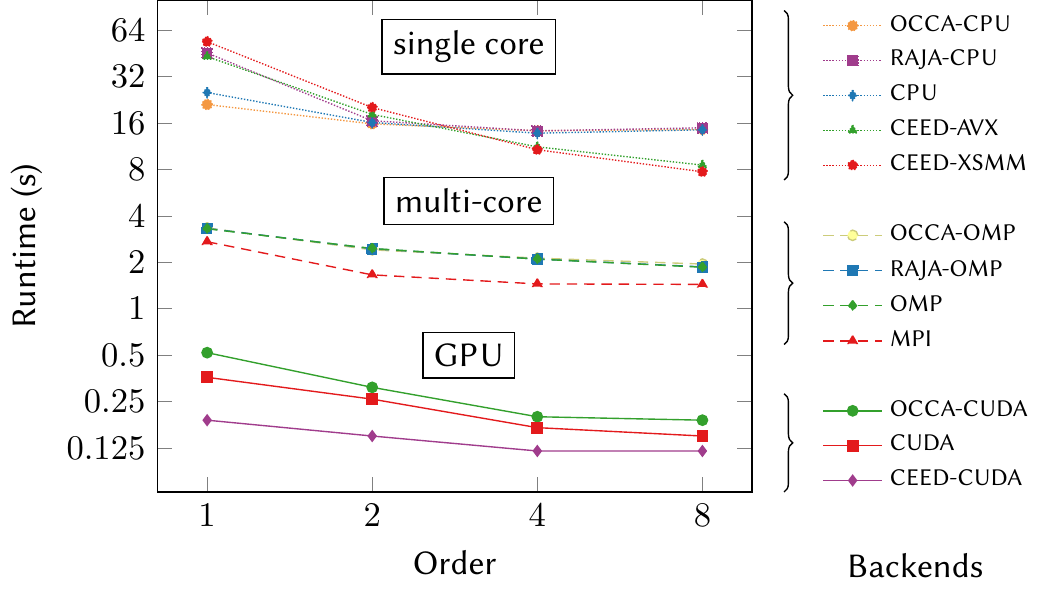}
\caption{\label{fig:orge744447}
Performance results with MFEM-4.0: Poisson problem (Example 1), 200 conjugate gradient iterations using partial assembly, 2D, 1.3M dofs, GV100, sm\_70, CUDA 10.1, Intel Xeon Gold 6130@2.1GHz}
\end{figure}

Figure \ref{fig:orge744447} and Table \ref{fig:org0021fee} present initial
performance results with MFEM v4.0 measured on a Linux desktop with a Quadro
GV100 GPU (Volta, 5120 cuda cores, 7.4 TFLOPS FP64 peak; 32 GB HBM2, 870 GB/s
peak), CUDA 10.1, and Intel Xeon Gold 6130 CPU (Skylake, 16 cores/32 threads,
970 GFLOPS FP64 peak; 128 GB/s memory bandwidth peak) @ 2.10GHz.

Single-core, multi-core CPU, and single-GPU performance for different
discretization orders is shown, keeping the total number of degrees of freedom
(DOFs) constant at 1.3 million in 2D. Results from backends supported in MFEM
4.0, as well as recent results based on the libCEED library (integrated with
MFEM) are included. \cgreen{The libCEED library itself includes several
backends, targeting, for example, CPUs using AVX instructions, Intel CPUs taking
advantage of the LIBXSMM library \cite{Heinecke2016}, and GPUs using CUDA.}
\autoref{fig:orge744447} shows that
GPU acceleration offers a significant gain in performance relative to
multi-core CPU.

We emphasize that these results are preliminary and additional performance
improvements in several of the backends are under active development. Therefore
these results illustrate only the current state of the MFEM backends, and should
not be viewed as a fair and exhaustive comparison of the specific CPU and GPU
hardware.

\begin{table}[!h]
\renewrobustcmd{\bfseries}{\fontseries{b}\selectfont}
\renewrobustcmd{\boldmath}{}
\newrobustcmd{\B}{\bfseries}
\tbl{Performance results with MFEM-4.0: Poisson problem (Example 1), 200 conjugate gradient iterations using partial assembly, 2D, 1.3M
dofs, GV100, sm\_70, CUDA 10.1, Intel Xeon Gold 6130@2.1GHz. The best performing
backends in each category (GPU, multicore, and CPU) are shown in bold.}{%
\begin{tabular}{rrrrrr}
\toprule
 & & $p=1$ & $p=2$ & $p=4$ & $p=8$ \\
\midrule
\parbox[t]{2mm}{\multirow{4}{*}{\rotatebox[origin=c]{90}{GPU}}}
& OCCA-CUDA & 0.52 & 0.31 & 0.20 & 0.19 \\
& RAJA-CUDA & 0.38 & 0.30 & 0.28 & 0.45 \\
& CUDA      & 0.36 & 0.26 & 0.17 & 0.15 \\
& CEED-CUDA & \B 0.19 & \B 0.15 & \B 0.12 & \B 0.12 \\
\midrule
\parbox[t]{2mm}{\multirow{4}{*}{\rotatebox[origin=c]{90}{Multicore}}}
& OCCA-OMP  & 3.34 & 2.41 & 2.13 & 1.95 \\
& RAJA-OMP  & 3.32 & 2.45 & 2.10 & 1.87 \\
& OMP       & 3.30 & 2.46 & 2.10 & 1.86 \\
& MPI       & \B 2.72 & \B 1.66 & \B 1.45 & \B 1.44 \\
\midrule
\parbox[t]{2mm}{\multirow{5}{*}{\rotatebox[origin=c]{90}{CPU}}}
& OCCA-CPU  & \B 21.05 & \B 15.77 & 14.23 & 14.53 \\
& RAJA-CPU  & 45.42 & 16.53 & 14.22 & 14.88 \\
& CPU       & 25.18 & 16.11 & 13.73 & 14.45 \\
& CEED-AVX  & 43.04 & 18.16 & 11.20 & 8.53  \\
& CEED-XSMM & 53.80 & 20.13 & \B 10.73 & \B 7.72  \\
\bottomrule
\end{tabular}}
\label{fig:org0021fee}
\end{table}

\section{Finite Element Adaptivity} \label{sec:adaptivity}
MFEM includes extensive support for serial and parallel finite element
adaptivity on general high-order unstructured meshes, including: local
conforming mesh refinement on triangular and tetrahedral meshes (conforming
$h$-adaptivity), non-conforming adaptive mesh refinement on quadrilateral and
hexahedral meshes (non-conforming $h$-adaptivity), and support for mesh
optimization by node movement ($r$-adaptivity).  The unified support for local
refinement on simplex and tensor-product elements is one of the distinguishing
features of the MFEM library.
These capabilities are described in the following subsections. Additional
parallel conforming mesh adaptivity and modification algorithms are available
via the integration with RPI's parallel unstructured mesh infrastructure (PUMI)
\cite{pumi}.
\subsection{Conforming Adaptive Mesh Refinement} \label{ssec:conf-amr}
The conforming $h$-adaptivity algorithm in MFEM is based on the bisection
procedure for tetrahedral meshes proposed in \cite{Bisection00}. This approach
supports both uniform refinement of all elements in the mesh, as well as local
refinement of only elements of interest with additional (forced) refinement of
nearby elements to ensure a conforming mesh. Note that in parallel these forced
refinements may propagate to neighboring processors, which MFEM handles
automatically for the user.

When a tetrahedral mesh is marked for refinement with
\code{Mesh::MarkForRefinement()} the vertices of each tetrahedron are permuted so
that the longest edge of the tetrahedron becomes the edge between vertices 0 and
1. MFEM ensures that the longest edge in each tetrahedron is chosen consistently
in neighbor tetrahedra based on a global sort of all edges (by length).  The
edge between vertices 0 and 1 becomes the marked edge, i.e. the edge that will
be bisected during refinement. Initially, this is the longest edge in the
element (with equal length edges ordered according to the global sort). However,
later, the bisection algorithm may choose to mark an edge that is not the
longest. When a tetrahedron is bisected, its type (M, A, etc., see
\cite{Bisection00}) determines which edges in the two children become marked, as
well as what types are assigned to them. The initial type of the tetrahedron is
also determined based on the globally sorted edges.

The bisection algorithm consists of several passes. For example, during {\em
green} refinement (cf.~\cite{Bisection00}), every tetrahedron is checked if it
``needs refinement'' by calling the method \code{Tetrahedron::NeedRefinement()}
and if it does, the element is bisected once. The method
\code{NeedRefinement()} returns true if any of its edges have been refined.
When a tetrahedron is bisected, it is replaced (in the list of elements) by one
of its children and the other child is appended at the end of the element list.
That way, the children will be checked if they need refinement in the next loop
over the elements. If no elements ``need refinement'', the green refinement
step is done.

In parallel, the tetrahedra are marked consistently across processors, as
inherited from the serial mesh before the parallel partitioning. The
consistently marked tetrahedra guarantee that a face between any two
tetrahedra will be refined the same way from both sides. This implies in
particular that uniform refinement can be performed in parallel without
communication. In the case of local refinement we need to know which of the five
possible cases of face refinement was actually performed on the other side of a
shared face.

\subsection{Non-Conforming Adaptive Mesh Refinement} \label{ssec:nc-amr}
Many high-order applications can be enriched by parallel adaptive mesh
refinement (AMR) on unstructured quadrilateral and hexahedral meshes.
Quadrilateral and hexahedral elements are attractive for their tensor product
structure (enabling efficiency, see \autoref{ssec:pa}) and for their
refinement flexibility (enabling e.g., {\em anisotropic} refinement). However,
as opposed to the bisection-based methods for simplices considered in the
previous section, {\em hanging} nodes that occur after local refinement of
quadrilaterals and hexahedra are not easily avoided by further refinement
\cite{Demkowicz89,Schonfeld95,Heuveline07}. We are thus interested in {\em
non-conforming} (irregular) meshes, in which adjacent elements need not share a
complete face or edge and where some finite element degrees of freedom (DOFs)
need to be constrained to obtain a conforming solution.

In this section we review MFEM's software abstractions and algorithms for
handling parallel non-conforming meshes on a general discretization level,
independent of the physics simulation. These methods support the entire de Rham
sequence of finite element spaces (see \autoref{ssec:derham}), at arbitrarily
high-order, and can support high-order curved meshes, as well as finite element
techniques such as hybridization and static condensation (see
\autoref{ssec:fels}). They are also highly scalable, easy to incorporate into
existing codes, and can be applied to complex, anisotropic 3D meshes with
arbitrary levels of non-conforming refinement. While MFEM's approaches can be
exclusively on non-conforming $h$-refinement with fixed polynomial degree.

\begin{figure}[!hbt]
\begin{center}
  \includegraphics[width=0.3\textwidth]{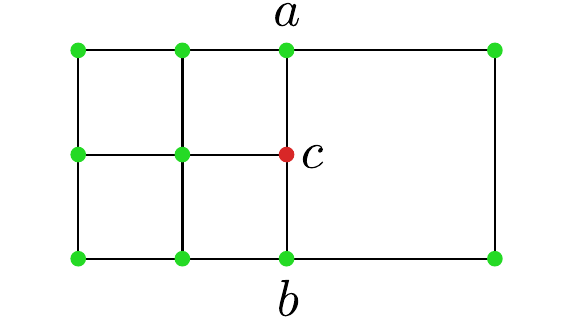}
  \hspace{1.2cm}
  \includegraphics[width=0.3\textwidth]{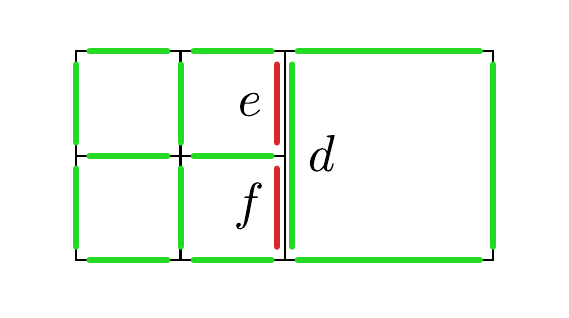}
\end{center}
\caption{Illustration of conformity constraints for lowest order nodal elements
in 2D.  Left: Nodal elements (subspace of $H^1$), constraint $c = (a+b)/2$.
Right: Nedelec elements (subspace of \Hcurl, constraints $e = f = d/2$.  In all
cases, fine degrees of freedom on a coarse-fine interface are linearly
interpolated from the values at coarse degrees of freedom on that interface.  }
\label{fig:local-simple}
\end{figure}

These approaches are based on a variational restriction approach to AMR,
described below. For more details, see \cite{2018-AMR}. Consider the weak
variational formulation \eqref{eq:abstractbvp} where for simplicity we assume
that the bilinear form $a(\cdot,\cdot)$ is symmetric.  To discretize the
problem, we cover the computational domain $\Omega$ with a mesh consisting of
mutually disjoint elements $K_i$, their vertices $V_j$, edges $E_m$, and faces
$F_n$. Except for the vertices, we consider these entities as open sets, so that
$\Omega = (\cup_i K_i) \cup (\cup_j V_j) \cup (\cup_m E_m) \cup (\cup_n F_n)$.
In the case of non-conforming meshes, there exist faces $F_s$ that are strict
subsets of other faces, $F_s \subsetneq F_m$, see \autoref{fig:local-simple}. We
call $F_s$ {\em slave faces} and $F_m$ {\em master faces}. The remaining
standard faces $F_c$ are disjoint with all other faces and will be referred to
as {\em conforming faces}. Similarly, we define {\em slave edges}, {\em master
edges} and {\em conforming edges}.

Non-conforming meshes in MFEM are represented by the \code{NCMesh} and
\code{ParNCMesh} classes. We use a tree-based data structure to represent
refinements which has been optimized to rely only on the following information:
1) elements contain indices of eight vertices, or indices of eight child
elements if refined; 2) edges are identified by pairs of vertices; 3) faces are
identified by four vertices. Edges and faces are tracked by associative maps
(see below), which reduce both code complexity and memory footprint. In the case
of a uniform hexahedral mesh, our data structure requires about 290 bytes per
element, counting the complete refinement hierarchy and including vertices,
edges, and faces.

To construct a standard finite dimensional FEM approximation space $V_h \subset
V$ on a given non-conforming mesh, we must ensure that the necessary conformity
requirements are met between the slave and master faces and edges so that we get
$V_h$ that is a (proper) subspace of $V$. For example, if $V$ is the Sobolev
space $H^1$, the solution values in $V_h$ must be kept continuous across the
non-conforming interfaces. In contrast, if $V$ is an \Hcurl\ space, the
tangential component of the finite element vector fields in $V_h$ needs to be
continuous across element faces. More generally, the conformity requirement can
be expressed by requiring that values of $V_h$ functions on the slave faces
(edges) are interpolated from the finite element function values on their master
faces (edges). Finite element degrees of freedom on the slave faces (and edges)
are thus effectively constrained and can be expressed as linear combinations of
the remaining degrees of freedom. The simplest constraints for finite element
subspaces of $H^1$ and \cred{$\Hcurl$} in 2D are illustrated
in \autoref{fig:local-simple}.

\begin{figure}[!hbt]
\begin{center}
  \includegraphics[width=0.31\textwidth]{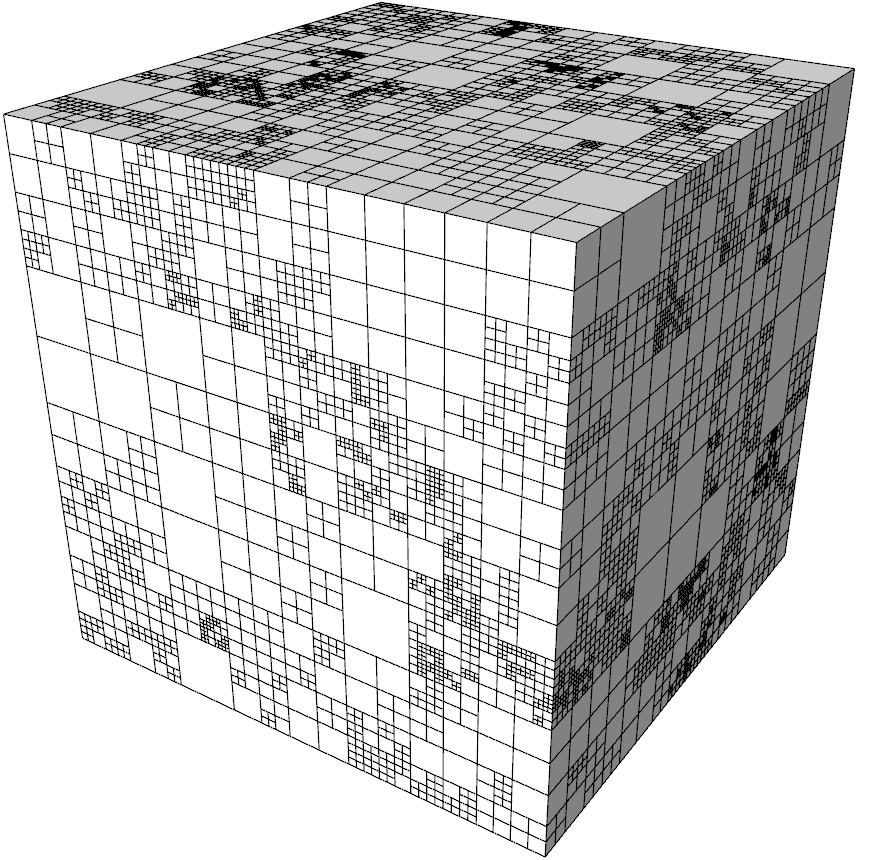} \hfill
  \includegraphics[width=0.31\textwidth]{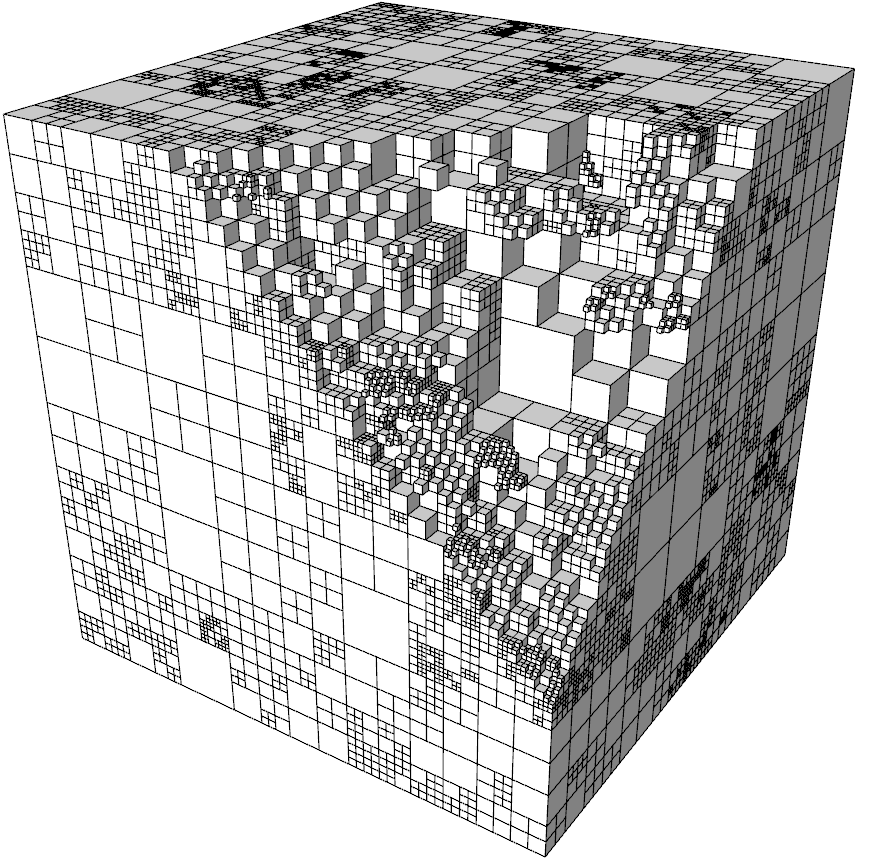} \hfill
  \includegraphics[width=0.31\textwidth]{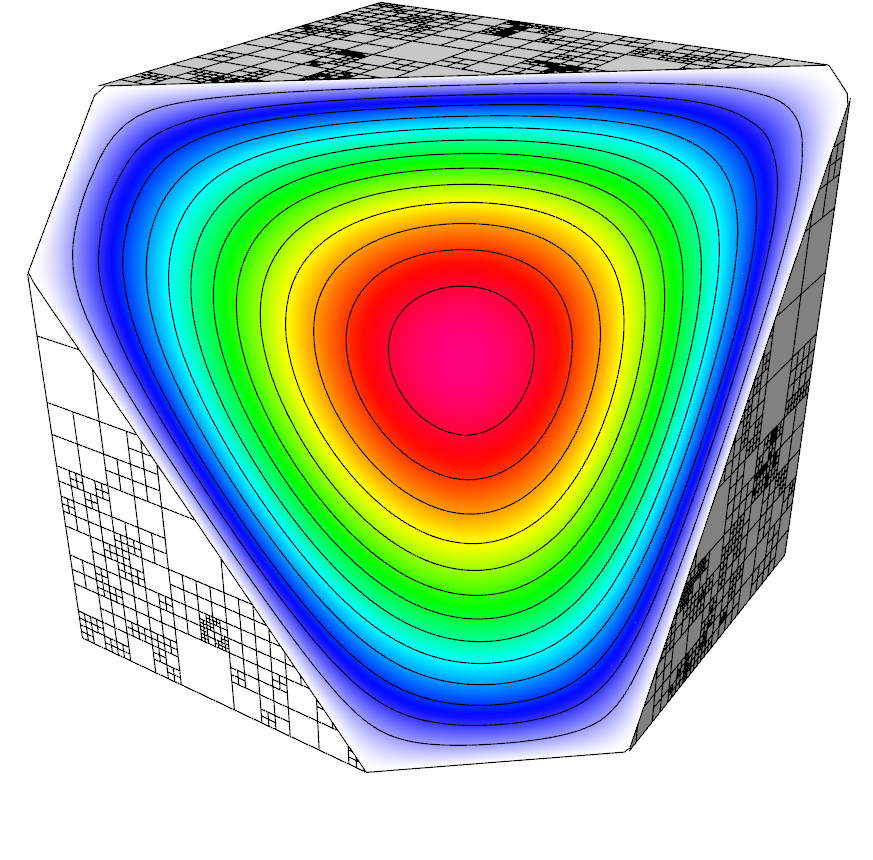}
\end{center}
\caption{Illustration of the variational restriction approach to forming the
  global AMR problem. Randomly refined non-conforming mesh (left and center)
  where we assemble the matrix $A$ and vector $b$ independently on each element.
  The interpolated solution $x = P x_c$ (right) of the system
  \eqref{eq:system} is globally conforming (continuous for an $H^1$ problem). }
\label{fig:PtAP}
\end{figure}

The degrees of freedom can be split into two groups: {\em unconstrained (or
true)} degrees of freedom and {\em constrained (or slave)} degrees of
freedom. If $z$ is a vector of all slave DOFs, then $z$ can
be expressed as $z = Wx$, where $x$ is a vector of all true DOFs and $W$ is a
global interpolation matrix, handling indirect constraints and arbitrary
differences in refinement levels of adjacent elements. Introducing the {\em
conforming prolongation matrix}
$$
P = \begin{pmatrix}I \\ W\end{pmatrix},
$$
we observe that the coupled AMR linear system can be written as
\begin{equation}
P^t A P x_c = P^t b, \label{eq:system}
\end{equation}
where $A$ and $b$ are the finite element stiffness matrix and load vector
corresponding to discretization of \eqref{eq:abstractbvp} on the ``cut'' space
(see \autoref{ssec:mesh-nc}) $\widehat{V}_h = \cup_i (V_h|_{K_i})$. After
solving for the true degrees of freedom $x_c$ we recover the complete set of
degrees of freedom, including slaves, by calculating $x = P x_c$. Note that in
MFEM this is handled automatically for the user via \code{FormLinearSystem()}
and \code{RecoverFEMSolution()}, see \autoref{ssec:fels}. An illustration of
this process is provided in Figure \ref{fig:PtAP}.

\begin{figure}[!hbt]
  \centering \includegraphics[width=0.315\textwidth]{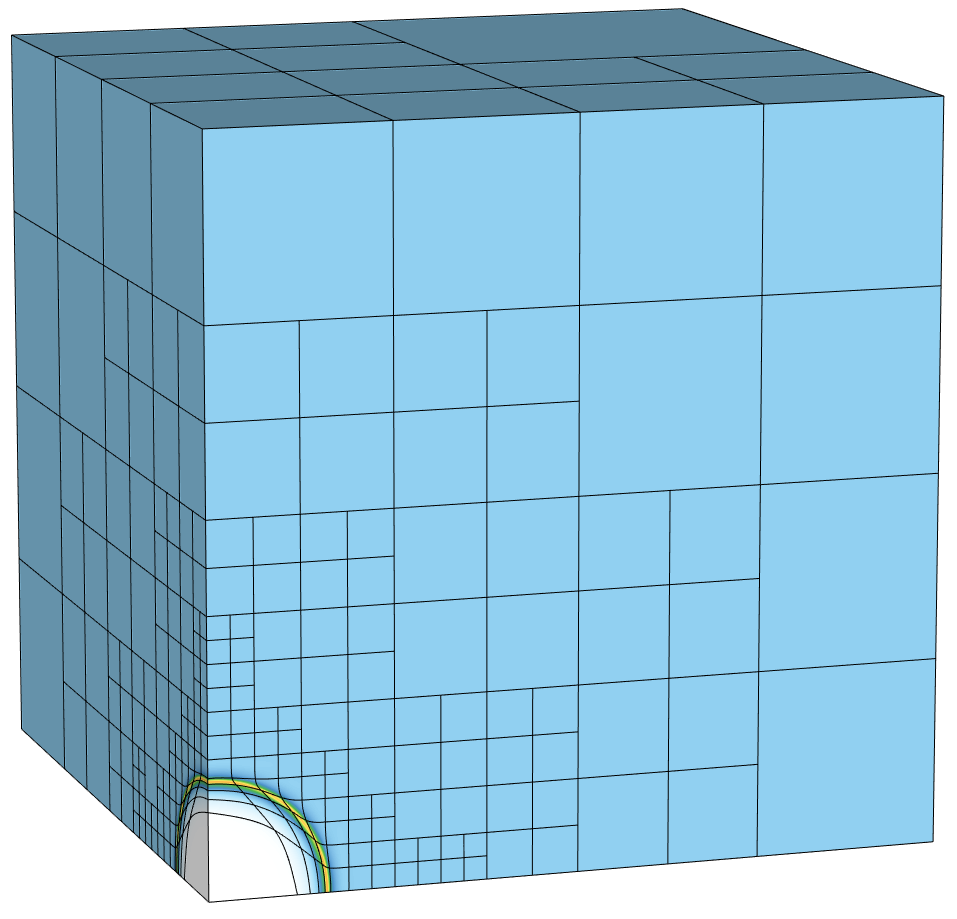}
  \hspace{1mm}
  \includegraphics[width=0.315\textwidth]{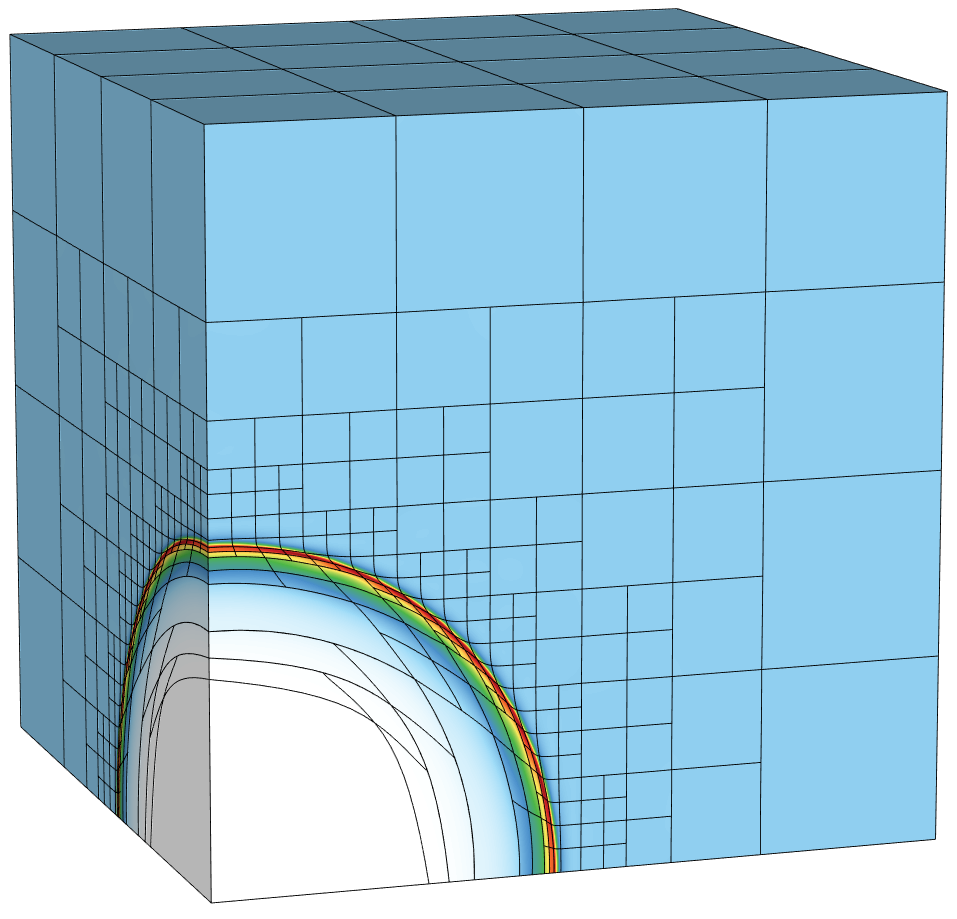}
  \hspace{1mm}
  \includegraphics[width=0.315\textwidth]{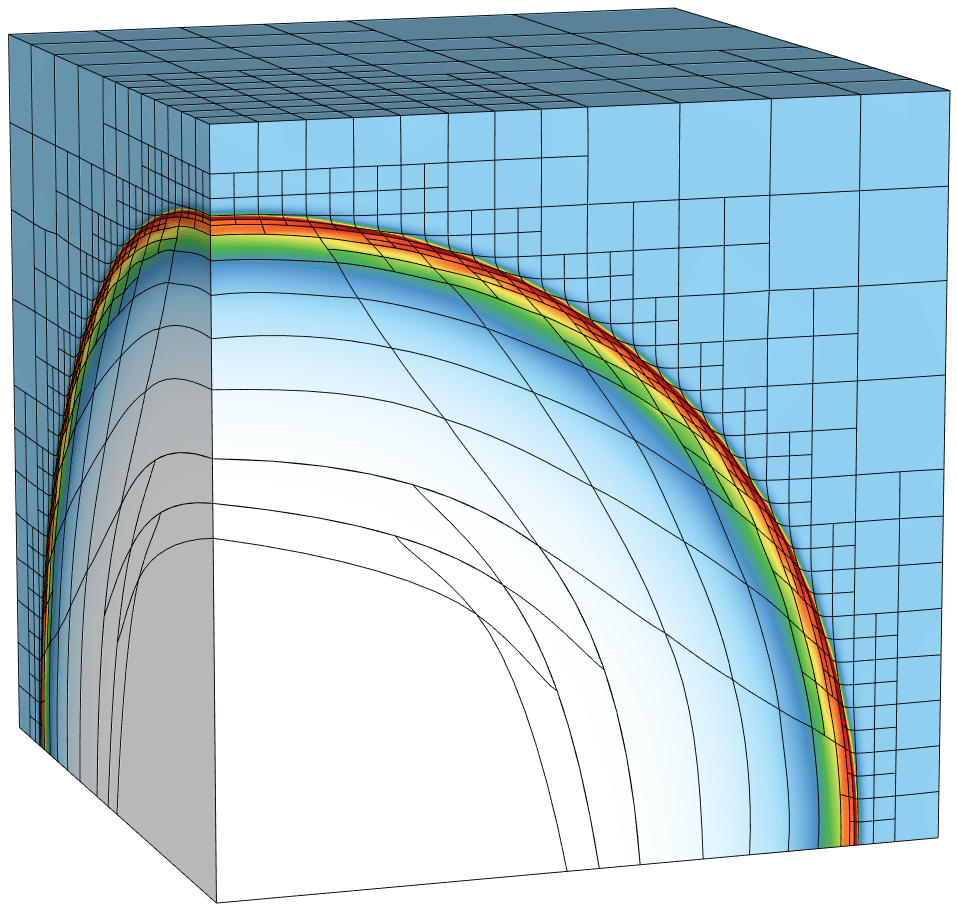}
  \caption{MFEM-based dynamic refinement/derefinement in the 3D Sedov blast problem.  Mesh
    and density shown at $t = 0.0072$ (left), $t = 0.092$ (center) and $t =
    0.48$ (right). Q3Q2 elements ($p = 3$ kinematic, $p = 2$ thermodynamic
    quantities).}
  \label{fig:amr-sedov-dynamic}
\end{figure}

In MFEM, given an \code{NCMesh} object, the conforming prolongation matrix can
be defined for each \code{FiniteElementSpace} class and accessed with the
\code{GetConformingProlongation()} method.  The algorithm for constructing this
operator can be interpreted as a sequence of interpolations $P = P_k P_{k-1}
\cdots P_1$, where for a $k$-irregular mesh the DOFs in $\widehat{V}_h$ are
indexed as follows: $0$ corresponds to true DOFs, $1$ corresponds to the first
generation of slaves that only depend on true DOFs, $2$ corresponds to second
generation of slaves that only depend on true DOFs and first generation of
slaves, and so on. $k$ corresponds to the last generation of slaves. We have
$$
P_1 =
\begin{pmatrix}
I \\ W_{10}
\end{pmatrix} ,\>
P_2 =
\begin{pmatrix}
I & 0 \\ 0 & I \\ W_{20} & W_{21}
\end{pmatrix} ,\> \ldots \>,\>
P_k =
\begin{pmatrix}
I & 0 & \cdots & 0 \\
0 & I & \cdots & 0 \\
  &   & \ddots &   \\
0 & 0 & \cdots & I \\
W_{k0} & W_{k1} & \cdots & W_{k(k-1)}
\end{pmatrix}
$$
are the local interpolation matrices defined only in terms of the edge-to-edge
and face-to-face constraining relations.
\cblue{Note that while MFEM supports meshes of arbitrary irregularity ($k \ge 1$),
the user can specify a limit on $k$ when refining elements, if necessary (an
example of a 1-irregular mesh is shown in Figure \ref{fig:amr-sedov-dynamic}).}

\begin{figure}[!hbt]
  \centering
  \includegraphics[width=0.3\textwidth]{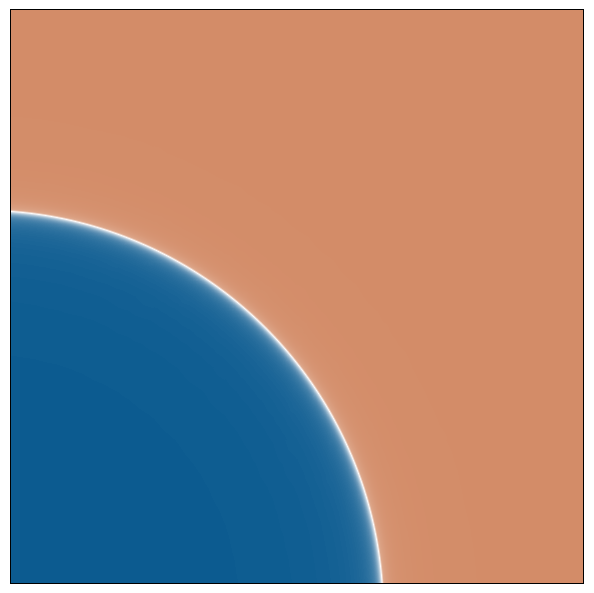} \hfill
  \includegraphics[width=0.3\textwidth]{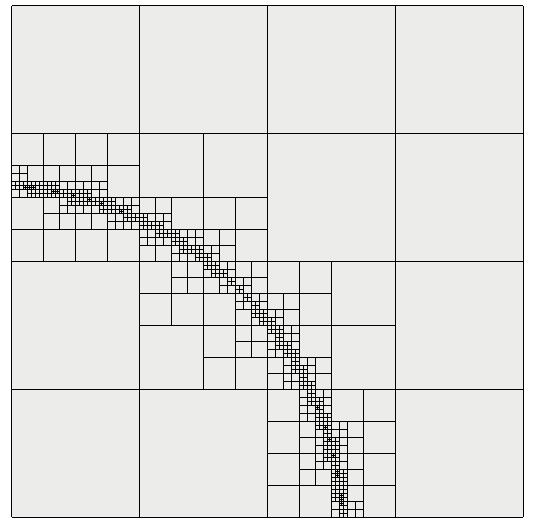} \hfill
  \includegraphics[width=0.3\textwidth]{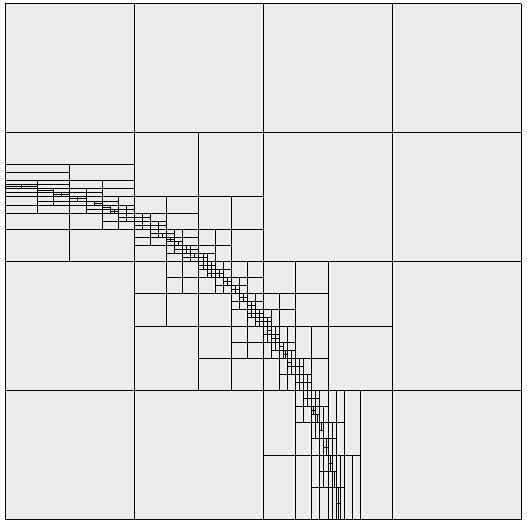}
  \caption{Left: 2D benchmark problem for a Poisson problem with a known exact
           solution. Center: Isotropic AMR mesh with 2197 DOFs. Right: Anisotropic
           AMR mesh with 1317 DOFs. Even though the wave front in the solution is not perfectly
           aligned with the mesh, many elements could still be refined in one direction
           only, which saved up to 48\% DOFs in this problem for similar error.}
  \label{fig:layer2d}
\end{figure}

The basis for determining face-to-face relations between hexahedra is the
function \code{FaceSplitType}, sketched below. Given a face $(v_1, v_2, v_3,
v_4)$, it tries to find mid-edge and mid-face vertices and determine if the face
is split vertically, horizontally (relative to its reference domain), or not
split.
\begin{lstlisting}
Split FaceSplitType(v1, v2, v3, v4)
{
   v12 = FindVertex(v1 , v2);
   v23 = FindVertex(v2 , v3);
   v34 = FindVertex(v3 , v4);
   v41 = FindVertex(v4 , v1);

   midf1 = (v12 != NULL && v34 != NULL) ? FindVertex(v12, v34) : NULL;
   midf2 = (v23 != NULL && v41 != NULL) ? FindVertex(v23, v41) : NULL;

   if (midf1 == NULL && midf2 == NULL)
      return NotSplit;
   else
      return (midf1 != NULL) ? Vertical : Horizontal;
}
\end{lstlisting}
The function \code{FindVertex} uses a hash table to map end-point vertices to
the vertex in the middle of their edge.  This algorithm naturally supports
anisotropic refinement, as illustrated in \autoref{fig:layer2d}.

The algorithm to build the $P$ matrix in parallel is more complex, but
conceptually similar to the serial algorithm. We still express slave DOF rows of
$P$ as linear combinations of other rows, however some of them may be located on
other MPI tasks and may need to be communicated first.

Unlike the conforming
\code{ParMesh} class, which is partitioned with METIS, the \code{ParNCMesh} is
partitioned between MPI tasks by splitting a space-filling curve obtained by
enumerating depth-first all leaf elements of all refinement trees
\cite{Aluru}.  The simplest traversal with a fixed order of children at each
tree level leads to the well-known Morton ordering, or the Z-curve. We use
instead the more efficient Hilbert curve that can be obtained just by changing
the order of visiting subtrees at each level \cite{Zoltan12}. The
use of space-filling curve partitioning ensures that balancing the mesh so that
each MPI task has the same number of elements ($\pm1$ if the total number of
elements is not divisible by the number of tasks) is relatively straightforward.

\begin{figure}[!hbt]
\begin{center}
  \includegraphics[width=0.35\textwidth]{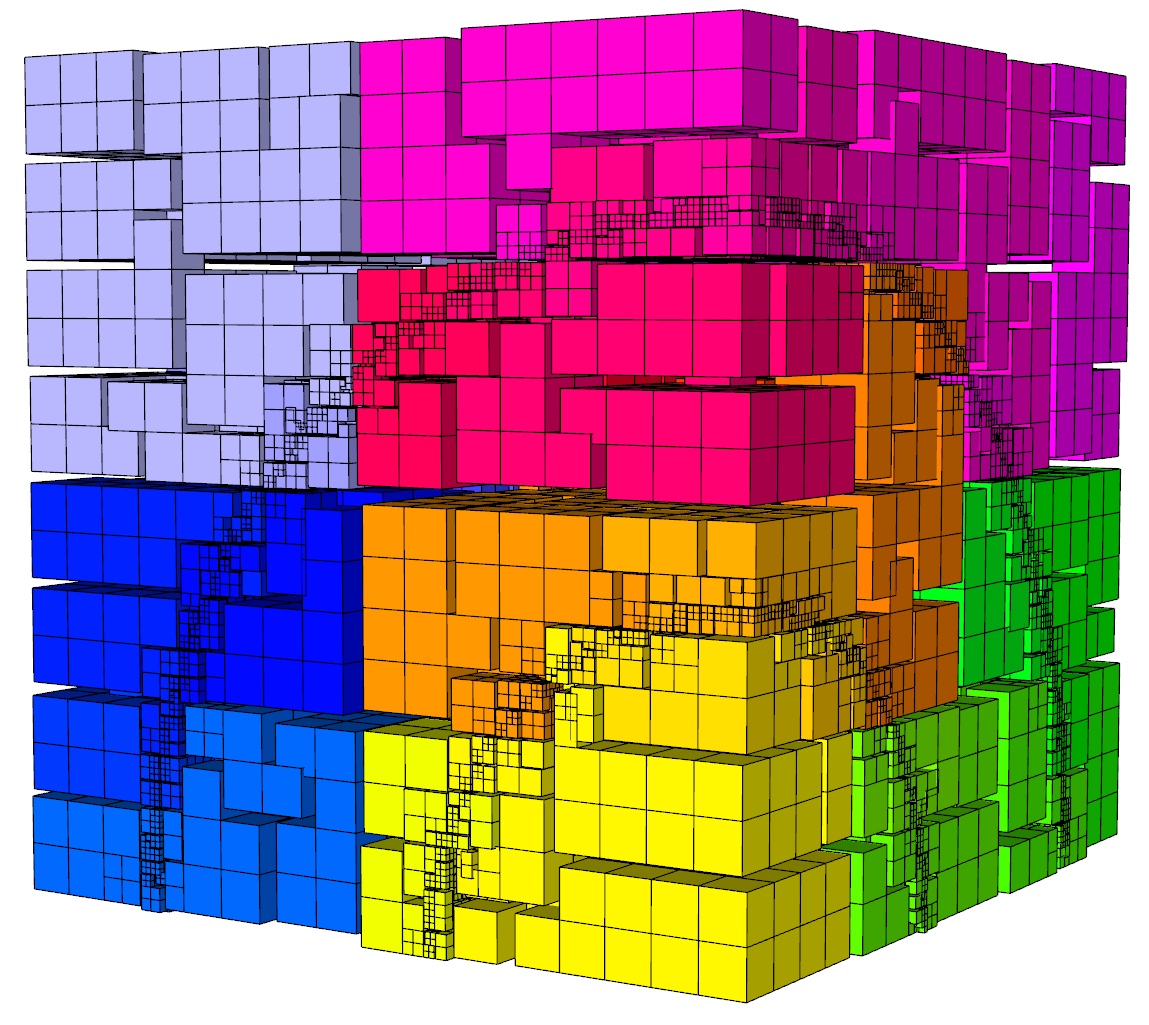}
  \hspace{0.05\textwidth}
  \includegraphics[width=0.50\textwidth]{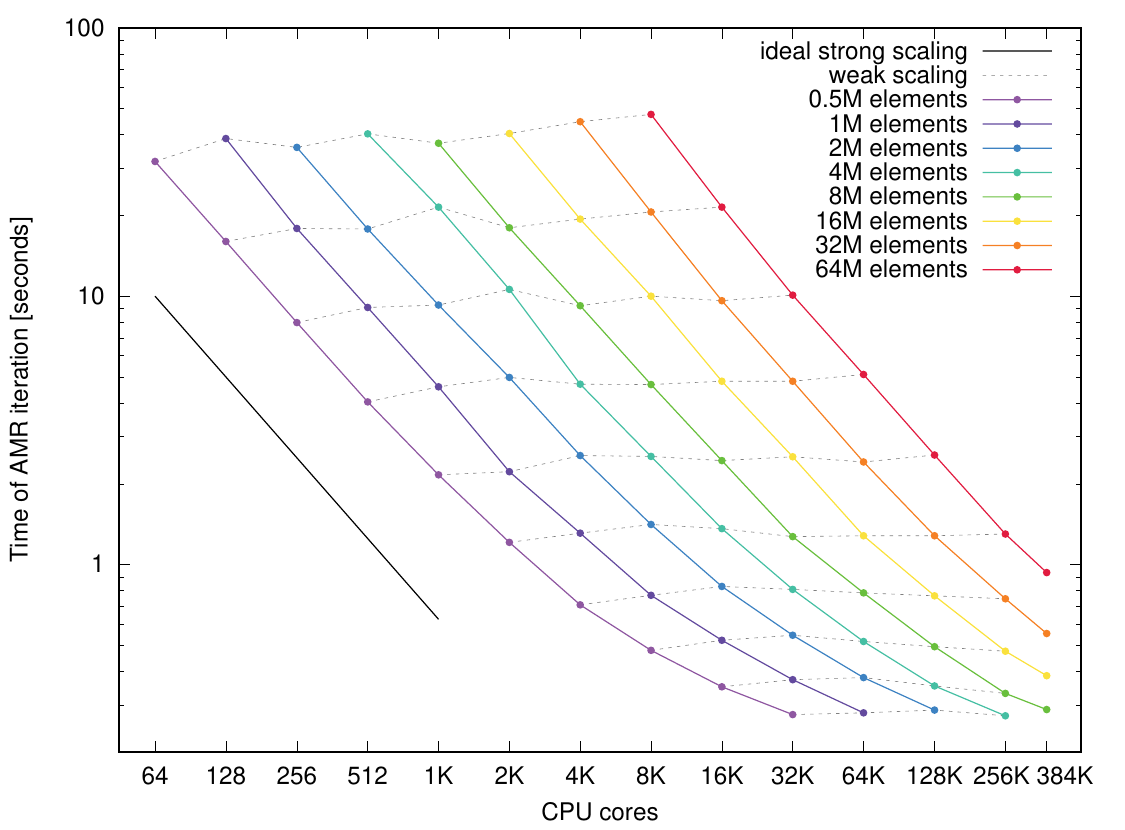}
\end{center}
\caption{Left: One octant of the parallel test mesh partitioned by the
      Hilbert curve (2048 domains shown). Right: Overall parallel weak and
      strong scaling for selected iterations of the AMR loop. Note that these
      results test only the AMR infrastructure, no physics computations are
      being timed.}
\label{fig:amr-scaling}
\end{figure}

These algorithms have been heavily optimized for both weak and strong parallel
scalability as illustrated in \autoref{fig:amr-scaling}, where we report results
from a 3D Poisson problem on the unit cube with exact solution having two
shock-like features. We initialize the mesh with $32^3$ hexahedra and repeat the
following steps, measuring their wall-clock times (averaged over all MPI ranks):
1) Construct the finite element space for the current mesh (create the $P$ matrix); 2)
Assemble locally the stiffness matrix $A$ and right hand side $b$; 3) Form the
products $P^t A P$, $P^t b$; 4) Eliminate Dirichlet boundary conditions from the
parallel system; 5) Project the exact solution $u$ to $u_h$ by nodal
interpolation; 6) Integrate the exact error $e_i = ||u_h - u||_{E,K_i}$ on each
element; 7) Refine elements with $e_j > 0.9 \max\{e_i\}$; 8) Load balance so
each process has the same number of elements ($\pm$1).  We run about 100
iterations of the AMR loop and select iterations that happen to have approximately
0.5, 1, 2, 4, 8, 16, 32 and 64 million elements in the mesh at the beginning.
We then plot the times of the selected iterations as if they were 8 independent
problems. We run from 64 to 393{,}216 (384K) cores on LLNL's Vulcan BG/Q machine.
The solid
lines in \autoref{fig:amr-scaling} show strong scaling, i.e. we follow the same
AMR iteration and its total time as the number of cores doubles. The dashed
lines skip to a double-sized problem when doubling the number of cores showing
weak scaling, and should ideally be horizontal.

\begin{figure}[!hbt]
  \centering
  \includegraphics[width=0.85\textwidth]{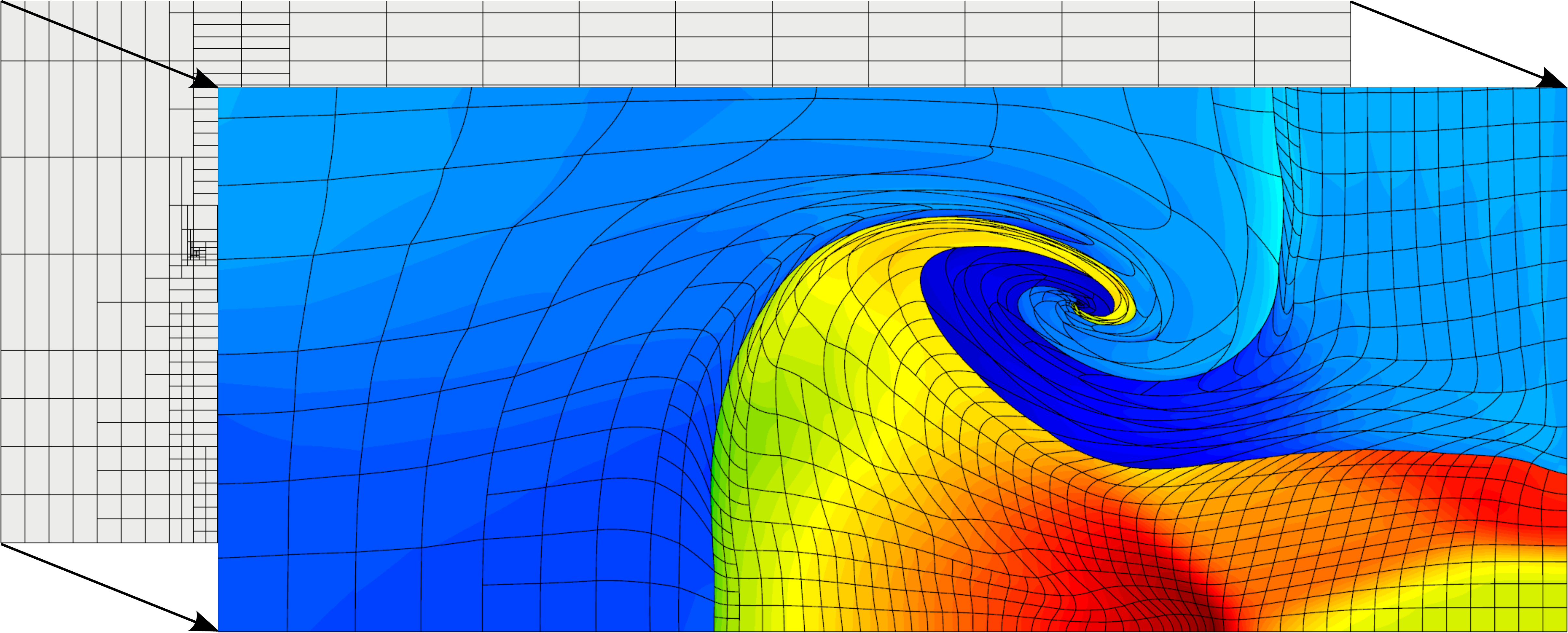}
  \caption{MFEM-based static refinement in a triple point shock-interaction problem.
           Initial mesh at $t = 0$ (background) refined anisotropically
           in order to obtain more regular element shapes at target time
           (foreground).}
  \label{fig:amr-triple-pt}
\end{figure}

MFEM's variational restriction-based AMR approach can be remarkably unintrusive
when it comes to integration in a real finite element application code. To
illustrate this point we show two results from the {\it Laghos} miniapp
(see \autoref{ssec:hydro}) which required minimal changes for static refinement
support (see \autoref{fig:amr-triple-pt}) and about 550 new lines of code for
full dynamic AMR, including derefinement (see \autoref{fig:amr-sedov-dynamic}).

\subsection{Mesh Optimization} \label{ssec:tmop}
A vital component of high-order methods is the use of high-order representation
not just for the {\em physics} fields, but also for the geometry, represented by
a high-order computational mesh.  High-order meshes can be relatively coarse and
still capture curved geometries with high-resolution, leading to equivalent
simulation quality for a smaller number of elements.  High-order meshes can also
be very beneficial in a wide range of applications, where e.g. radial symmetry
preservation, or alignment with physics flow or curved model boundary is
important \cite{sljs10, 2012-SISC-BLAST, MeshMod}.  Such applications can
utilize {\em static} meshes, where a good-quality high-order mesh needs to be
generated only as an input to the simulation, or {\em dynamic} meshes, where the
mesh evolves with the problem (e.g.\ following the motion of a material) and its
quality needs to be constantly controlled. In both cases, the quality of
high-order meshes can be difficult to control, because their properties vary in
space on a sub-zonal level. Such control is critical in practice, as poor mesh
quality leads to small time step restrictions or simulation failures.

The MFEM project has developed a general framework for the optimization of
high-order curved meshes based on the node-movement techniques of the
Target-Matrix Optimization Paradigm (TMOP) \cite{2012-TMOP,2018-TMOP}. This
enables applications to have precise control over local mesh quality, while
still optimizing the mesh globally. Note that while our new methods are
targeting high-order meshes, they are general, and can also be applied to
low-order mesh applications that use linear meshes.

TMOP is a general approach for controlling mesh quality, where mesh nodes
(vertices in the low-order case) are moved so-as to optimize a multi-variable
objective function that quantifies global mesh quality.
Specifically, at a given point of interest
(inside each mesh element), TMOP uses three Jacobian matrices:
\begin{itemize}
\item The Jacobian matrix $A_{d \times d}$ of the transformation from reference
  to physical coordinates, where $d$ is the space dimension.
\item The \textit{target matrix}, $W_{d \times d}$, which is the Jacobian of the
  transformation from the reference to the {\em target} coordinates.
  The target matrices are defined according to a user-specified method prior to
  the optimization; they define the desired properties in the optimal mesh.
\item The \textit{weighted Jacobian} matrix, $T_{d \times d}$, defined by
  $T = A W^{-1}$, represents the Jacobian of the transformation
  between the target and the physical (current) coordinates.
\end{itemize}
The $T$ matrix is used to define the \textit{local quality measure}, $\mu(T)$.
The quality measure can evaluate shape, size, or alignment of the region around
the point of interest.  The combination of targets and quality metrics is used
to optimize the node positions, so that they are as close as possible to the
shape/size/alignment of their targets.  This is achieved by minimizing a global
\textit{objective function}, $F(x)$, that depends on the local quality measure
throughout the mesh:
\begin{equation}
\label{eq_vm}
 F(x) := \sum_{E \in \mathcal{E}} \int_{E_t} \mu(T(x)) dx_t =
 \sum_{E \in \mathcal{E}} \sum_{x_q \in Q_E} w_q\,\det(W(x_q))\, \mu(T(x_q)) \,,
\end{equation}
where $E_t$ is the target element corresponding to the physical element $E$,
$Q_E$ is the set of quadrature points for element $E$, $w_q$ are the
corresponding quadrature weights, and both $T(x_q)$ and $W(x_q)$ are evaluated
at the quadrature point $x_q$ of element $E$.  The objective function can be
extended by using combinations of quality metrics, space-dependent weights for
each metric, and limiting the amount of allowed mesh displacements.  As $F(x)$
is nonlinear, MFEM utilizes Newton's method to solve the critical point
equations, $\partial F(x) / \partial \boldsymbol{x} = 0$, where $\boldsymbol{x}$
is the vector that contains the current mesh positions.  This approach involves
the computation of the first and second derivatives of $\mu(T)$ with respect to
$T$.  Furthermore, boundary nodes are enforced to stay fixed or move only in the
boundary's tangential direction.  Additional modifications are performed to
guarantee that the Newton updates do not lead to inverted meshes, see
\cite{2018-TMOP}.

\begin{figure}[!hbt]
\centerline{
  \includegraphics[height=.27\textwidth]{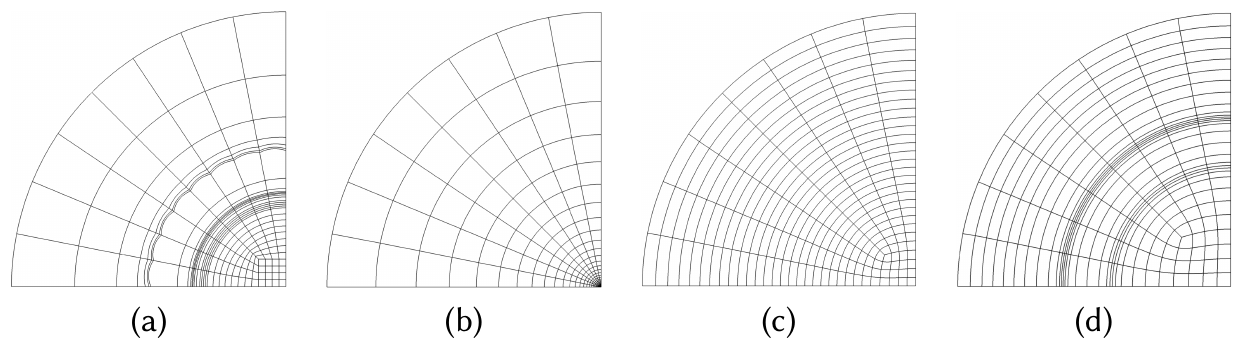}
}
\caption{A perturbed fourth order 2D mesh (a) is being optimized by targeting
         shape-only optimization (b), shape and equal size (c), and finally
         shape and space-dependent size (d).}
\label{fig_tmop}
\end{figure}

\begin{figure}[!hbt]
\centering
\includegraphics[height=.35\textwidth]{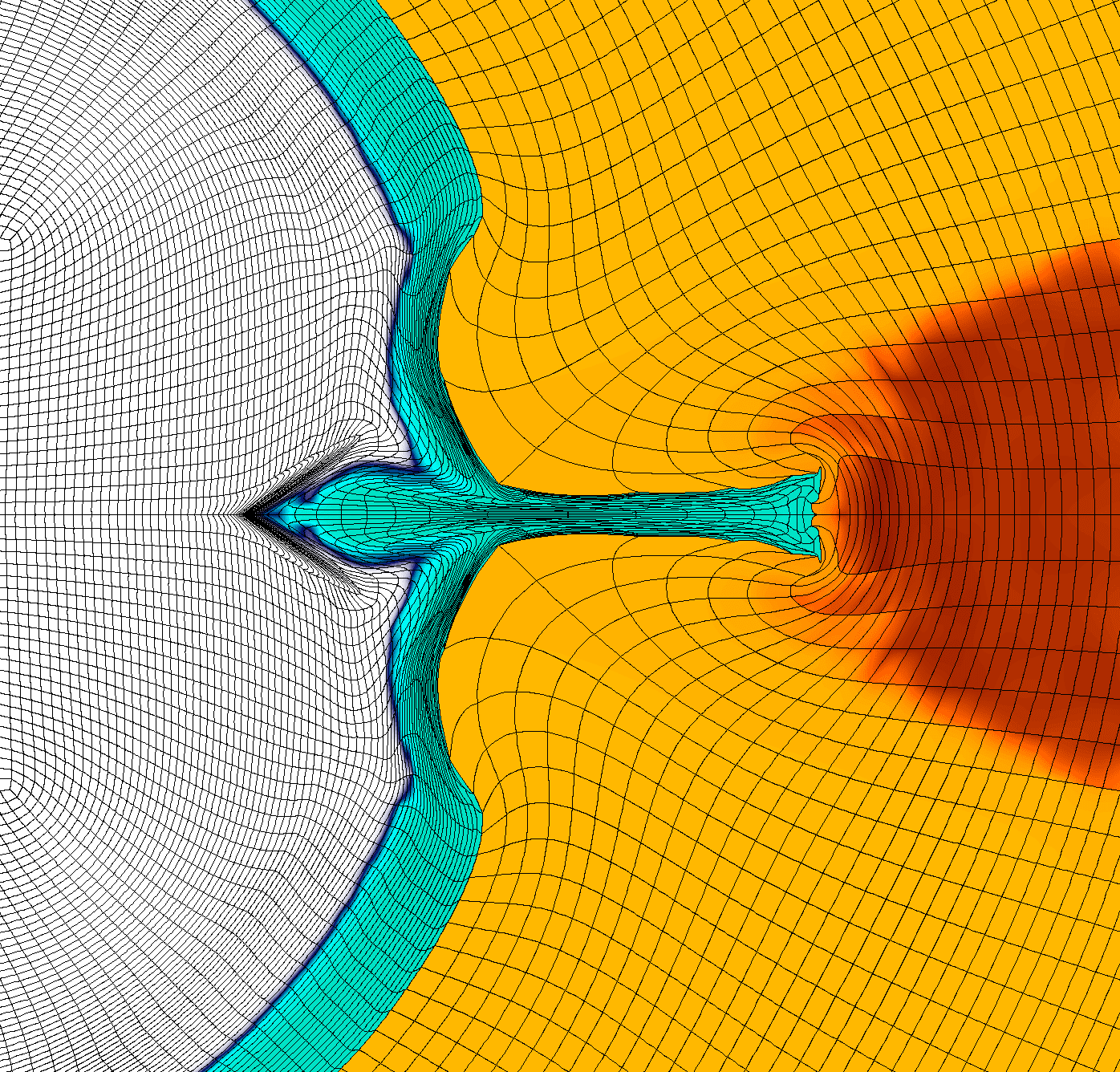} \hfil
\includegraphics[height=.35\textwidth]{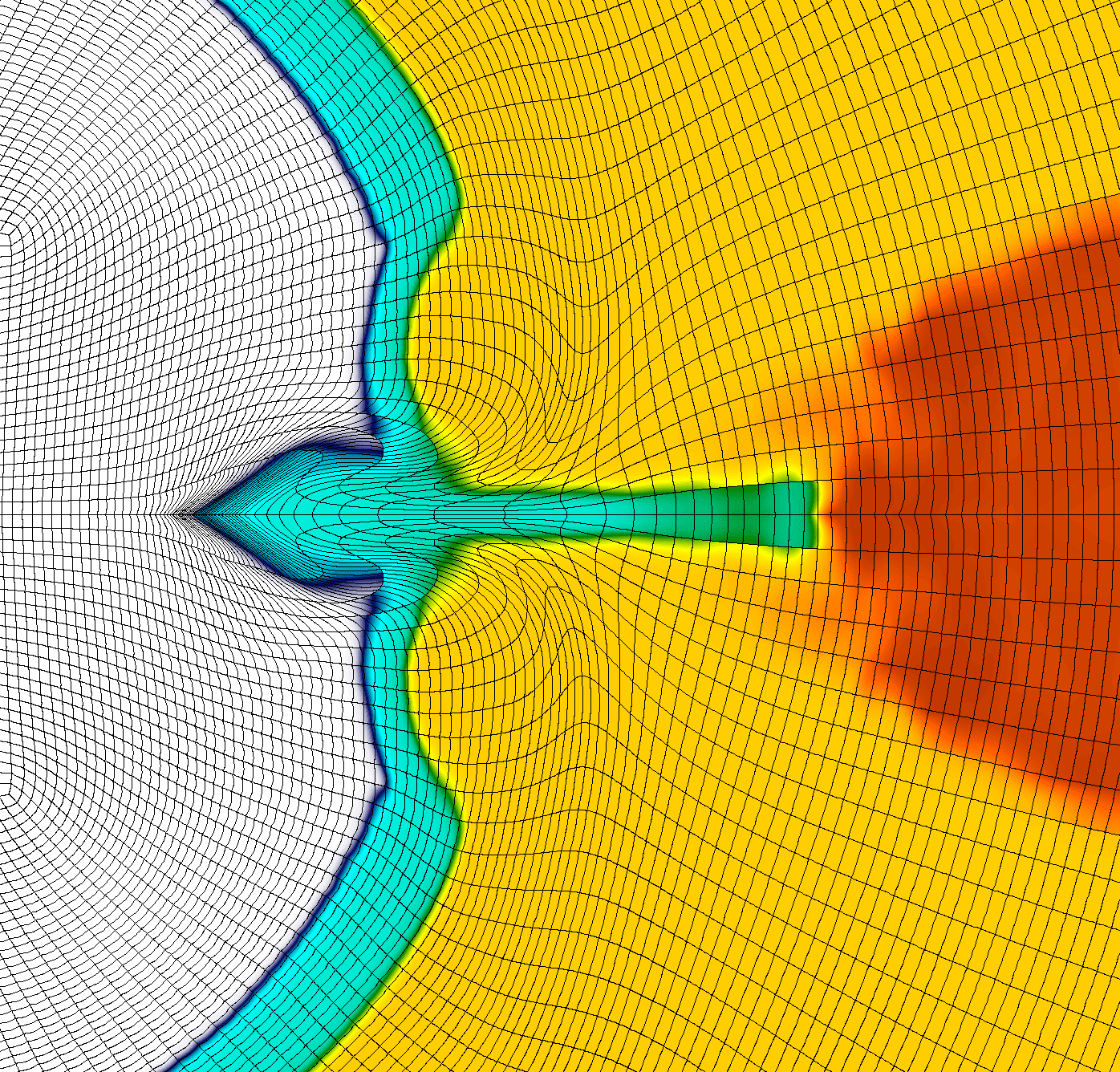}
\caption{Optimized meshes in parallel inertial confinement fusion simulation,
         see \cite{2018-SISC-ALE} and \autoref{ssec:hydro}. Shown is the
         region around the capsule's fill tube. Both meshes are optimized
         with respect to shape, size, and amount of mesh displacement.
         On the left, material interfaces are kept
         fixed. On the right, interfaces are relaxed later in the simulation.}
\label{fig_blast}
\end{figure}

The current MFEM interface provides access to 12 two-dimensional mesh quality
metrics, 7 three-dimensional metrics, and 5 target construction methods,
together with the first and second derivatives of each metric with respect to
the matrix argument. The quality metrics are defined by the inheritors of the
class \code{TMOP\_QualityMetric}, and target construction methods are defined
by the class \code{TargetConstructor}. MFEM supports the computation of matrix
invariants and their first and second derivatives (with respect to the matrix),
which are then used by the \code{NewtonSolver} class to solve $\partial F(x) /
\partial \boldsymbol{x} = 0$. The library interface allows users to choose
between various options concerning target construction methods and mesh quality
metrics and adjust various parameters depending on their particular problem.
The mesh optimization module can be easily extended by additional mesh quality
metrics and target construction methods. Illustrative examples are presented in
the form of a simple mesh optimization miniapp, \code{mesh-optimizer}, in the
\code{miniapps/meshing} directory, which includes both serial and parallel
implementations. Some examples of simulations that can be performed by this
miniapp are shown in \autoref{fig_tmop}. MFEM's mesh optimization capabilities
are also routinely used in production runs for many of the ALE simulation
problems in the BLAST code, see \autoref{ssec:hydro}, and the example in
\autoref{fig_blast}.

\begin{figure}[!hbt]
\centerline{
  \includegraphics[width=0.32\textwidth]{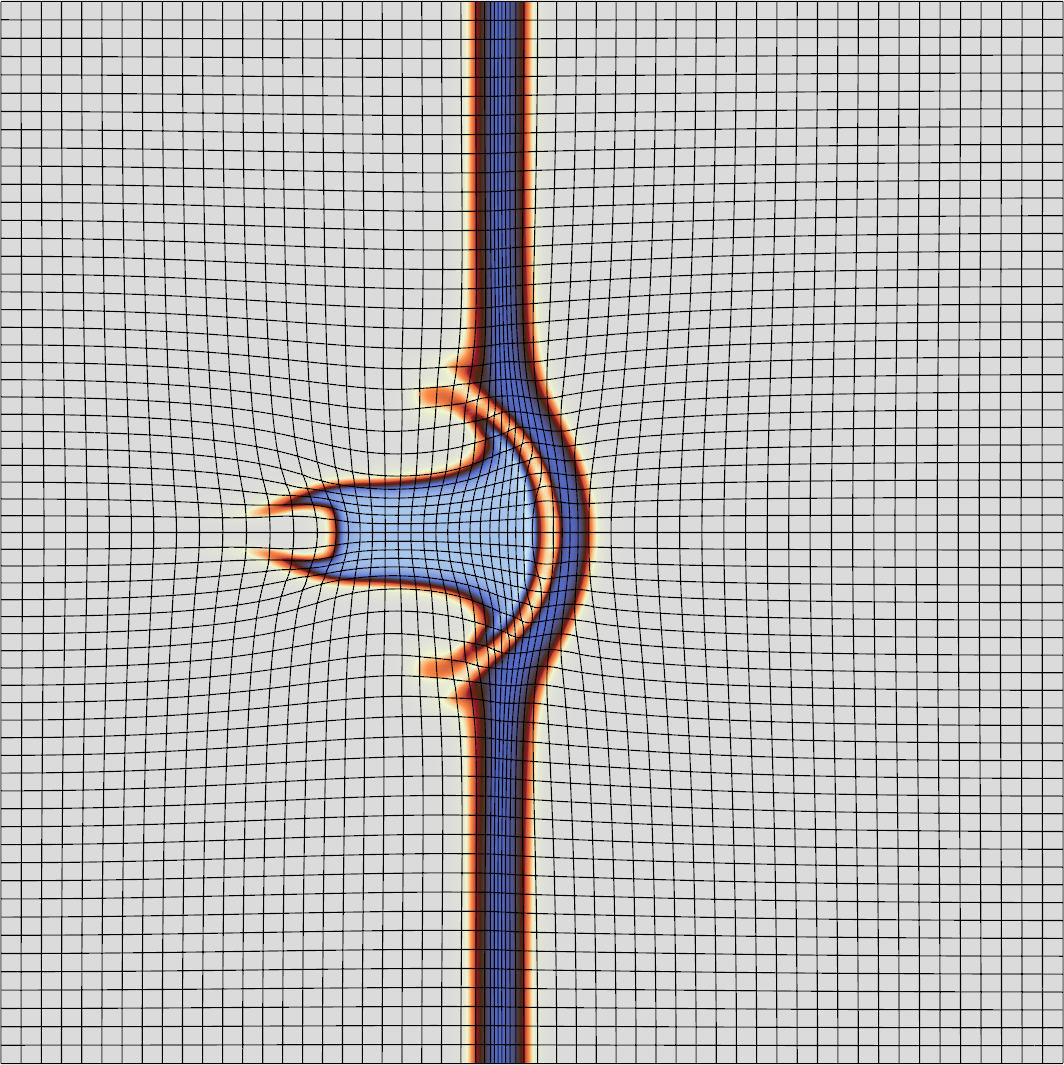}
  \includegraphics[width=0.32\textwidth]{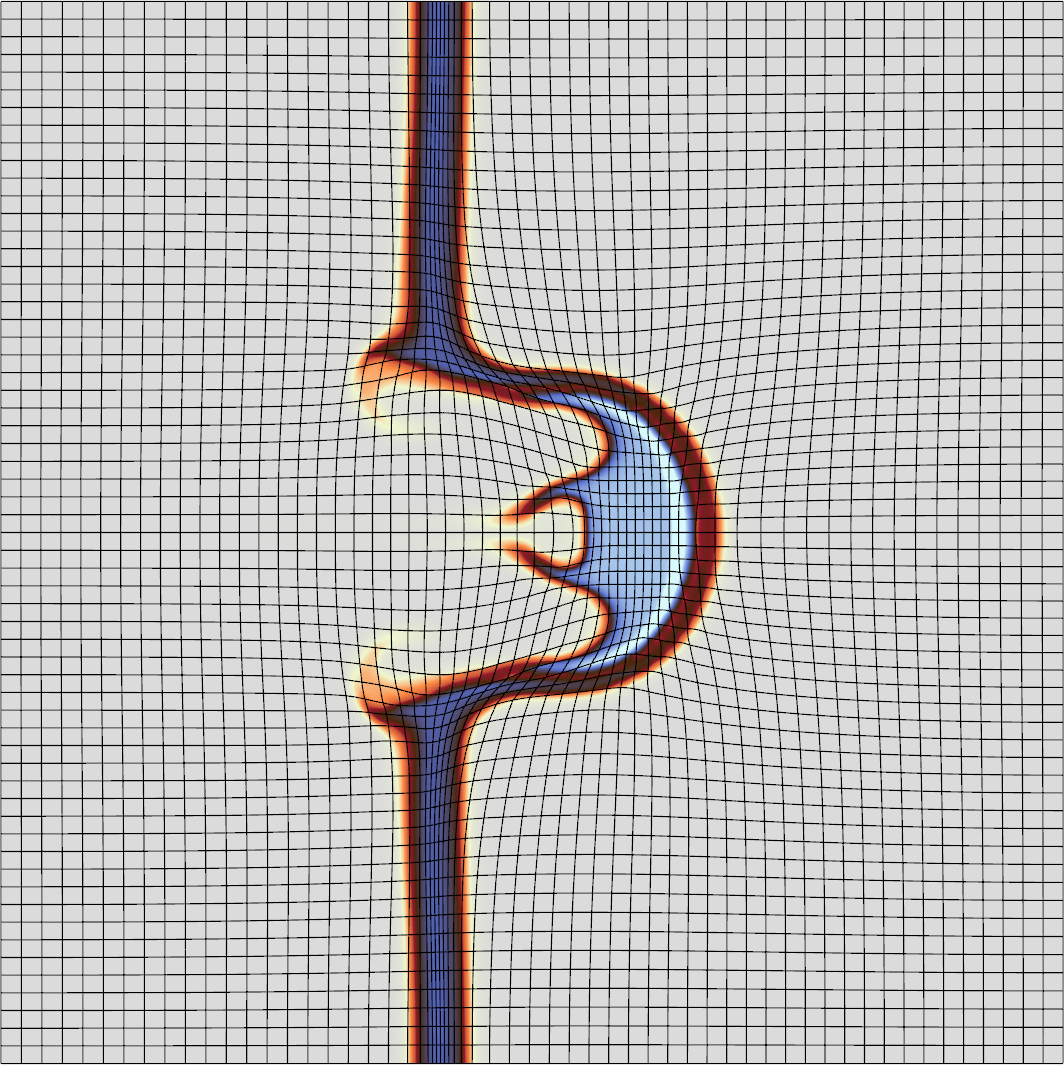}
  \includegraphics[width=0.32\textwidth]{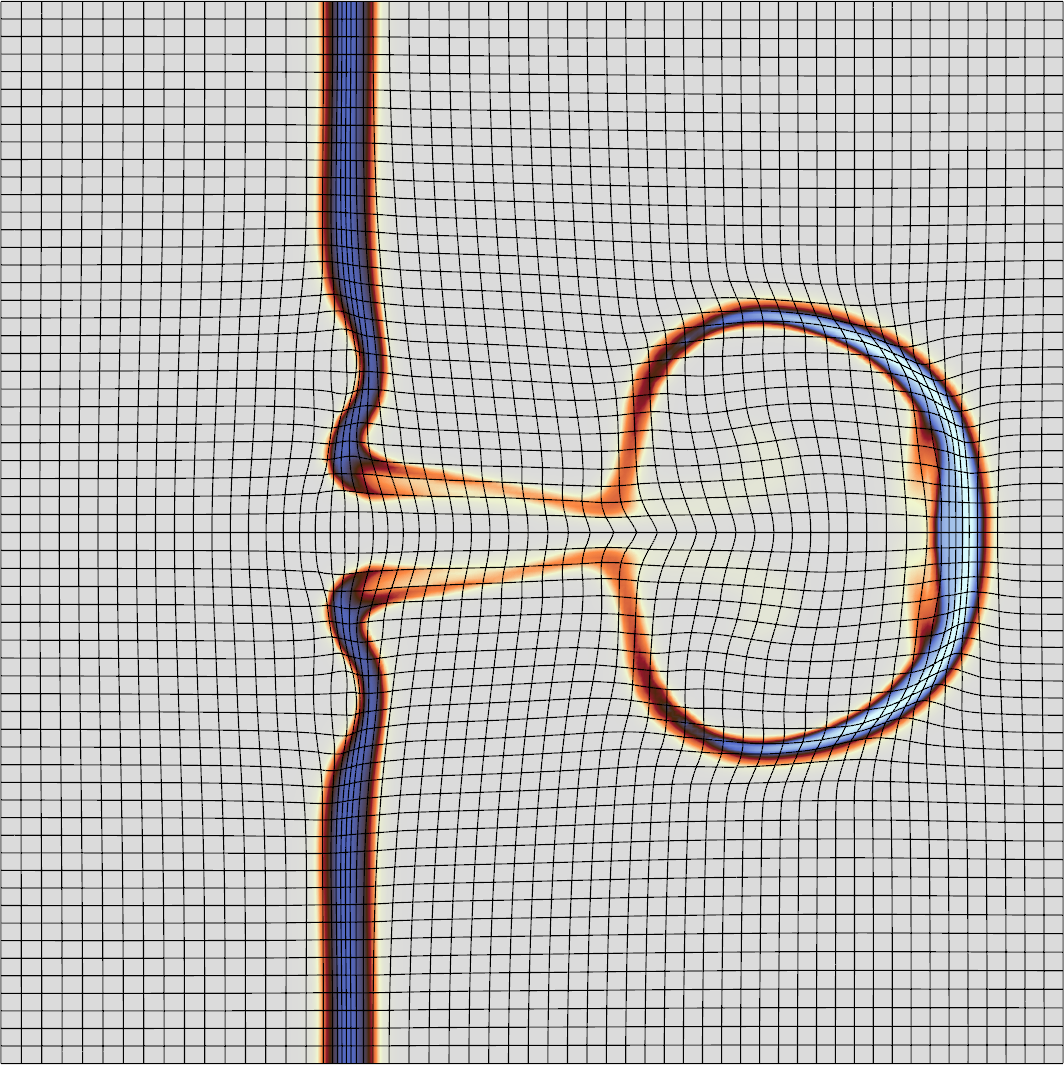}}
\caption{Example of MFEM $r$-adaptivity to align the mesh with materials in a
         multi-material ALE simulation of high velocity gas impact,
         cf. \cite{Barlow14}. Time evolution of the materials and mesh positions
         at times 2.5 (left), 5 (center), and 10 (right). See \cite{2018-TMOP-ADAPT} for details.}
\label{fig_gas_impact}
\end{figure}

Work to extend MFEM's mesh optimization capabilities to simulation-driven
adaptivity (a.k.a. $r$-adaptivity) \cite{2018-TMOP-ADAPT}, and coupling $h$- and
$r$-adaptivity of high-order meshes by combining the TMOP and AMR concepts is
ongoing. See \autoref{fig_gas_impact} for some preliminary results in that
direction.

\section{Applications} \label{sec:apps}

MFEM has been used in numerous applications and research publications, a
comprehensive list of which is available on the project website at
\url{https://mfem.org/publications}. In this section we illustrate a small
sample of these applications.

\subsection{Examples and Miniapps} \label{ssec:examples}

The MFEM codebase includes a wide array of example applications that utilize
numerous MFEM features and demonstrate the finite element discretization of many
PDEs.  The goal of these well documented example codes is to provide a
step-by-step introduction to MFEM in simple model settings.  Most of the
examples have both serial and parallel versions (indicated by a {\tt p} appended
to the filename) which illustrate the straightforward transition to parallel
code and the use of the {\em hypre} solvers and \cred{preconditioners}.  There
are also variants of many example codes in the \code{petsc}, \code{sundials} and
\code{pumi} subdirectories that display integration with those packages. Each
example code has the flexibility to change the order of the calculations, switch
various finite element features on or off, and utilize different meshes through
command line options.  Once the example codes are built, their options can be
displayed by running the code with \code{--help} as a command line option.  The
outputs of the examples can be visualized with GLVis, see \autoref{ssec:vis} and
\url{https://glvis.org}. Basic tutorials for running the examples can be found
online at \url{https://mfem.org} under the links ``Serial Tutorial'' and
``Parallel Tutorial''.

The example codes are simply named \code{ex1}--\code{ex21}, roughly in order of
complexity, so it is recommended that users start with earlier numbered examples
in order to learn the basics of interfacing with MFEM before moving on to more
complicated examples. More details can be found in our online documentation at
\url{https://mfem.org/examples}.

\begin{figure}[!hbt]
\centering
\includegraphics[height=.30\textwidth]{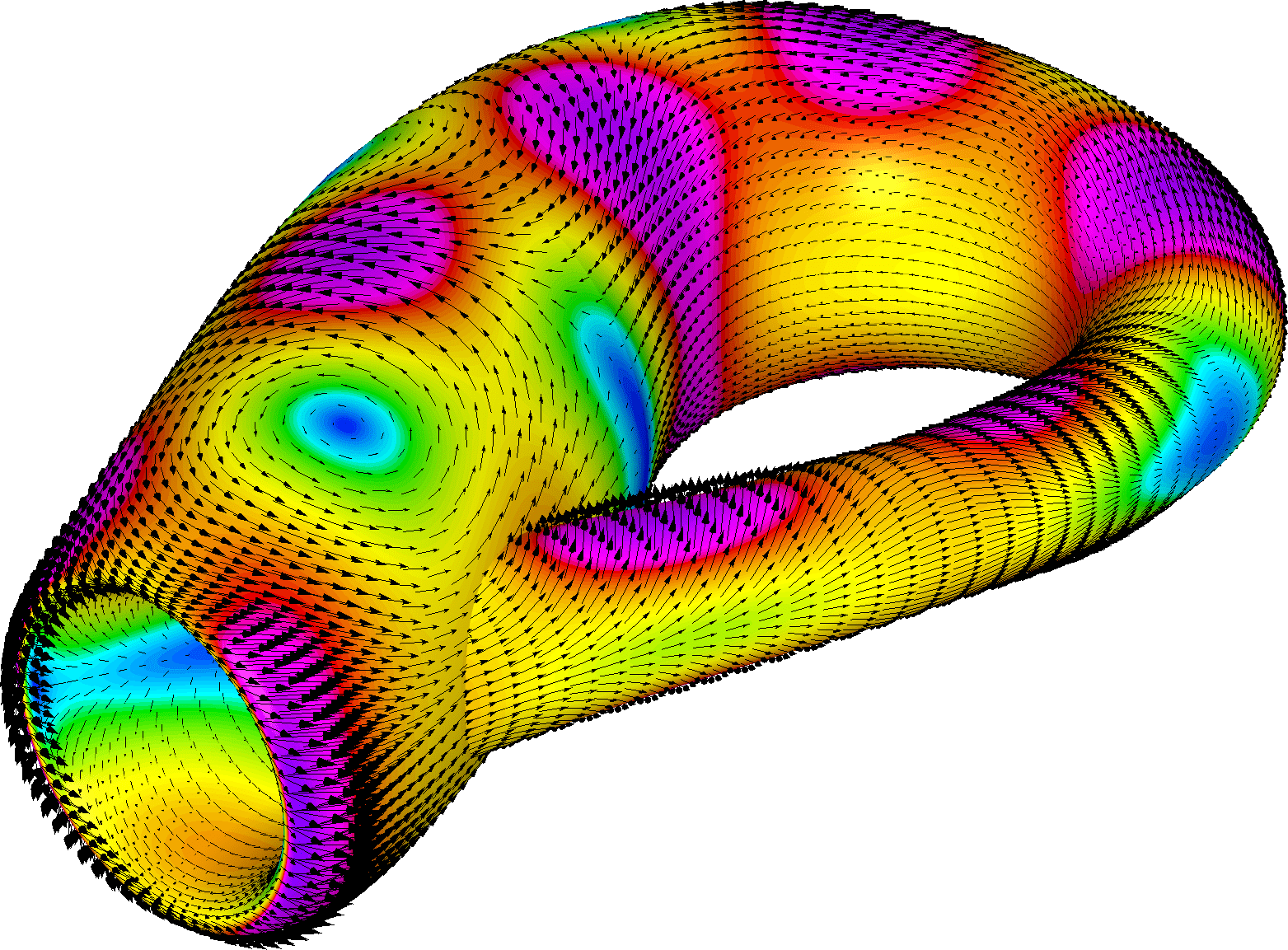} \hfil
\includegraphics[height=.36\textwidth]{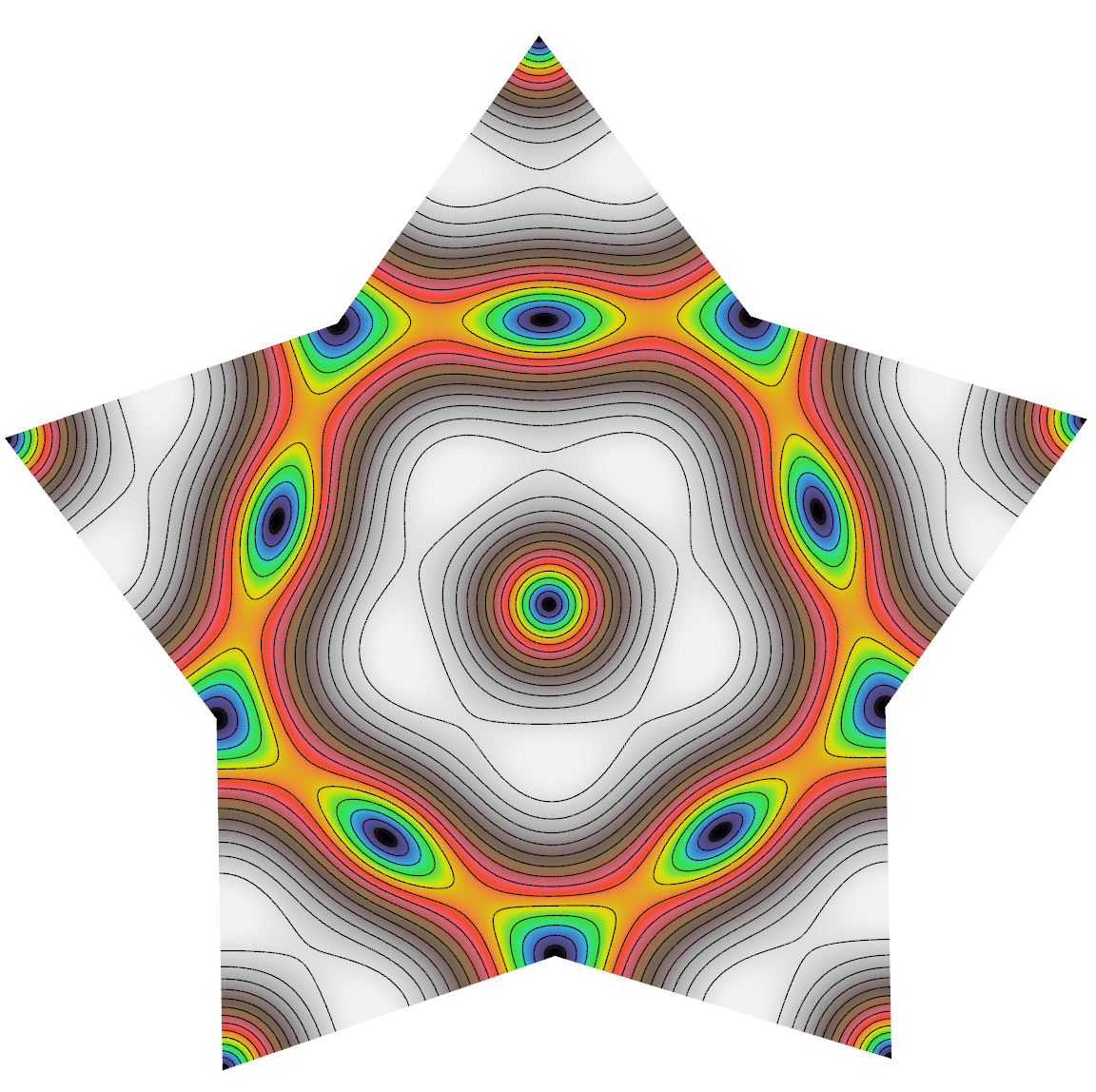}
\caption{Left: Solution of a Maxwell problem on a Klein bottle surface with \code{ex3}; mesh generated with the \code{klein-bottle} miniapp in \code{miniapps/meshing}. Right: An electromagnetic eigenmode of a star-shaped domain computed with 3rd order finite elements computed with \code{ex13p}.}
\label{fig_examples}
\end{figure}

The first example \code{ex1} begins with the solution of the Laplace problem
with homogeneous Dirichlet boundaries utilizing nodal $H^1$ elements. Examples
\code{ex6}, \code{ex8}, and \code{ex14} also solve the Poisson problem, but they
also highlight AMR, DPG and DG formulations, respectively. Examples \code{ex2}
and \code{ex17} solve the equations of linear elasticity with Galerkin and DG
formulations, respectively, while \code{ex10} provides an implementation of
nonlinear elasticity utilizing a Newton solver; the interface to PETSc's
nonlinear solvers is described in \code{petsc/ex10p}, which also showcases
the support for a Jacobian-free Newton Krylov approach. An elementary introduction to
utilizing $\Hcurl$ vector elements to solve problems arising from Maxwell's
equations can be found in \code{ex3} and \code{ex13}.  An example of utilizing
surface meshes embedded in a 3D space can be found in \code{ex7} while more
advanced dynamic AMR is explored in \code{ex15}.  Time-dependent simulations
are considered in examples \code{ex9}, \code{ex10}, \code{ex16}, and \code{ex17};
users interested in the usage of the SUNDIALS and PETSc ODE solvers are referred to
examples \code{sundials/ex9p} and \code{petsc/ex9p} respectively.
Finally, \code{ex11}, \code{ex12}, and \code{ex13} tackle frequency
domain problems solving for eigenvalues of their respective systems.
Results from some of the example runs are shown in \autoref{fig_examples}.

\subsection{Electromagnetics} \label{ssec:em}
The electromagnetic miniapps in MFEM are designed to provide a starting point
for developing real-world electromagnetic applications.  As such, they cover a
few common problem domains and attempt to support a variety of boundary
conditions and source terms.  This way application scientists can easily adapt
these miniapps to solve particular problems arising in their research.

The \textit{Volta} miniapp solves Poisson's equation with boundary conditions and
sources tailored to electrostatic problems.  This miniapp supports fixed voltage
or fixed charge density boundary conditions which correspond to the usual
Dirichlet or Neumann boundary conditions, respectively.  The volumetric source
terms can be derived from either a prescribed charge density or a fixed
polarization field.

The \textit{Tesla} miniapp models magnetostatic problems.  Magnetostatic boundary
conditions are more complicated than those for electrostatic problems due to the
nature of the curl-curl operator.  We support two types of boundary conditions:
the first leads to a constant magnetic field at the boundary, the second arises
from a surface current.  The surface current is itself the solution of a Poisson
problem restricted to the surface of the problem domain.  The motivation for
this surface current boundary condition is to approximate the magnetic fields
surrounding current carrying conductors.  \textit{Tesla} also supports volumetric
sources due to current densities or materials with a fixed magnetization
(i.e. permanent magnets).  Note that the curl-curl operator cannot be solved
with an arbitrary source term; the source must be a solenoidal vector field.  To
ensure this, the \textit{Tesla} miniapp must remove any irrotational components by
performing a projection operation known as ``divergence cleaning''.

The \textit{Maxwell} miniapp simulates full-wave time-domain electromagnetic
wave propagation.  This miniapp solves the Maxwell equations as a pair of
coupled first order partial differential equations using a symplectic
time-integration algorithm designed to conserve energy (in the absence of lossy
materials).  The simulation can be driven by a time-varying applied electric
field boundary condition or by a volumetric current density.  Perfect electrical
conductor, perfect magnetic conductor, and first order Sommerfeld absorbing
boundary conditions are also available.  A frequency-domain version of this
miniapp is currently under development.

One of the most practical applications of electromagnetics is the approximation
of the Joule heating caused by an alternating electrical current in an imperfect
conductor.  The MFEM miniapp \textit{Joule} models this behavior with a system of
coupled partial differential equations which approximate low-frequency
electromagnetics and thermal conduction.  The boundary conditions consist of a
time-varying voltage, used to drive a volumetric current, and a thermal flux
boundary condition, which can approximate a thermal insulator.  This miniapp is
a good example of a simple multi-physics application which could be modified to
simulate a variety of important real-world problems in electrical engineering.

\subsection{Compressible Hydrodynamics} \label{ssec:hydro}
The MFEM-based \textit{Laghos} \cite{laghos_web,CEED-MS6} (short for Lagrangian
high-order
solver) miniapp models time-dependent, compressible, inviscid gas dynamics via
the Euler equations in the Lagrangian form.  The Euler equations describe the
conservation of mass, momentum, and energy of an inviscid fluid. In the
Lagrangian setting, the elements represent regions of fluid that move with the
flow, resulting in a moving and deforming mesh.  The high-order curved mesh
capabilities of MFEM provide a significant advantage in this context, since
curved meshes can describe larger deformations more robustly than meshes using
only straight segments. This in turn mitigates problems with the mesh
intersecting itself when it becomes highly deformed.

The \textit{Laghos} miniapp uses continuous finite element spaces to describe
the position and velocity fields, and a discontinuous space to describe the
energy field.  The order of these fields is determined by runtime parameters,
making the code arbitrarily high order.  The assembly of the finite elements in
\textit{Laghos} can be accomplished using either standard {\em full assembly} or
as {\em partial assembly}, see \autoref{ssec:pa}.  With partial assembly, the
global matrices are never fully created and stored, but rather only the local
action of these operators is required.  This reduces both memory footprint and
computational cost.

\textit{Laghos} is also a simplified model for a more complex multi-physics code
known as \textit{BLAST} \cite{2018-SISC-ALE, 2012-SISC-BLAST,
2013-JCP-Strength}, which features mesh remapping, arbitrary Lagrangian-Eulerian
(ALE) capabilities, solid mechanics, and multi-material zones.  The remapping
capability allows arbitrarily large deformations to be modeled, since the mesh
can be regularized at intervals sufficient to continue a simulation
indefinitely.  The remap capability in {\it BLAST} is accomplished with a
high-order discontinuous Galerkin method, which is both conservative and
monotonic \cite{2014-MonoRemap}. The DG component of the remap algorithm is very
similar to the DG advection in Example 9.

\begin{figure}
  \begin{center}
  \includegraphics[width=0.32\textwidth]{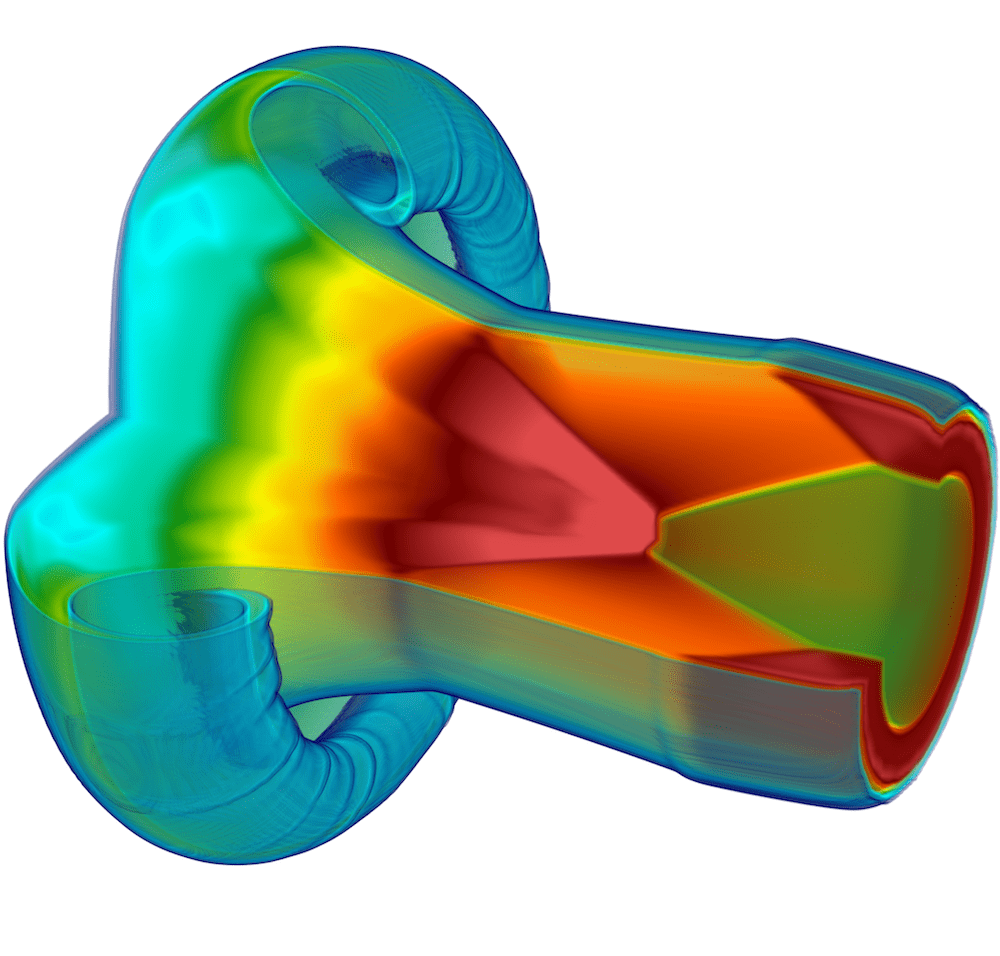}
  \includegraphics[width=0.32\textwidth]{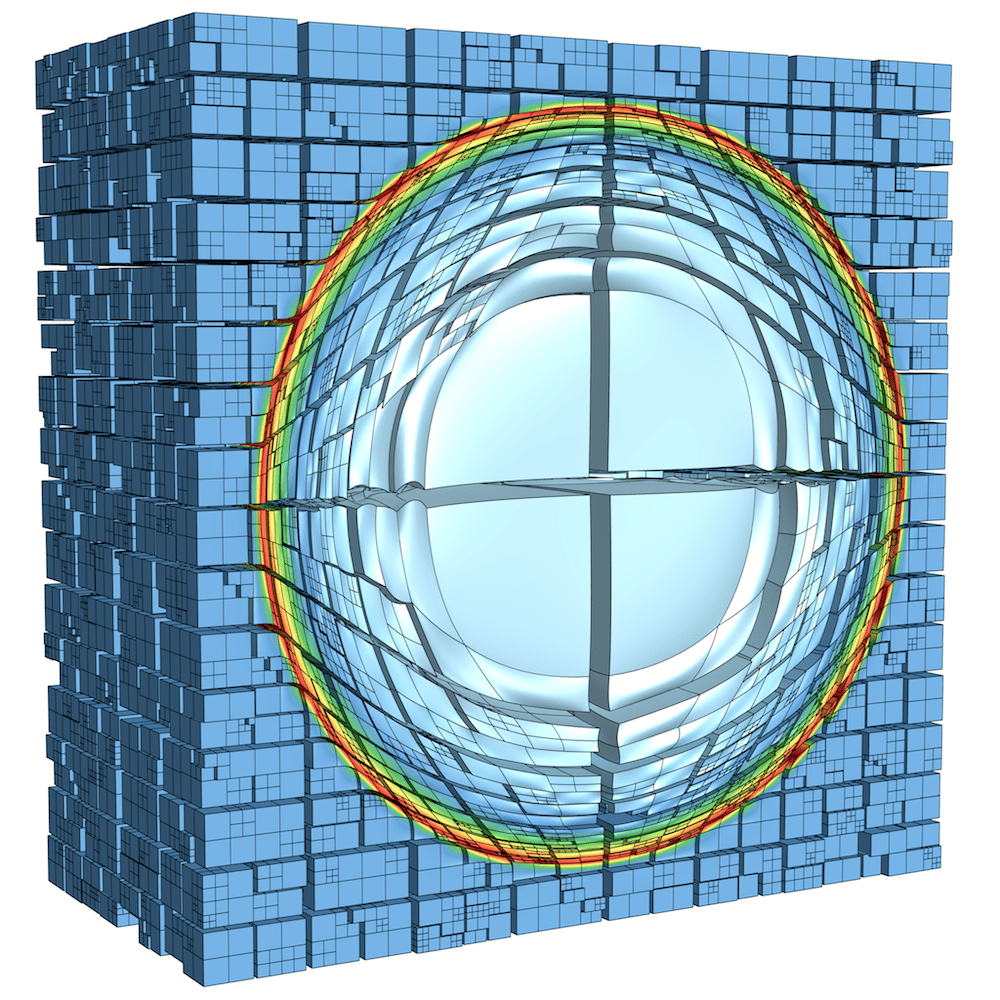}
  \includegraphics[width=0.32\textwidth]{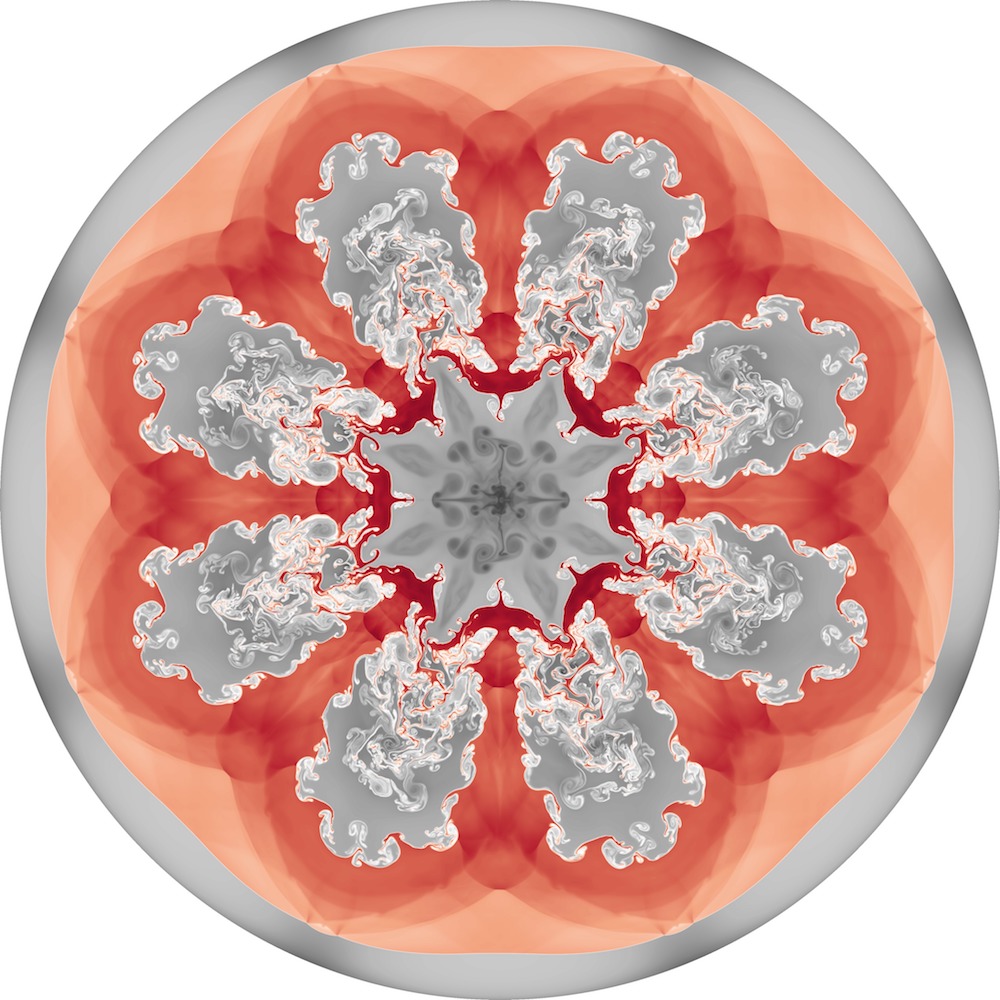}
\end{center}
\caption{Left: A shock interface 3D hydrodynamics calculation on 16,384 processors with
    BLAST. Center: Static non-conforming AMR in a Sedov blast simulation,
    see \autoref{ssec:nc-amr}. Right: High-order axisymmetric multi-material
    inertial confinement fusion (ICF)-like implosion in
    BLAST.}  \label{fig:triplept}
\end{figure}

\textit{BLAST} uses a general stress tensor formulation which allows for the
simulation of elasto-plastic flows in 2D, 3D, and in axisymmetric coordinates.
Multi-material elements are described using high-order material indicator
functions, which describe the volume fractions of materials at all points in the
domain.  A new, high-order multi-material closure model was developed to solve
the resulting multi-material system of equations \cite{2016-Closure}.  This
capability has been used to model many types of hydrodynamic systems, such as
Rayleigh-Taylor instability, shock-interface interactions, solid impact
problems, and inertial confinement fusion dynamics.  Some examples of {\it
BLAST} calculations are shown in \autoref{fig:triplept}.

\subsection{Other Applications} \label{ssec:otherapps}
MFEM has been applied successfully to a variety of applications including
radiation-diffusion, additive manufacturing, topology optimization, heart
modeling applications, linear and nonlinear elasticity, reaction-diffusion,
time-domain electromagnetics, DG advection problems, Stokes/Darcy flow, and
more. Two examples of such applications are the Cardioid and ParElag projects
described below.

\begin{figure}
  \centering
  \includegraphics[width=0.3\textwidth]{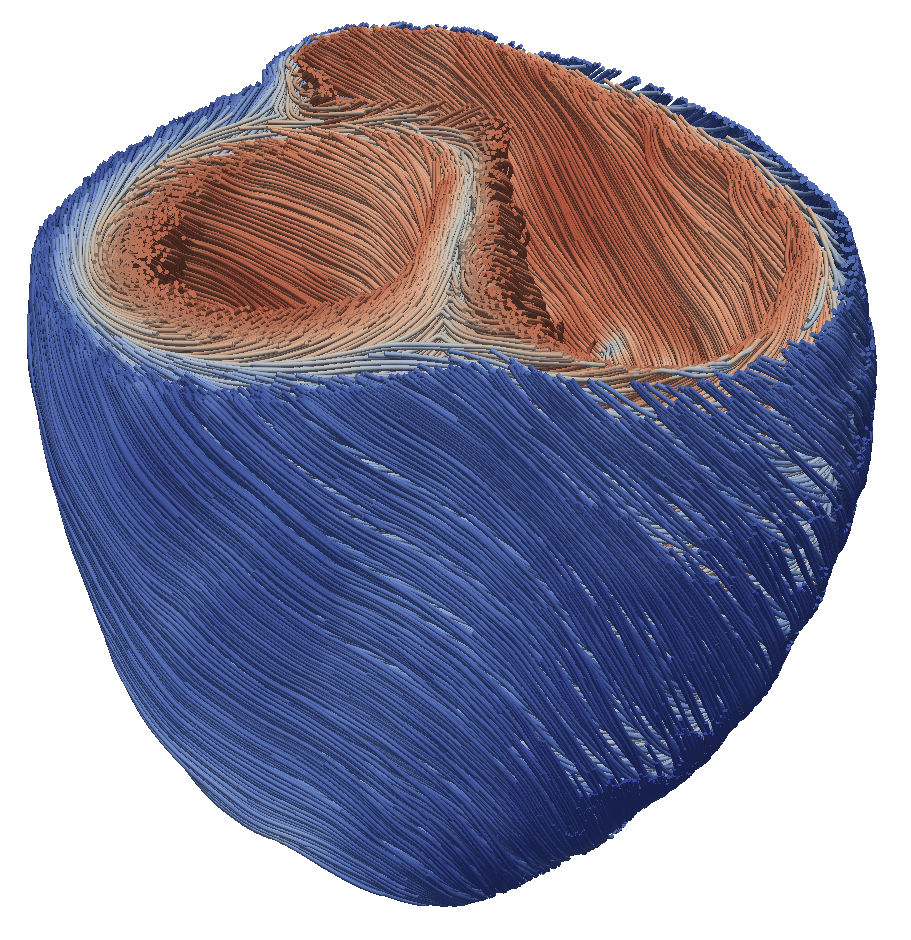}
  \hspace{10mm}
  \raisebox{8mm}{\includegraphics[width=0.5\textwidth]{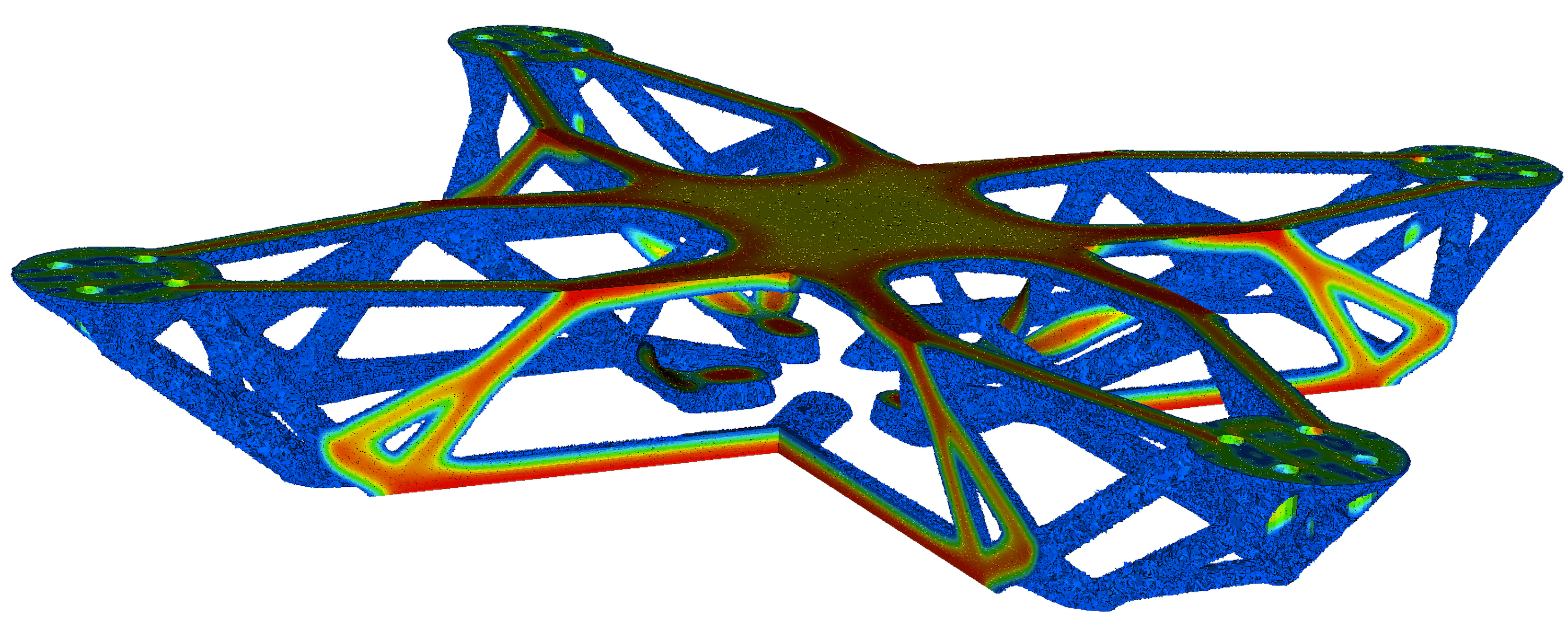}}
  \caption{\cred{Left: Heart fiber orientations computed in Cardioid using MFEM. Cardioid is
    being developed for virtual drug screening and modeling heart activity in
    clinical settings. Right: Drone body optimized for maximum strength in a given
    mass based on MFEM discretizations in LLNL's LiDO code. LiDO enables engineers
    to optimize immensely complex systems in HPC environments---in this case using
    100 million elements, well beyond the capability of commercial software.}}
  \label{fig:cardioid}
\end{figure}

The Cardioid project at LLNL \cite{Cardioid} recently used MFEM to rewrite and
simplify two cardiac simulation tools. The first is a fiber generation code
which solves a series of Poisson problems to compute cardiac fiber orientations
on a given mesh. See \autoref{fig:cardioid} for sample output. Additionally, a
deformable cardiac mechanics code which solves incompressible anisotropic
hyperelasticity equations with active tension has also been developed. The
methods implemented are outlined in \cite{Bayer2012} and \cite{Gurev2015}.  A
second MFEM-based code to generate electrocardiograms using simulated
electrophysiology data is also under development.

MFEM has also been applied to topology optimization for additive manufacturing (3D
printing) by LLNL's Center for Design and Optimization, which develops the Livermore
Design Optimization (LiDO) software. LiDO is used to solve challenging structural
engineering problems consisting of millions of design variables. See
\cite{2018-STR-LIDO} and \autoref{fig:cardioid}.
Another project that is built extensively around MFEM is ParElag \cite{parelag},
which is a library organized around the idea of algebraic coarsening of the de
Rham complex introduced in \autoref{ssec:derham}.  ParElag leverages MFEM's
high-order Lagrange, Nedelec, and Raviart-Thomas finite element spaces as well
as its auxiliary space solvers (\autoref{ssec:solvers}) to systematically
provide a de Rham complex on a coarse level, even for unstructured grids with no
geometric hierarchy.  Its algorithms and approach are described in
\cite{lashuk-vassilevski, pasciak-vassilevski}.

\section{Conclusions} \label{sec:conclusions}
In this paper we provided an overview of the algorithms, capabilities and
applications of the MFEM finite element library as of version \cred{4.1,
released in March 2020}. Our goal was to emphasize the mathematical ideas and
software design choices that enable MFEM to be widely applicable and highly
performant from a relatively small and lightweight code base.

While this manuscript covers all major MFEM components, it is really just an
introduction to MFEM, and readers interested in learning more should consult the
additional material available on the website \url{https://mfem.org} and in the
MFEM code distribution.

In particular, new users should start with the interactive documentation of the
example codes, available online as well as in the \code{examples/} directory,
and may be interested in reading some of the references in \autoref{sec:apps},
e.g. \cite{2018-SISC-ALE, 2018-SMO-LiDO, 2018-CompGeo-Reservoirs,
 2016-IJNMF-NonlocalTransport, 2018-SISC-Distance,
 2017-SISC-SpectralUpscaling}.

Researchers interested in learning mathematical details about MFEM's finite
element algorithms and potentially contributing to the library can follow up
with \cite{2018-AMR,2018-TMOP,2018-SISC-DPG,2018-Hybridization,2018-TMOP-ADAPT}
and the instructions/developer documentation in the \code{CONTRIBUTING.md} file.

\section*{Acknowledgments}
The MFEM project would not have been possible without the help and advice of
Joachim Sch\"oberl, Panayot Vassilevski, and all the contributors
in the MFEM open-source community, see \url{https://mfem.org/about} and
\url{https://github.com/orgs/mfem/people}.

MFEM has been supported by a number of U.S.\ Department of Energy (DOE) grants,
including the Applied Math and SciDAC programs in the DOE Office of Science, and
the ASC and LDRD programs in NNSA. MFEM is also a major participant in the
co-design Center for Efficient Exascale Discretizations (CEED) in the DOE's
Exascale Computing Project.

This work was performed under the auspices of the U.S. Department of Energy by
Lawrence Livermore National Laboratory under Contract DE-AC52-07NA27344,
LLNL-JRNL-795849.

This document was prepared as an account of work sponsored by an agency of the
United States government. Neither the United States government nor Lawrence
Livermore National Security, LLC, nor any of their employees makes any warranty,
expressed or implied, or assumes any legal liability or responsibility for the
accuracy, completeness, or usefulness of any information, apparatus, product, or
process disclosed, or represents that its use would not infringe privately owned
rights. Reference herein to any specific commercial product, process, or service
by trade name, trademark, manufacturer, or otherwise does not necessarily
constitute or imply its endorsement, recommendation, or favoring by the United
States government or Lawrence Livermore National Security, LLC. The views and
opinions of authors expressed herein do not necessarily state or reflect those
of the United States government or Lawrence Livermore National Security, LLC,
and shall not be used for advertising or product endorsement purposes.

\bibliographystyle{abbrv}
\bibliography{mfem-paper,mfem}

\end{document}